International

Virtual

Observatory

Alliance

# Space-Time Coordinate Metadata for the Virtual Observatory

# Version 1.33

## *IVOA Recommendation 30 October 2007*

**Author:**
A. H. Rots

## Abstract

This document provides a complete design description of the Space-Time Coordinate (STC) metadata for the Virtual Observatory. It explains the various components, highlights some implementation considerations, presents a complete set of UML diagrams, and discusses the relation between STC and certain other parts of the Data Model. Two serializations are discussed: XML Schema (STC-X) and String (STC-S); the former is an integral part of this Recommendation.

## Status of This Document

This document has been produced by the Data Model Working Group.
It has been reviewed by IVOA Members and other interested parties, and has been endorsed by the IVOA Executive Committee as an IVOA Recommendation. It is a stable document and may be used as reference material or cited as a normative reference from another document. IVOA's role in making the Recommendation is to draw attention to the specification and to promote its widespread deployment. This enhances the functionality and interoperability inside the Astronomical Community.

*A list of [current IVOA Recommendations and other technical documents](http://www.ivoa.net/Documents/) can be found at http://www.ivoa.net/Documents/.*

## Acknowledgements

I gratefully acknowledge many helpful discussions with Wil O'Mullane, Alex Szalay, Jonathan McDowell, Ray Plante, Ed Shaya, Pat Dowler, David Berry, Anita Richards, Steve Allen, Gerard Lemson, Tamas Budavari, François Bonnarel, and many others. I am indebted to SAO, ULP, CDS, and the NSF NVO grant for support.

## Contents

# 1 Introduction

This document attempts to explain and document the design and implementation of the Space-Time (and related coordinate axes) metadata for the Virtual Observatory. The concepts represented in this model are not particularly profound. We need to allow users to specify the spatial coordinates they work in; these will, for most astronomical users, be some flavor of equatorial coordinates. However, we also need to be cognizant of the many variations that exist, not only in terms of different equatorial systems, either from historical collections (FK1-4) or in the uses of Galactic or ecliptic coordinates, but also geographic, barycentric, planetocentric, and instrumental detector coordinates, most of them in spherical as well as Cartesian form. In addition, high-accuracy requirements and special situations such as spacecraft-based observatories create the need for specifying the origin of such coordinate frames – in most cases the location of the observatory. The same is true for time and spectral coordinates: for many applications it may not matter, but there are situations where it is crucial to know what timescale was used, where time was measured, or what inertial standard of rest was used to express the frequency.

What this amounts to is that for spatial coordinates we want to know the coordinate system (its type and orientation) and the origin that were used, for time the timescale (UTC, TT, TAI, TDB, etc.) and the spatial reference position, for the spectral coordinate the origin in phase space, and for redshifts (Doppler velocity)[1] the definition as well as the phase space origin. In the past, many of these particulars were obvious to the sub-communities in which data were collected and distributed, but in the context of the VO we cannot assume anymore that these "obvious defaults" are indeed obvious (or even known) and we have to be explicit about our default assumptions. The issue here is that no single set of defaults is intuitive anymore for the entire community of users; and defaults that are not intuitive are too dangerous to use. Nevertheless, there are a few common coordinate systems that will serve the bulk of our data and we will make an effort to make their use as simple as possible.

An added complication is that users need to be able to specify the transformation between coordinate systems, not only for custom systems, but also, for instance, between the sky and a detector, or between a world coordinate system and a pixel coordinate system, analogous to the FITS WCS standards.

Much of the seemingly complicated content in this design is necessitated by three requirements, but can be ignored for more traditional astronomical use. Those requirements are that it *(a)* support the inclusion of planetary data, *(b)* allow the definition of arbitrary coordinate orientations, and *(c)* be extensible.

In order to make this specification optimally useful for the VO, we have generalized the Coordinate System concept and we have added a generic Coordinate Frame. A Coordinate System consists of one or more Coordinate

[1] The term "Redshift coordinate" is used in this document in a generic sense, referring to measured red- and blueshifts as well as (positive and negative) Doppler velocities.

Frames. A Coordinate Frame is usually defined by a Coordinate Reference Frame (the "direction" of the axes, if you like) and a Coordinate Reference Position (the "origin"). Most Coordinate Frames will be one-dimensional, but Spatial Frames may have up to three dimensions and Pixel Frames may have up to three, while certain other generic frames may also be multi-dimensional (e.g., Visibility with amplitude/phase or sine/cosine). Derived from these are the AstroCoordSystem and the time, space, redshift, spectral frames that together make up the Space-Time Coordinate (STC) system. Coordinate and Coordinate Area also have generic as well as Astro versions. We shall use "STC" as a proper name in this document which will lead to terms like "STC coordinates". Although we agree that such usage is linguistically deplorable, we will use it nevertheless because of the clarity it provides.
We present two implementations of STC. The fundamental implementation is provided in the XML schema and is known as STC-X; this implementation is an integral part of this standard. A string implementation, based on STC-X, is known as STC-S. A Utypes implementation will be specified in a separate document.
There are six companion files to this document:

- http://www.ivoa.net/xml/STC/v1.30
  is the schema defining the XML implementation STC-X.
- http://www.ivoa.net/Documents/latest/STC-Model.html
  is an IVOA Note clarifying the STC model and may serve as a glossary.
- http://www.ivoa.net/Documents/latest/STC-X.html
  is an IVOA Note clarifying the STC-X (XML) implementation.
- http://hea-www.harvard.edu/~arots/nvometa/STC/
  contains more detailed documentation on the STC schema.
- http://www.ivoa.net/Documents/latest/STC-S.html
  is an IVOA Note clarifying the STC-S (string) implementation.
- http://www.ivoa.net/xml/Xlink/xlink.xsd
  is the standard Xlink schema.

The schema is fully up-to-date.
In the following sections we shall discuss the justification and scope for this metadata specification, the requirements and use context, a detailed design description, and some implementation considerations. Some concepts are only provided in their simplest form and we expect that more sophisticated classes will be derived (and will fit in seamlessly) as work progresses. The emphasis of this document is on the framework of the STC metadata.

# 2 Justification and Scope

Before we plunge into the requirements and design of these metadata structures, it may be helpful to discuss what is driving this exercise and to provide some context to the design. Our intent in this is to explain why the design needs to be fully general and extensible, leading to a certain level of complexity.

The basic justification for considering the following coordinate axes in a single metadata specification is that they are closely intertwined:

- Space (including its time-derivatives, i.e., velocities)
- Time
- Spectral (frequency, wavelength, energy)
- Redshift (Doppler velocity)

It should be emphasized that the STC design distinguishes between the spectral coordinate and the redshift coordinate (one might, after all, have to deal with a dataset containing Doppler velocities measured in different parts of the spectrum) and between redshifts and velocities (the latter are physical quantities – derivatives of spatial positions; the former represent a spectral property that is transformed via one of several formalisms to what is commonly referred to as a Doppler velocity – when we get to relativistic situations the connection with physical velocities becomes tenuous). In the presence of a redshift coordinate, the values of a spectral coordinate, if present, will contain rest frequencies. Please note that the term *Redshift* is to be taken generically and really stands for *Dopplershift*.
We shall refer to these as the *STC coordinates*. The issue is that time is bound to a position and positions are time-variable. Similarly, spectral and redshift data are tied to reference frames that may or may not be time-variable.

In the past, most data archives have not been extremely concerned with dotting every *i* and crossing every *t*, in a meticulous specification of all the details of all coordinate axes. They assumed sets of coordinate system-related defaults. Such context-dependent defaults are quite acceptable: issues are generally well-defined, obvious, and clear for single-observatory observations, even when not all is explicitly specified. However, there are no global defaults in the VO. All implicit assumptions need to be made explicit since they will not be "obvious" anymore. One must be able to transform the coordinates of two observations to a common coordinate system; this means that every little tidbit of information needs to be documented in the metadata.

Here are some simple examples of situations and questions that need an answer, when such context-dependent defaults do not work anymore:

- Are Doppler velocities optical or radio?
- What is the spatial reference position for the time stamps; if corrected for pathlength, what direction was assumed?
- Does "J2000" mean "FK5/J2000" or "ICRS"?
- At which spatial position was this observation made?
- How can we get information on Galactic background objects from solar system observations, and vice versa?

Reality dictates, though, that we allow for cases where these values are simply unknown. In the XML implementation *nillable* elements that are nil should be interpreted as unknown values; the burden is then on the client to determine and adopt a sensible default value. This holds true for elements named “UNKNOWN” as well. We sincerely hope that the existence of such *nillable* and UNKNOWN elements will not be an excuse to provide incomplete metadata in cases where the values are known.
This document defines the metadata items (and their structure) that specify coordinate properties in the VO. It identifies four major contexts in which these metadata play a crucial role and it shows how the metadata description can be constructed in these contexts, in a form that is preferred from the perspective of STC. However, the precise syntax of the incorporation of the STC metadata in the various VO interfaces and applications is left to their controlling documents.

# 3 Requirements and Use-Context

The top-level requirement for the design of the Space-Time Coordinate metadata can be formulated in a very simple way. It is that they:

- provide **sufficient and necessary** information
- are **self-consistent**

That is really all. However, to fit it into the larger VO Data Model context, we should leverage two more requirements:

- the system should be extensible to include other coordinates and accommodate coordinate metadata descriptions that include axes other than the STC ones
- it must be possible to define relocatable frames to accommodate, for instance, simulation data

From this one can derive secondary requirements: the metadata must include complete descriptions of the following three concepts:

- the coordinate frames that anchor the STC coordinate axes, collectively referred to as the *Coordinate System*
- a specific position in these frames, i.e., a coordinate object, that includes not only coordinate values, but also essential attributes such as errors and resolutions
- the volume in coordinate space that is taken up by the data object that these metadata are attached to

In all, there are four different contexts for the use of the STC metadata:

- STC Resource profile – to specify the spatial, temporal, redshift, spectral coverage of a data collection which is ultimately used by clients to determine whether or not the resource is of interest for their purposes
- Query – to specify the STC constraints of a particular request
- Catalog information – clearly, the response to a catalog query may contain coordinate information (e.g., positions), but it should also contain coverage information: specifically, the intersection of the catalog's coverage and the query's STC constraints
- Observational data – here we need the volume in coordinate space that is occupied by the observational data, as well as the coordinates of the observatory; note that these may not necessarily be given in the same Coordinate System; a Pixel Coordinate System is needed for pixelated data

We will return to the use of STC in these contexts in Section 7. As commented before, the emphasis of this document is on the structure of the STC metadata. Hence, though the classes and attributes are laid out in detail, little is said about methods.

Notwithstanding our emphasis on the necessity of the completeness of STC metadata, it will be possible for servers to provide incomplete metadata. The universal default rule in STC is:

> **Any missing component in STC metadata shall be considered to have the value `UNKNOWN`. It is up to the *client* to decide whether it is willing to accept such metadata and to assign a suitable default value.**

For detailed implementation requirements, see Section 6.1.

# 4 Design

It may be helpful to refer to the UML diagrams in Appendix A: UML Representation, while reading this section. The detailed implementation requirements that follow from this design are presented in Section 6.1.
An important part of the XML implementation is the referencing mechanism; it is described in Section 6.2.1. In addition, all entities described in this section, and their individual components, MAY optionally contain a UCD.

## *4.1 The Metadata Components*

There are three metadata components in the design, as already noted in the derived requirements:

- **Coordinate system**
  This object contains the coordinate frames that together define a complete coordinate system. There may be one or more per document.
- **Coordinate values**
  This is an object that specifies a specific position in the space defined by the coordinate system it refers to. The object contains values, units, errors, resolution, size, and pixel size, though not all of these need be present.
- **Coordinate areas or ranges**
  This is an object that defines a volume in the coordinate space specified by the coordinate system it refers to. Regions are a special case, specifically designed for 2-D (and 3-D unit-sphere) spatial coordinates.

## *4.2 Generic Coordinates*

Generic coordinates are primarily intended for all coordinate axes that are not temporal, spatial, spectral, redshift, or pixel-like.

### 4.2.1 Coordinate System (CoordSys)

The CoordSys class is nothing more than a collection of one or more Coordinate Frames, where each frame describes a set of one or more coordinate axes that belong together. A CoordSys MUST contain at least one Coordinate Frame.

#### 4.2.1.1 Coordinate Frame (CoordFrame)

A CoordFrame object provides the metadata about one or more coordinate axes. Typically, it consists of a reference frame and a reference position, as well as a coordinate “flavor”, in particular for multi-dimensional frames. Usually, we will be dealing with one axis. The three exceptions are the spatial frame (which may control up to three position coordinate axes, as well as their time-derivatives – velocity), the pixel frame (which may, to support these spatial axes, also contain up to three axes), and certain generic axes (e.g., amplitude and phase, or polarization percentage and angle). In the XML schema CoordFrame is really the base class for coordinate frames. To ensure its full generality, this base class should only contain a name. Real-life coordinate frames should be derived from this class. If the name defines the coordinate unambiguously (e.g., “FluxDensity”), no further specification may be needed. A frame_id is required for all but the astronomical frames, in order to allow tying coordinates to frames.
If the Coordinate Frame is defined as a transformation from a known coordinate system, that transformation is specified through the Coordinate Reference Frame and the Coordinate Reference Position. The frame’s structure is defined by the Coordinate Flavor.

#### 4.2.1.2 Coordinate Reference Frame (CoordRefFrame)

The reference frame relates the Coordinate Frame to a known frame. There are three basic types of reference frames: scalar, Cartesian (2-D or 3-D), and spherical.

##### *4.2.1.2.1 Scalar Reference Frame*

This is the simplest case, only containing a scale factor and a projection. The projection may be linear or logarithmic.

#### *4.2.1.2.2 Cartesian Reference Frame*

The 2-D or 3-D Cartesian reference frame may specify a projection and a transform. The latter may consist of a set of scale-and-rotate parameters or a proper transformation matrix. The projections include a list of spherical-to-Cartesian projections (see Table 4).

#### *4.2.1.2.3 Spherical Reference Frame*

The transformations for spherical coordinate systems are specified through a pole and an X-axis direction, expressed as coordinates in another, known spherical coordinate frame.

### 4.2.1.3 Coordinate Reference Position (CoordRefPos)

The Coordinate Reference Position specifies the origin of the Coordinate Frame and thereby completes the transformation.

### 4.2.1.4 Coordinate Flavor (CoordFlavor)

The Coordinate Flavor specifies the number of axes in the coordinate frame (1, 2, or 3) and the structure of the coordinate system: Cartesian, Spherical, Polar, UnitSphere, or Healpix[xi]. The Healpix specification includes the parameters H and K (with default values 4 and 3, respectively). Handedness ("left" or "right") is well defined for known coordinate frames, but is an optional attribute for the benefit of generic and custom frames.

## 4.2.2 Coordinates (Coords)

The Coords object specifies a particular position in coordinate space. It requires a reference (IDREF in the XML implementation) to a CoordSys object and it is made up of one or more Coordinate objects, corresponding to the Coordinate Frames defined in the Coordinate System. All objects are optional. However, in order to constitute a meaningful Coords object, one would expect at least one of the Coordinate objects to be present.

### 4.2.2.1 Coordinate

A Coordinate is an aggregate of up to 6 components: CoordName, CoordValue, CoordError, CoordResolution, CoordSize, and CoordPixSize. Of these, only CoordName is a member of the Coordinate base class. The others, which in their simplest forms consist of a scalar numerical Value, are all optional, though one may doubt whether a Coordinate object that contains none of these has any meaning. Although at first sight it would seem that at least CoordValue is indispensable, we shall show that there are legitimate cases where the value may be absent. Coordinate has a single Unit attribute that applies to all components, as well as optional Unit attributes for the individual components that would override the global one. Classes that are derived from Coordinate may differ in the data types of the objects, their structure and meaning, as well as restrict allowable values for Unit. There are currently two derived classes that have only a Name and a Value: StringCoordinate and PixelCoordinate.
The numeric elements, other than the Coordinate Value, may appear singly or in pairs. In the former case it will be obvious that a single specific value is indicated. In the latter case the pair indicates a range of values.
A generic Coordinate SHALL contain a reference to a specific Coordinate Frame. A numeric Coordinate may be 1-, 2-, or 3-dimensional; all components of a Coordinate object SHALL be consistent in their dimensionality.

#### 4.2.2.1.1 Name

This member of Coordinate is a simple string that acts as a label; usage should indicate whether this should be restricted to an enumerated list in a future version.

#### 4.2.2.1.2 Value

The value of the Coordinate consists of a scalar numerical, a 2- or 3-dimensional vector, or a string value and a Unit string.

#### 4.2.2.1.3 Error

Again, this consists of a scalar Value for scalar coordinates, representing an rms[2] error. However, we expect a variety of derived classes to appear with more sophisticated error descriptions. For multi-dimensional coordinates more information is required: error circle, error ellipse, error matrix, and their 3-D equivalents.

#### 4.2.2.1.4 Resolution

The resolution along the Coordinate is expressed as a FWHM[3] Value. In cases where this is a poor description, a more sophisticated class should be derived. For multi-dimensional coordinates more information is required: circle, ellipse, matrix, and their 3-D equivalents.

#### 4.2.2.1.5 Size

The size along the Coordinate is also expressed as a FWHM[3] Value, though one might expect derived classes with more precise definitions (e.g., Holmberg[4] diameter). This object is used in three contexts: in a resource profile it indicates the typical size (e.g., FOV[5]) of datasets in the resource; in queries it serves the same purpose; and in catalog datasets it allows the server to provide Coordinate positions of objects as well as their sizes. For multi-dimensional coordinates more information is required: circle, ellipse, matrix, and their 3-D equivalents.

[2] Root-mean-square
[3] Full Width at Half Maximum
[4] A photometric measure of the diameter of galaxies
[5] Field Of View

##### *4.2.2.1.6 PixSize*

The pixel size (scalar Value) is often a useful figure to characterize resources, queries, and datasets, even if it only provides a typical value. It is intended to be a quick characterization and should not replace a proper PixelCoords and PixelCoordSys pair that rigorously defines the transformation between world coordinates and pixel space. For multi-dimensional coordinates more information is required: pixel size vectors, position angles, CD matrix.

### 4.2.3 Coordinate Area (CoordArea)

The Coordinate Area class defines the volume in coordinate space that is occupied by the object it is attached to. It is an aggregate of one or more ranges in individual coordinates and is required to have a reference (IDREF in the XML implementation) to a CoordSys. Multiple ranges along each coordinate axis are OR-ed together; the ranges of different axes are logically AND-ed together.

#### 4.2.3.1 ScalarInterval

A ScalarInterval consists of a LoLimit and/or a HiLimit, both of which contain a double precision value and a unit. Note that they do not both have to be present, but at least one of them needs to be. If only one is present, an upper or lower limit (rather than an interval) is indicated. ScalarInterval has Boolean attributes LoInclude and HiInclude and a FillFactor attribute with a value between 0 and 1 that indicates what fraction of the interval is actually covered by the data.

#### 4.2.3.2 Multi-dimensional Areas

Rectangles or boxes are the equivalent of the Scalar Interval in two and three dimensions. The rules and properties are the same. In addition, Spheres are added in three dimensions and Regions in two. The latter are described in more detail in Section 4.5.

## *4.3 Pixel Coordinates*

The three top-level classes (PixelCoordSys, PixelCoords, PixelCoordArea) are derived from their generic counterparts and therefore contain all the (generic) elements enumerated in the previous section. In addition, they contain what is described below.

### 4.3.1 Pixel Coordinate System

The Pixel Frames in PixelCoordSys are like their generic numeric counterparts (1-D, 2-D, or 3-D; see Section 4.2.1.1), with the addition of two important components: axisorder (for 2-D and 3-D) and ReferencePixel. The former is required because it will not be a priori obvious what the order of the pixel axes is, even though there is a link through the frame_id. The Reference Pixel is required since pixel frames typically do not put their origin at zero.
Note that a PixelCoordSys is a collection of Pixel Frames, where the dimensionality needs to correspond with the number of frames and their dimensionality in physical space. The sum of all Pixel Frame axes should equal the dimensionality of the associated pixel array. For instance, a three-axis pixel grid mapping RA, Dec, and Doppler velocity should be built up from two frames: a 2-dimensional frame that corresponds with the 2-D spatial frame, and a scalar frame that corresponds with the redshift coordinate; the axis order attributes tell which one is which.

### 4.3.2 Pixel Coordinates

Pixel coordinates are simpler than others in that they only have a name and a value, and do not have units associated with them.

### 4.3.3 Pixel Coordinate Area

The Pixel Coordinate Area has the same properties as the generic Coordinate Area (see Section 4.2.3).

## *4.4 Astronomical Coordinates*

The three top-level classes (AstroCoordSystem, AstroCoords, AstroCoordArea) are derived from their generic counterparts and therefore may contain all the (generic) elements enumerated in Section 4.2. In addition, they contain what is described below.

### 4.4.1 AstroCoordSystem

The AstroCoordSystem is derived from CoordSys. It still is a collection of coordinate frames, but it SHOULD contain one or more specific frame classes derived from the CoordFrame base class, in addition to zero or more additional generic frames. These special frames are TimeFrame, SpaceFrame, SpectralFrame, and RedshiftFrame. See Appendix C.1 for examples.

#### 4.4.1.1 TimeFrame

The Time Frame contains two or three objects: a reference position, a time scale (the equivalent of a reference frame), and, optionally, a time reference direction.

#### 4.4.1.1.1 ReferencePosition

The Reference Coordinate Position is similar to the ones defined for generic coordinates (see Section 4.2.1.3), but in addition we have defined a number of Standard Reference Positions which have special meaning in an astronomical context.

The reference position, or spatial coordinate origin, may be chosen from a standard list of such origins (StdRefPosition) or specified as a particular coordinate position (CoordOrigin). In the latter case, one may nest the origins, but eventually they should be referenced to a known standard origin. The list of standard origins includes (see Table 1): `GEOCENTER, BARYCENTER, HELIOCENTER, TOPOCENTER, LSR, GALACTIC_CENTER, LOCAL_GROUP_CENTER, EMBARYCENTER, MOON, MERCURY, VENUS, MARS, JUPITER, SATURN, URANUS, NEPTUNE, PLUTO, RELOCATABLE`; additional values (e.g., planetary satellites, non-planet solar system objects) are allowed, provided they are identified in a referenced ephemeris or other authoritative source (see: Section 4.4.1.1.4). Ultimately, one must be able to tie the coordinate origin down to the geocenter, be it through a geographic position, an IAU resolution, an orbit ephemeris, or a solar system ephemeris – unless one deals with simulation data that have a `RELOCATABLE` origin. Planetary Ephemeris is required for any position related to a solar system entity other than the geocenter. ReferencePosition is a standard class that is being used for all four STC frames, though there are different restrictions for the four usages. For the Time Frame `LSRK`, `LSRD`, `GALACTIC_CENTER`, `LOCAL_GROUP_CENTER`, and `RELOCATABLE` are not allowed values; for simulations one may want to use a custom position (or "`LOCAL`") at the origin of the relocatable spatial frame. `TOPOCENTER` refers to the location of the observatory, if appropriate. If there is no such location defined then a `TOPOCENTER` time reference position refers to the spatial reference position.

Table 1. Standard Reference Positions

| Reference Position | Description | Comments |
|---|---|---|
| GEOCENTER | Center of the earth | |
| BARYCENTER | Center of the solar system barycenter | |
| HELIOCENTER | Center of the sun | |
| TOPOCENTER | "Local"; in most cases this will mean: the location of the telescope | |
| LSR or LSRK | Kinematic Local Standard of Rest: 20 km $s^{-1}$ in the direction of GALACTIC_II(56,+23) | Only to be used for redshifts and Doppler velocities, and spectral coordinate |

| Reference Position | Description | Comments |
|---|---|---|
| LSRD | Dynamic Local Standard of Rest: 16.6 km $s^{-1}$ in the direction of GALACTIC_II(53,+25) | Only to be used for redshifts and Doppler velocities, and spectral coordinate |
| GALACTIC_CENTER | Center of the Galaxy: 220 km $s^{-1}$ in the direction of GALACTIC_II(90,0) wrt LSRD | |
| LOCAL_GROUP_CENTER | Center of the Local Group: 300 km $s^{-1}$ in the direction of GALACTIC_II(90,0) wrt BARYCENTER | Only to be used for redshifts and Doppler velocities, and spectral coordinate |
| EMBARYCENTER | Earth-moon barycenter | |
| MOON | Center of the moon | |
| MERCURY | Center of Mercury | |
| VENUS | Center of Venus | |
| MARS | Center of Mars | |
| JUPITER | Center of Jupiter | |
| SATURN | Center of Saturn | |
| URANUS | Center of Uranus | |
| NEPTUNE | Center of Neptune | |
| PLUTO | Center of Pluto | |
| RELOCATABLE | Relocatable center; for simulations | Only to be used for spatial coordinates |
| UNKNOWNRefPos | Unknown reference position | Only to be used as a last resort.<br>The client is responsible for assigning a suitable default |

#### *4.4.1.1.2 TimeScale*

The time scale must be chosen from among the list recognized by the IAU (see Table 2): `TT`, `TDT`, `ET`, `TAI`, `IAT`, `UTC`, `GPS`, `TDB`, `TEB`, `TCG`, `TCB`, `LST`. `TT` is the default; `TDT` and `ET` are obsolete synonyms for `TT`; `IAT` is an unofficial synonym for `TAI`; and use of `LST` is to be discouraged. At some point other time scales may need to be added as they become recognized, such as planet- or moon-based time scales. In most cases where `TEB` is specified, the likely intent was `TDB`, so `TEB` may be considered a synonym of `TDB`; its use requires the presence of a Planetary Ephemeris. For further details, see Seidelmann & Fukushima (1992)[i] and Standish (1998)[ii].
We will allow the value `LOCAL`, only to be used with `RELOCATABLE` space frames (see: Section 4.4.1.1.1), specifically for simulations.
For a handy explanation of time scales, see:
http://tycho.usno.navy.mil/systime.html

For a glossary of fundamental astronomy terms, see:
http://syrte.obspm.fr/iauWGnfa/NFA5_B.html
In the XML implementations the Timescale element is *nillable*.

Table 2. Timescales

| Timescale | Description | Comments |
|---|---|---|
| TT | Terrestrial Time | |
| TDT | Terrestrial Dynamic Time; synonym for TT | |
| ET | Ephemeris Time: predecessor of, and continuous with, TT | |
| TAI | International Atomic Time; 32.184 s behind TT | |
| IAT | Synonym for TAI | |
| UTC | Coordinated Universal Time; 32 s behind TAI in 2000-2005 | Includes leap seconds<br>Pre-1972 times will be assumed to be UT/GMT |
| GPS | Global Positioning System's time scale; 19 s behind TAI, 51.184 s behind TT | This time scale may become important in the future |
| TDB | Barycentric Dynamical Time; synchronous with TT, except for variations in earth's orbital motion | Requires specification of the solar system and planetary ephemeris used |
| TEB | Barycentric Ephemeris Time; independent variable in solar system ephemeris, linear function of TT | In most cases where TEB is specified, TDB is really the one used |
| TCG | Geocentric Coordinate Time; properly relativistic time, running a factor $7 \cdot 10^{-10}$ faster than TT | |
| TCB | Barycentric Coordinate Time; properly relativistic time, running a factor $1.5 \cdot 10^{-8}$ faster than TDB | |
| LST | Local Siderial Time | Ground-based observations only |
| LOCAL | "Local" time | Only to be used for simulations, in conjunction with RELOCATABLE spatial coordinates |

#### 4.4.1.1.3 TimeRefDirection

If the Time Frame's Reference Position is not `TOPOCENTER`, times have clearly been transformed from observed time at the location where the observation was made to some other spatial location. In order to effect such a transformation, a direction of origin must have been assumed for the observed phenomenon; that direction is given in TimeRefDirection.

#### *4.4.1.1.4 Planetary ephemeris (if needed)*

The planetary ephemeris object PlanetaryEphem indicates which solar system ephemeris was used in the AstroCoordSystem. In general, this will be JPL's DE200 or DE405 (default)[iii]. It should only be present if a solar system ephemeris was used, for instance because of a planet-based ReferencePosition or the application of barycenter corrections. In addition, the information in Sections 7.2-7.5 of the *Explanatory Supplement to the Astronomical Almanac*[iv] may be considered a known authoritative source for the purpose of spatial reference frames. If the SpaceRefFrame (see: Section 4.4.1.2.3) is defined on a body, ephemeris information on its coordinate system (pole, primary meridian) must be included in the PlanetaryEphem.
In the XML implementations this element is *nillable*.

### 4.4.1.2 SpaceFrame

The Space Frame contains a reference position, a reference frame, and a coordinate flavor object. Simulations may want to use a `RELOCATABLE` reference position in combinations with an `UNKNOWNFrame` reference frame.

#### *4.4.1.2.1 ReferencePosition*

This object is of the same class as described in Section 4.4.1.1.1, with the following restrictions. The value `RELOCATABLE` is allowed, but `LSRD`, `LSRK`, `GALACTIC_CENTER`, and `LOCAL_GROUP_CENTER` are not. Obviously, if an explicit coordinate origin is provided, those coordinates should be referenced to a different CoordSys; one may still nest them, but eventually they should be referenced to a known standard origin.

#### *4.4.1.2.2 CoordFlavor*

This class is especially relevant for multi-dimensional coordinates, in particular spatial coordinates and pixel coordinates. It indicates the intrinsic dimensionality of the coordinate system (1, 2, or 3), and the type of coordinates: `CARTESIAN`, `SPHERICAL`, `UNITSPHERE`, `POLAR`, `CYLINDRICAL`, or `HEALPIX`. For 3-dimensional `SPHERICAL` (because of geographical applications) we allow the special Units value "deg deg m", although the preferred way is to assign separate units to the individual vector components. The UnitSphere has default units "".
This list will likely be expanded with, for instance, cylindrical coordinates.
CoordFlavor may also be used to identify string coordinates.

#### *4.4.1.2.3 SpaceRefFrame*

While the Reference Position pins down the origin of the spatial coordinate system, the Space Reference Frame specifies its orientation. In most cases this will be a StdFrame, consisting of a RefSystem (`FK4, FK5, ECLIPTIC, ICRS, GALACTIC_I, GALACTIC_II, SUPER_GALACTIC, AZ_EL, BODY`, etc.; the first three require a CoordEquinox, which is *nillable* in the XML implementation). The Standard Frame may be `BODY` if the Reference Position is specified as a solar system object. In addition, we allow solar, lunar, planetary, and planetary satellite coordinate frames (planetocentric and planetographic) as defined by Sections 7.2-7.5 of the *Explanatory Supplement to the Astronomical Almanac* (Seidelmann, 1992)[iv]; see also Fränz & Harper (2002)[v] and Thompson (2006)[vi]. For the most current update of cartographic coordinates and rotation rates, see Seidelmann *et al.* (2002)[vii].
A full list of Standard Reference Frames is provided in Table 3.
Alternatively, a CustomFrame may be specified through a transformation from an existing frame. In the case of spherical coordinates this might be done by way of the pole (Z) axis and the longitude zero-point (X-axis) directions. For Cartesian systems one may specify scale and (optionally) rotation angle, or a rotation matrix. The transformation includes the ability to specify a projection and (for pixel frames) the ordering of the axes. The allowable projection types are listed in Table 4.

We want to provide some caveats on subtle differences between various coordinate systems that may look similar.
First, one should note that the 2-dimensional spherical versions of all celestial coordinate systems are left-handed, but that their 3-dimensional Cartesian versions are right-handed.
Second, because of the definition of longitude (IAU 1970), following astronomical tradition, longitude increases westward in the *planetographic* coordinate frames for Mercury, Mars, Jupiter, Saturn, and Neptune (`MERCURY_G`, `MARS_G`, `JUPITER_G`, `SATURN_G`, and `NEPTUNE_G` – ignoring planetary satellites), since their rotation is direct (prograde); hence, these coordinate systems are also left-handed. Note that earth, sun, and moon are exempted from this rule because of even older conventions. In addition, many coordinate systems that are based on video devices are left-handed. All other coordinate systems (including all *planetocentric* systems) are right-handed.
Finally, beware that planeto*graphic* and planeto*centric* coordinates may differ in three respects:

- For some, the longitudinal directions are opposite
- Planeto*centric* latitude is based on the vector from the planet's center, while planeto*graphic* latitude is measured with respect to the local vertical (and hence a reference spheroid needs to be adopted)
- Planeto*centric* coordinates are longitude, latitude, and radius; planeto*graphic* coordinates are longitude, latitude, and elevation (with respect to the reference spheroid)
- The origin of planeto*graphic* longitude is tied to a particular feature on the surface; planeto*centric* longitude's origin is defined purely in terms of the rotational ephemeris of that feature, which may result in slight differences.

For further details, see Seidelmann (1992, Sections 4.2 and 7)[iv], Seidelmann *et al.* (2002)[vii], and Duxbury *et al.* (2002)[viii].

Table 3. Standard Reference Frames

| Reference Frame | Description | Comments |
|---|---|---|
| FK4 | Fundamental Katalog, system 4; Besselian | Requires Equinox; default B1950.0<br>Left-handed in spherical coordinates |
| FK5 | Fundamental Katalog, system 5; Julian | Requires Equinox; default J2000.0<br>Left-handed in spherical coordinates |
| ECLIPTIC | Ecliptic coordinates | Left-handed in spherical coordinates |
| ICRS | International Celestial Reference System | Left-handed in spherical coordinates |
| GALACTIC_I | Old Galactic coordinates | Left-handed in spherical coordinates |
| GALACTIC[_II] | "New" Galactic coordinates | Left-handed in spherical coordinates |
| SUPER_GALACTIC | Super-galactic coordinates:<br>pole at GALACTIC_II (47.37,+6.32)<br>origin at GALACTIC_II (137.37,0) | Left-handed in spherical coordinates |
| AZ_EL | Local azimuth and elevation | Ground-based observatories<br>Azimuth: from north through east |
| BODY | Generic "BODY" coordinates | |
| GEO_C | Geographic (geocentric) coordinates: longitude, latitude, geocentric distance | 3-D spherical or 3-D Cartesian |
| GEO_D | Geodetic coordinates: longitude, latitude, elevation | Semi-major axis and inverse flattening of the reference spheroid may need to be provided; default is IAU 1976 (6378140 m, 298.2577) |
| MAG | Geomagnetic coordinates | See F&H (2002) |
| GSE | Geocentric Solar Ecliptic coordinates | See F&H (2002) |
| GSM | Geocentric Solar Magnetic coordinates | See F&H (2002) |
| SM | Solar Magnetic coordinates | See F&H (2002) |
| HGC | Heliographic coordinates (Carrington) | See Explanatory Supplement, Section 7.2<br>Thompson (2006), Section 2.2 |
| HGS | Heliographic coordinates (Stonyhurst) | See Explanatory Supplement, Section 7.2<br>Thompson (2006), Section 2.2 |
| HEEQ | Heliographic Earth Equatorial coordinates | See F&H (2002); related to Heliographic (Stonyhurst), see Thompson (2006), Section 2.1 |

| Reference Frame | Description | Comments |
|---|---|---|
| HRTN | Heliocentric Radial-Tangential-Normal coordinates | See F&H (2002) |
| HPC | Helioprojective Cartesian coordinates | See Thompson (2006), Section 4.1, 2- or 3-dimensional (angular coordinates); left-handed |
| HPR | Helioprojective Polar coordinates | See Thompson (2006), Section 4.1, 2-dimensional (angular coordinates); left-handed |
| HCC | Heliocentric Cartesian coordinates | See Thompson (2006), Section 3.1 (linear coordinates); right-handed |
| HGI | Heliographic Inertial coordinates | See F&H (2002) |
| MERCURY_C | Planetocentric coordinates on Mercury | See Explanatory Supplement, Section 7.4 |
| VENUS_C | Planetocentric coordinates on Venus | See Explanatory Supplement, Section 7.4 |
| LUNA_C | Selenocentric coordinates | See Explanatory Supplement, Section 7.3 |
| MARS_C | Planetocentric coordinates on Mars | See Explanatory Supplement, Section 7.4 |
| JUPITER_C _III | Planetocentric coordinates on Jupiter, system III | See Explanatory Supplement, Section 7.4 |
| SATURN_C _III | Planetocentric coordinates on Saturn, system III | See Explanatory Supplement, Section 7.4 |
| URANUS_C _III | Planetocentric coordinates on Uranus, system III | See Explanatory Supplement, Section 7.4 |
| NEPTUNE_C _III | Planetocentric coordinates on Neptune, system III | See Explanatory Supplement, Section 7.4 |
| PLUTO_C | Planetocentric coordinates on Pluto | See Explanatory Supplement, Section 7.4 |
| MERCURY_G | Planetographic coordinates on Mercury | See Explanatory Supplement, Section 7.4 Left-handed |
| VENUS_G | Planetographic coordinates on Venus | See Explanatory Supplement, Section 7.4 |
| LUNA_G | Selenographic coordinates | See Explanatory Supplement, Section 7.3 |
| MARS_G | Planetographic coordinates on Mars | See Explanatory Supplement, Section 7.4 Left-handed |
| JUPITER_G _III | Planetographic coordinates on Jupiter, system III | See Explanatory Supplement, Section 7.4 Left-handed |
| SATURN_G _III | Planetographic coordinates on Saturn, system III | See Explanatory Supplement, Section 7.4 Left-handed |

| Reference Frame | Description | Comments |
|---|---|---|
| URANUS_G _III | Planetographic coordinates on Uranus, system III | See Explanatory Supplement, Section 7.4 |
| NEPTUNE_G _III | Planetographic coordinates on Neptune, system III | See Explanatory Supplement, Section 7.4<br>Left-handed |
| PLUTO_G | Planetographic coordinates on Pluto | See Explanatory Supplement, Section 7.4 |
| UNKNOWNFrame | Unknown reference frame | Only to be used as a last resort or for simulations<br>The client is responsible for assigning a suitable default |

Table 4. Projection Types

| Projection Code | Description |
|---|---|
| "" (blank) | Planar (i.e., linear cartesian-to-cartesian) projection |
| LOG | Linear-to-logarithmic cartesian-to-cartesian projection |
| TAN | Tangent plane projection |
| SIN | Sine projection |
| STG | Stereographic projection |
| ARC | Zenithal equidistant projection |
| ZEA | Zenithal equal-area projection |
| AIR | Airy projection |
| CEA | Cylindrical equal-area projection |
| CAR | Plate Carree projection |
| MER | Mercator projection |
| SFL | Sanson-Flamsteed projection |
| PAR | Parabolic projection |
| MOL | Mollweide projection |
| AIT | Hammer-Aitoff projection |
| COE | Conic equal-area projection |
| COD | Conic equidistant projection |
| COO | Conic orthomorphic projection |
| BON | Bonne equal-area projection |
| PCO | Polyconic projection |

| Projection Code | Description |
|---|---|
| TSC | Tangential spherical cube projection |
| CSC | COBE quadrilateralized spherical cube projection |
| QSC | Quadrilateralized spherical cube projection |

##### 4.4.1.2.4 OffsetCenter

In order to accommodate positions that are given as offsets in Right Ascension and Declination from some specific position, the optional OffsetCenter object is provided. Note that this is purely for numeric offsets and values are not corrected for cos(dec). True angle offsets should be handled through a CustomFrame.

##### 4.4.1.2.5 Position Angles

Position angle definitions can be a great source of confusion. To reduce ambiguity, we will attach a Reference attribute to position angles that may have the values “North”, “X” (default), or “Y”. If the reference is “North”, position angles will be counted from north through east; this should only be used with coordinate systems where “North” has meaning. If the reference is “X” (the default), the position angle will be counted from the positive first coordinate axis through the positive second coordinate axis. If the reference is “Y”, the position angle will be counted from the positive second coordinate axis through the positive first coordinate axis. These definitions are independent of the handedness of the coordinate system. For spherical systems, the longitude is considered the first axis in this context, and latitude the second. Note that in most (though not all) cases “North” is equivalent to “Y”.

#### 4.4.1.3 SpectralFrame

The Spectral Frame only contains a reference position.

##### 4.4.1.3.1 ReferencePosition

This object is of the same class as described in Section 4.4.1.1.1, without any restrictions.

#### 4.4.1.4 RedshiftFrame

The Redshift Frame contains two objects: a reference position and a Doppler definition object.

#### 4.4.1.4.1 ReferencePosition

This object is of the same class as described in Section 4.4.1.1.1, without any restrictions.

#### 4.4.1.4.2 DopplerDefinition

This object specifies what the definition of redshift is and how it should be translated to Doppler velocity. Allowed values are `OPTICAL`, `RADIO`, and `RELATIVISTIC`.

`OPTICAL`: $$V_{opt} = c \cdot \frac{\Delta\lambda}{\lambda_0} = -c \cdot \frac{\Delta\nu}{\nu}$$

`RADIO`: $$V_{rad} = -c \cdot \frac{\Delta\nu}{\nu_0} = c \cdot \frac{\Delta\lambda}{\lambda}$$

`RELATIVISTIC`: $$V_{rel} = c \cdot \frac{\lambda^2 - \lambda_0^2}{\lambda^2 + \lambda_0^2} = -c \cdot \frac{\nu^2 - \nu_0^2}{\nu^2 + \nu_0^2}$$

It should be emphasized that Doppler velocities are formal velocities; i.e., defined by a formalism and not necessarily physical in nature. One should also note here that `RELATIVISTIC` is not strictly correct; we expect in a future version to move toward compliance with the third FITS WCS paper[ix].
An attribute indicates what the type of the values is, `VELOCITY` or `REDSHIFT`.
We strongly advise against using `REDSHIFT` with anything but `OPTICAL`.
In the XML implementations this element is *nillable*.

### 4.4.2 AstroCoords

The AstroCoords object specifies a particular position in STC coordinate space. It requires a reference to an AstroCoordSystem object and it is made up of 1 to 5 Coordinate objects, at most one each of the following:

- Time of class TimeCoord
- Position of class SpatialCoord
- Velocity of class SpatialCoord
- Redshift of class RedshiftCoord
- Spectrum of class SpectralCoord

Note that all objects are optional. However, in order to constitute a meaningful AstroCoords object, one would expect at least one of them to be present.
The four specialized STC Coordinate classes, TimeCoord, SpatialCoord, RedshiftCoord, and SpectralCoord, are all derived from a generic base-class Coordinate (see: Section 4.2.2.1). Instead of providing the actual coordinate elements, the object may refer to a FITS file that provides some or all of them.
Also note that if both a Redshift coordinate and a Spectral coordinate are present, the latter will contain rest frequency values.

#### 4.4.2.1 TimeCoord

The error, resolution, size, and pixel size components of this class (derived from Coordinate) are fully generic, i.e., consisting of a scalar value (double) and unit string. The allowed time units are ‘s’ (second), ‘d’ (day = 86400 s), ‘yr’ or ‘a’ (Julian year = 365.25 d), ‘cy’ (Julian century = 36525 d).
The CoordValue object in TimeCoord is of class AstronTime.

##### *4.4.2.1.1 AstronTime*

A time (i.e., a particular instant in history) is an aggregate of one, two, or three objects: a TimeScale (Section 4.4.1.1.2; unless already provided in an associated AstroCoordSystem), an AbsoluteTime, and, optionally, a RelativeTime (or ElapsedTime). Note that, in principle, all times are really elapsed times. The difference is that for absolute times there is an implicit agreement as to what the reference instant in time is.

###### 4.4.2.1.1.1 AbsoluteTime

AbsoluteTime is instantiated as one of its four subclasses:

1. ISOTime: a time instant expressed in a subset of the ISO-8601 format; i.e., a string of format `yyyy-mm-ddThh:mm:ss.sss…`; the integer part of the seconds is in the range 0 to 60 for time scales that include leap seconds, 0 to 59 for all other time scales. We shall assume that ISOTime always relates to an (extrapolated) Gregorian date, independent of date and place, though caution should be exercised when using it for dates before, say, 1800 CE (some might argue 1918 CE). The earliest date that can be expressed in this format is 0001-01-01; prior to this date one should use JDTime. The same restrictions apply as for the FITS use of this format; this means in particular that time zone indicators are not supported and that UTC shall be indicated by specifying a TimeScale, not through the ‘Z’ indicator.
2. JDTime: a decimal number (note: in principle unlimited precision!) representing the Julian Day of a particular instant.
3. MJDTime: a decimal number (note: in principle unlimited precision!) representing the Modified Julian Day of a particular instant (MJDTime = JDTime – 2,400,000.5).
4. TimeOrigin: a string with the value `RELOCATABLE`, primarily for simulation work, to be used in conjunction with TimeScale `LOCAL` and a `RELOCATABLE` SpaceFrame.

###### 4.4.2.1.1.2 RelativeTime

Relative time is a decimal number representing the amount of time elapsed since AbsoluteTime. It may include a unit.

#### 4.4.2.2 SpatialCoord

Spatial coordinates are special in that they can be explicitly multi-dimensional and that they serve positions as well as their first time derivatives (physical velocities – not to be confused with redshifts or Doppler velocities). Not only does this mean that scalar values are to be replaced by vectors, but components like errors, resolutions, and sizes become more complicated. SpatialCoord is an aggregate of CoordName, Dimensionality, CoordValue (class SCValue), CoordError, CoordResolution, CoordSize, CoordPixSize (all class MultiCoord). We stress that Dimensionality should force all components of SpatialCoord to be consistent in type and content. This is, of course, similar to the situation described for multi-dimensional generic coordinates (see Section 4.2.2.1). There is merit in sub-classing a VelocityCoord class, since a time unit needs to be added. We suggest that the units for velocity be the same as for position, but that a (standard) time unit be added for the denominator.

##### *4.4.2.2.1 Dimensionality*

A single object of this class controls all components of the SpatialCoord class by specifying how many coordinate axes need to be represented. Note that this is distinct from the CoordFlavor object described in Section 4.4.1.2.2: CoordFlavor specifies the number of spatial axes defined in the AstroCoordSystem; Dimensionality specifies how many of those are actually provided in the AstroCoords object – one could envision a catalog with only Right Ascensions, where the CoordFlavor would be spherical, 2-dimensional, but the dimensionality just be 1.

##### *4.4.2.2.2 CoordName*

This object now consists of ‹Dimensionality› (separate) strings that contain the names of all axes present.

##### *4.4.2.2.3 SCValue*

The spatial coordinate value is a vector (length determined by Dimensionality) of numbers (SValue). In the XML implementation vectors are constructed as lists of elements, not as XML Schema list types.

##### *4.4.2.2.4 MultiCoord*

Expressing errors and sizes for vectors is slightly more complicated. If the errors are independent for each vector component, then a simple SValue object (vector of decimal values with a Unit) will do. If they are not independent (or if the resolution is tilted), one can either add a PosAngle (see also Section 4.4.1.2.5) to the SValue object, or specify the MultiCoord object through a Matrix of decimal values with a Unit. Matrices are implemented as lists of elements in XML, similar to vectors (see previous subsection). Position angles are measured following the definition in Section 4.4.1.2.5. In the case of spherical coordinates, all sizes, resolutions, and errors will be expressed in terms of arc lengths, for longitudes as well as for latitudes; i.e., the error in right ascension is expressed as a real arc angle, independent of declination.

##### 4.4.2.2.5 Orbital Elements

As a special case we allow spatial coordinates (position and velocity) to be expressed through Keplerian orbital elements. Such an orbit should be defined in a 3-dimensional spherical coordinate system centered on the barycenter of the system under consideration. The following parameters need to be provided:

- Semi major axis *a* (only for closed orbits: $0 \le e < 1$) or periapsis distance *q* both positive
- Eccentricity *e*: $0 \le e$
- Inclination *i*: $0^\circ \le i \le 180^\circ$
- Longitude of ascending node *N*: $0^\circ \le N < 360^\circ$
- Argument of periapsis *A*: $0^\circ \le A < 360^\circ$
- Epoch *T* of mean anomaly *M* (if present) or periapsis (if *M* is not present)

The following parameters may be provided:

- Mean anomaly *M*: $0^\circ \le M < 360^\circ$; if *M* is present, *T* will be considered to represent the epoch of *M*
- Orbital period *P*

In addition, it is strongly recommended to pair the orbit description with a Time Instant coordinate value indicating the coordinate epoch.
It is further worth noting that there are two alternative methods of communicating orbital information: a simple time series of time and position coordinate values; and through the two unit vectors *P* and *Q*, indicating the directions of periapsis and pole of the orbital plane, in addition to e and *T*.

#### 4.4.2.3 SpectralCoord

The SpectralCoord is used for the spectral component of the AstroCoords object. It is a generic scalar Coordinate class (see: Section 4.2.2.1) where Name is a string and all other components consist of a double precision Value and a string Unit, restricted to values appropriate for a spectral coordinate.

#### 4.4.2.4 RedshiftCoord

The Redshift Coordinate class differs from the Spectral Coordinate class only in the values that are allowed for the Unit.

### 4.4.3 AstroCoordArea

The Coordinate Area class defines the volume in coordinate space that is occupied by (or: the coverage of):

- Resource

- Query
- Catalog data set
- Observation

It is an aggregate of one or more ranges in individual coordinates and is required to have a reference to an AstroCoordSystem. Multiple ranges along each coordinate axis are OR-ed together; the ranges of different axes are logically AND-ed together. Each range has associated with it a fillfactor that indicates the fraction of the range, interval, or region that is actually occupied by the data.

#### 4.4.3.1 TimeInterval

A TimeInterval consists of a StartTime and/or a StopTime, both of which are instantiations of AstronTime. Note that they do not both have to be present, but at least one of them needs to be. TimeInterval has Boolean attributes LoLimInclude and HiLimInclude, and attribute FillFactor (see Section 4.2.3.1).

#### 4.4.3.2 SpatialInterval

The spatial position area has more options:

- Simple intervals on a per-axis basis
- Sphere
- Regions (see Section 4.5):
    - Shapes: all sky, polygon, box, sector, circle, ellipse, convex, convex hull, sky index
    - Operations on regions: intersection, union, negation, difference

##### *4.4.3.2.1 CInterval*

CInterval is the vector equivalent of ScalarInterval (see: Section 4.2.3.1). LoLimit and HiLimit are of type SCValue (see Section 4.4.2.2.3).

##### *4.4.3.2.2 Sphere*

Since spherical regions are quite common in three dimensions these are made available as special cases. The class is an aggregate of a Center position (of type SCValue; see: Section 4.4.2.2.3) and a Radius (consisting of a double and a unit). The spatial coordinates of the center should be in the AstroCoordSystem associated with AstroCoordArea, of course. Note that the 2-D equivalent (circle) is available in the Region class (see: Section 4.5).

##### *4.4.3.2.3 SpatialRegion*

This is an object of class Region (see: Section 4.5).

#### 4.4.3.3 VelocityInterval

The velocity interval is an instantiation of CInterval (see: Section 4.4.3.2.1).

#### 4.4.3.4 SpectralInterval

The SpectralInterval is derived from the generic ScalarInterval class (see: Section 4.2.3.1), restricting Unit to appropriate values.

#### 4.4.3.5 RedshiftInterval

The RedshiftInterval is derived from the generic ScalarInterval class (see: Section 4.2.3.1), restricting Unit to appropriate values.

## 4.5 Region

The Region specification is a rather elaborate construct that allows a very flexible definition of arbitrary 2-dimensional spatial shapes. It has 13 derived classes: 9 shapes and 4 operations. Note that the result of these operations performed on one or more regions is a region. All spatial coordinates should be in the AstroCoordSystem associated with AstroCoordArea (or Region), of course. The Region object has an important attribute, fill_factor (between 0.0 and 1.0; default 1.0) that indicates what fraction of the region is really filled or covered. This is particularly useful when describing a resource's coverage. In addition, a region has an optional element Area providing, for convenience, the area occupied by the region. Attributes are validArea (a Boolean indicating whether Area's value is actually valid) and linearAreaUnit (the proper units of Area's value are linearAreaUnit squared).

### 4.5.1 Shapes

The first six shapes are 2-dimensional, while the remaining two are 3-dimensional, associated with the unit sphere. All boundaries are considered part of the inside of a shape. Different shapes in the same Region object may refer to different Coordinate Systems, although this should not be encouraged.

#### 4.5.1.1 AllSky

This is just a short-hand object to conveniently indicate "All".

#### 4.5.1.2 Circle

A Circle (2-dimensional) shape (region) is an aggregate of a Center position (type SCValue; see: Section 4.4.2.2.3) and a Radius (consisting of a double and a unit).

#### 4.5.1.3 Ellipse

The Ellipse (2-dimensional) is similar to the Circle but has, in addition, a minor radius and a position angle. Position angles are measured following the definition in Section 4.4.1.2.5 and refer to the first axis. The definition of an ellipse in a Cartesian coordinate system is unambiguous, but this is not the case for spherical coordinates. In a spherical coordinate system the ellipse shall be defined as the intersection of an elliptical cone with the unit sphere, where the axes and position angle describe the geometry of the cone.

#### 4.5.1.4 Polygon

A Polygon (2-dimensional) is an ordered list of one or more vertices. Its area is defined as that contained within (i.e., to the left of) the lines connecting neighboring (that is, in the ordered list) vertices. If the CoordFlavor is CARTESIAN, these lines are truly lines. The last vertex in the list connects back to the first.
If the CoordFlavor is SPHERICAL, the lines are, by default, great-circles, unless the Vertex object contains a SmallCircle object; in that case the line connecting that vertex with its predecessor is a small-circle (parallel). The curvature of the parallel is determined by the pole of the SpatialFrame in the current AstroCoordSystem, unless a PolePosition is explicitly specified in the Vertex object. It is the responsibility of the server to ensure that the positions of the two sequential vertices actually lie on a parallel that is consistent with the implied or specified pole. In order to avoid ambiguities in direction, vertices need to be less than 180° apart in both coordinates. Great circles or small circles spanning 180° require specification of an extra intermediate vertex.
The area *A* of a polygon with *n* vertices *x* in Cartesian space may be calculated as:

$$A = 0.5 \times \sum_{i=1}^{n} \left| \underline{x}_i \, \underline{x}_{i+1} \right|$$

The summation is over determinants of matrices formed by the position vectors $\boldsymbol{x}_i$ of successive vertices; $\boldsymbol{x}_{n+1} = \boldsymbol{x}_1$. In spherical space (left-handed coordinates) the area is:

$$A = -\sum_{i=1}^{n} \alpha_i - (n-2)\pi$$

$\alpha_i$ are the polygon's angles at the vertices. Reverse the sign for right-handed coordinates. If the coordinate system is defined on the surface of a body (rather than the celestial or unit sphere), one should multiply by the square of the radius. The boundaries are considered part of the region. The inside of the region is defined as that part of coordinate space that is encircled by the polygon in a counter-clockwise sense. What this means is that, in a plane, if $A > 0$, the "inside" of the polygon is included; if $A < 0$, the "outside" is selected. On a sphere with a left-handed (celestial) coordinate system, if $A > 0$, one has identified the inside of the polygon; if $A < 0$, one used the "outside" angles of the polygon and the area is really $4\pi - A$. Note that the negation operation is a simple matter of reversing the order of the vertex list.

#### 4.5.1.5 Box

A Box is a special case of a Polygon, defined purely for convenience. It is specified by a center position and size (in both coordinates) defining a cross centered on the center position and with arms extending, parallel to the coordinate axes at the center position, for half the respective sizes on either side. The box's sides are line segments or great circles intersecting the arms of the cross in its end points at right angles with the arms.

#### 4.5.1.6 Sector

A Sector (2-dimensional) is defined by a position of type SCValue (see: Section 4.4.2.2.3) and two position angles. The inside of the sector is that part of the plane or sphere that is swept by a half-line or great-circle rotating counter-clockwise from the first to the second position angle, centered on the specified position. Position angles are measured following the definition in Section 4.4.1.2.5.

#### 4.5.1.7 Convex

A Convex is the convex region that is defined by an aggregate of one or more HalfSpaces. It is defined on the surface of the unit sphere and is hence in 2-dimensional spherical space. Each HalfSpace selects a circular region (or point) on the unit sphere. The Convex is therefore the intersection of a collection of circles and its sides are great-circles and/or small-circles. A HalfSpace is specified by a 3-dimensional vector (a point on the unit sphere) and an Offset along that vector. The HalfSpace's circular area is defined by the intersection of the unit sphere and the plane normal to the specified vector, intersecting it at the Offset distance from the origin: it is the HalfSpace on the UnitSphere that contains the point specified by the vector. The valid range for Offset is -1 to +1, with the boundaries included. Note that a HalfSpace is fully equivalent to a Circle shape; hence, a Convex is equivalent to the intersection of one or more Circles. Similarly, a Convex can be described by a Polygon.

#### 4.5.1.8 ConvexHull

A ConvexHull is defined by an unordered list of one or more points. It is the smallest convex polygon that contains all points in the list. It may be constructed as the union of the triangles that can be formed from all possible triplets of points in the list. Note that, consequently, a ConvexHull on a sphere is limited to great circle sides only.

#### 4.5.1.9 SkyIndex

This is currently a place holder to allow integration of the specification of regions through sky indexing schemes.

### 4.5.2 Operations

There are four operations defined on or between Regions. The result of each of those operations is also a Region.

#### 4.5.2.1 Intersection

The Intersection of two Regions is the Region that is common to both (logical AND). We allow Intersections of more than two Regions.

#### 4.5.2.2 Union

The Union of two Regions is the Region that is contained in either or both of them (logical OR). We allow Unions of more than two Regions.

#### 4.5.2.3 Negation

The negation of a Region is the Region that is not contained in the original (logical NOT). Note that this logically leads to a Region that does not include its boundaries.

#### 4.5.2.4 Difference

This operation is mainly added for convenience. It can be expressed as a combination of Intersection and Negation ($R_1 – R_2 = R_1$ AND (NOT $R_2$)), but there is an operational advantage to having the ability to specify a Difference explicitly. Note that Difference is not a symmetrical operation, unlike Intersection and Union. We shall assume that the boundaries are included in the difference region.

# 5 Projection and Transformation

The metadata described in this document specify a position or the volume occupied by a dataset. In order to be able to transform the coordinates for individual Data Model components (such as pixels) one needs a Projection object as well. Such a Projection or Mapping metadata object is very much like the substance in the FITS WCS[ix] [x] [xi]. Originally, this was to be specified in a separate document that David Berry was working on. However, it was realized that the custom reference frame that was already allowed for spherical systems represented a transformation object and hence it was generalized for other coordinate flavors.
A CoordRefFrame object that is not a standard reference frame (as in Table 3) may be 1-, 2-, or 3-dimensional, Cartesian or Spherical. For the 1-dimensional case it contains a projection and a scale factor; for the 2- and 3-dimensional Cartesian coordinates it contains a projection code and either a scale-and-rotate object or a transformation (scale-and-rotate) matrix; for the spherical case it contains a pole and an X-axis direction. The second part of the transformation definition is found in the CoordRefPos object which allows one to specify the origin. This works for all coordinates. In the case of pixel coordinates there are extra attributes that define the axis order.

For further details, see Section 4.2.1.2. We emphasize that the description used in STC is compatible with FITS WCS.

# 6 Implementation

## *6.1 Implementation Requirements*

This section contains the rules that implementations need to follow.
All objects MAY contain:

- A UCD attribute
- A Unit
- An Identifier

All objects MUST contain one of the following:

- A content body
- A reference to a matching object within the scope of the subject's context
- A reference to a matching object outside the scope of the subject's context

### 6.1.1 Generic Coordinates

#### 6.1.1.1 Coordinate System

A Coordinate System MUST contain:

- At least one Coordinate Frame.

A Coordinate Frame MAY contain:

- A Coordinate Reference Frame that specifies the orientation of the frame, its scaling, and/or rotation with respect to another, well-defined Coordinate Frame; a Coordinate Reference Frame MUST contain a projection specification if different from LINEAR
- A Coordinate Reference Position that defines the origin, c.q., translation, of such a transformation

A Coordinate Frame MUST contain:

- A Coordinate Flavor that specifies its dimensionality and flavor, and that may indicate its handedness
- An Identifier

#### 6.1.1.2 Coordinates

A Coordinates object MUST contain:

- A reference to a Coordinate System object that covers all coordinates contained in the subject
- At least one Coordinate

A Coordinate object MUST be of one of the following types:

- String
- Scalar (1-D numeric)
- 2-D
- 3-D

A Coordinate object MUST contain:

- A reference to a Frame that is contained in the Coordinate System that the subject's Coordinates object refers to
- At least one of the following components: Name, Value, Error, Resolution, Size, Pixel size; each component, except for Name and Value, MAY occur:
    - Zero times
    - Once: a definite value
    - Twice: providing a range of possible values
- String Coordinates SHALL only contain Name and/or Value
- Multi-dimensional errors, resolutions, sizes, pixel sizes SHOULD be specified in one of the following ways:
    - Separate values in each dimension
    - Radius
    - As an ellipsoid
    - Through a matrix

#### 6.1.1.3 Coordinate Area

A Coordinate Area object MUST contain:

- A reference to a Coordinate System object that covers all coordinates contained in the subject
- At least one Coordinate Interval

A Coordinate Area MAY contain:

- More than one Coordinate Interval per Coordinate Frame

A Coordinate Interval object MUST be of one of the following types, commensurate with the dimensionality of the Coordinate Frame it refers to:

- Scalar
- 2-D
- 3-D

A Coordinate Interval object MUST contain:

- A reference to a Frame that is contained in the Coordinate System that the subject's Coordinate Area object refers to
- One or both of the following components:
  - Lower limit
  - Upper limit

A Coordinate Interval MAY contain the following attributes:

- Fill factor (default: 1.0)
- Include lower limit (default: true)
- Include upper limit (default: true)

### 6.1.2 Astronomical Coordinates

#### 6.1.2.1 Astronomical Coordinate System

An Astro Coordinate System MAY contain:

- A Generic Coordinate System

An Astro Coordinate System SHOULD contain at least one of the following (but SHALL NOT contain more than one of each):

- Time Frame
- Spatial Frame
- Spectral Frame
- Redshift Frame

A Time Frame MUST contain:

- A Time Scale, as specified in Section 4.4.1.1.2
- A Time Reference Position, as specified in Section 4.4.1.1.1

A Time Frame SHOULD contain:

- A Time Reference Direction if meaningful (see Section 4.4.1.1.3)
- A Planetary Ephemeris if meaningful (see Section 4.4.1.1.4)

A Spatial Frame MUST contain:

- A Spatial Reference Frame: either a standard frame as listed in Table 4 or through a transformation from a known frame; see Section 4.4.1.2.3
- A Spatial Reference Position, as specified in Section 4.4.1.2.1
- A Coordinate Flavor, as specified in Section 4.4.1.2.2

A Spatial Frame MAY contain:

- An Offset Center, as specified in Section 4.4.1.2.4

A Spectral Frame MUST contain:

- A Spectral Reference Position selected from Table 1; note that not only position, but also space velocity of this reference position MUST be provided, in particular when using the value TOPOCENTER

A Redshift Frame MUST contain:

- Doppler definition, as specified in Section 4.4.1.4.2
- A Redshift or Doppler Reference Position with the same properties as the reference position for spectral frames

#### 6.1.2.2 Astronomical Coordinates

An Astronomical Coordinates object MUST contain:

- A reference to an Astronomical Coordinate System that covers all coordinates contained in the subject
- At least one Astronomical Coordinate

It SHOULD be possible to provide all Astronomical Coordinate Values in a FITS file.

An Astronomical Coordinates object MAY contain one or more of the following Astronomical Coordinates:

- Generic Coordinate (see Section 6.1.1.2)
- Astronomical Time Coordinate
- Spatial Position Coordinate
- Spatial Velocity Coordinate
- Spectral Coordinate
- Redshift (or Doppler) Coordinate

All of these coordinates are subject to the same requirements as a Generic Coordinate, with the following restrictions and modifications.

- Non-Generic Coordinates do not need a reference to a Frame Id
- The Time Instant (Section 4.4.2.1.1) in a Time Coordinate (Section 4.4.2.1) MUST contain:
    - Absolute time value, in JD, MJD, or ISO-8601, as specified in Section 4.4.2.1.1.1
- A Time Instant MAY contain:
    - A Time Scale
    - A Relative Time Offset (or Elapsed Time) from the Absolute Time with appropriate Unit specification
- A Spatial Coordinate (Section 4.4.2.2) MUST be of one of the following types:
    - 1-dimensional
    - 2-dimensional
    - 3-dimensional
- A Spatial Coordinate MUST have a Spatial Unit
- A Spatial Velocity Coordinate MUST, in addition, have Time Unit
- A Spatial Coordinate MAY have an Epoch
- A Spatial Coordinate MAY represent a linear structure of positions (such as a curve)
- It SHALL be possible to specify Spatial Coordinate values (Position and Velocity) through Keplerian orbital elements (Section 4.4.2.2.5)
- A Spectral Coordinate (Section 4.4.2.3) MUST be 1-dimensional and MUST have a Spectral Unit (frequency, wavelength, or energy); in the presence of a Redshift Coordinate, the Spectral Coordinate value is to be interpreted as the rest frequency.
- A Redshift Coordinate (Section 4.4.2.4) MUST be 1-dimensional and MUST have a Spatial Position Unit and a Time Unit if it is expressed as a Doppler velocity

#### 6.1.2.3 Astronomical Coordinate Area

An Astronomical Coordinate Area object MUST contain:

- A reference to an Astronomical Coordinate System object that covers all coordinates contained in the object
- At least one Astronomical Coordinate Interval

An Astronomical Coordinate Area object MAY contain:

- Any number of Coordinate Intervals for each Coordinate Frame
- Coordinate Intervals for the following Coordinate Frames:
  - Generic (Section 6.1.1.3)
  - Time
  - Spatial Position
  - Spatial Velocity
  - Spectral
  - Redshift
- All non-Generic Coordinate Areas MUST have appropriate Unit attributes and MAY not have Frame Ids
- Time, Spectral, and Redshift Intervals MUST be 1-dimensional
- Spatial Intervals MUST be one of the following:
  - 1-dimensional
  - 2-dimensional
  - 3-dimensional
- 2-dimensional Spatial Position Coordinate Areas MAY be specified through a Region, in two dimensions (Spherical, Cartesian, Polar) or three dimensions (UnitSphere)
- A Region (see Section 4.4.3.2.3) MUST be specified as:
  - Either one of the following shapes: All Sky, Circle, Ellipse, Box, Polygon, Sector, HalfSpace, Convex, Convex Hull, Sky Index
  - Or the result of one of the following operations on one or more Regions: Union, Intersection, Negation, Difference
- 3-dimensional Spatial Coordinate Areas MAY be specified as a Sphere

### 6.1.3 Pixel Coordinates

#### 6.1.3.1 Pixel Coordinate System

A Pixel Coordinate System MAY contain:

- A Generic  Coordinate System

A Pixel Coordinate System SHOULD contain:

- At least one Pixel Frame which MAY be 1-, 2-, or 3-dimensional

A Pixel Frame MUST contain:

- An ID
- An axis order specification that maps the frame’s axes to the axes of the pixel array
- A Pixel Reference Frame that specifies the transformation from a World Coordinate Frame to this Pixel Frame: projection, rotation, scaling

- A Pixel Reference Position that specifies the translation part of this transformation ("Reference Value" in FITS WCS parlance)
- A Coordinate Flavor that includes the dimensionality of the frame
- A Reference Pixel

Note that there must be a one-to-one mapping between Pixel Frames and (World) Coordinate Frames. A single pixel array will be described by a single Pixel Coordinate System, but within that System there will usually be multiple Pixel Frames, one for each World Coordinate Frame. Multi-dimensional Pixel Frames are only to be used in cases where the corresponding frame is truly multi-dimensional. Spatial frames are an obvious case, but additional examples would include: complex visibility (real/imaginary or amplitude/phase), linear polarization intensity and angle, magnetic field strength and direction, wind velocity and direction.

#### 6.1.3.2 Pixel Coordinates

A Pixel Coordinates object MUST contain:

- A reference to a Pixel Coordinate System

A Pixel Coordinates object MAY contain:

- A Generic Coordinate

A Pixel Coordinates object SHOULD contain:

- One or more Pixel objects which MAY be 1-, 2-, or 3-dimensional

A Pixel object MUST contain:

- A reference to a Pixel Frame that has the same dimensionality as the Pixel and is contained in the referenced Pixel Coordinate System

A Pixel MAY contain:

- A Name for each of its axes
- A Value of appropriate dimensionality

Units are generally not to be used in Pixel Space.

#### 6.1.3.3 Pixel Coordinate Area

A Pixel Coordinate Area object MUST contain:

- A reference to a Pixel Coordinate System

A Pixel Coordinate Area object MAY contain:

- A Generic Coordinate Area

A Pixel Coordinate Area object SHOULD contain:

- One or more Pixel Interval objects which MAY be 1-, 2-, or 3-dimensional

A Pixel Interval object MUST contain:

- A reference to a Pixel Frame that has the same dimensionality as the Pixel Interval and is contained in the referenced Pixel Coordinate System
- A lower and upper limit value of appropriate dimensionality

#### 6.1.4 Observational Metadata

The metadata describing the data obtained in a particular observation SHALL contain:

- A specification of the location of the observatory
- The volume in coordinate space taken up by the observation; this SHOULD typically cover the following coordinate frames:
    - o Time
    - o Space
    - o Spectrum
- The properties of the data along the coordinates, such as errors (uncertainties) and resolutions
- If the data are pixelated: a specification of, and mapping to, the pixel space

The metadata SHOULD contain:

- A specification of the coordinate space of the observable and its properties

### *6.2 STC-X: XML Schema Implementation*

The design of Section 4 has been implemented in an XML schema, as stated in Section 1. http://www.ivoa.net/Documents/latest/STC-X.html provides documentation for this implementation; the full documentation may be found at http://hea-www.harvard.edu/~arots/nvometa/v1.30/
The present document contains examples in Appendix B: Examples.
One important thing to note is that the schema implementation will allow all elements to contain actual values or to refer to another element; this will allow integration into VOTable where such references would be to table columns.
Another notable implementation detail is that coordinate components are allowed to be provided in FITS file. The schema includes a mechanism to associate particular elements with specific binary table columns. This is particularly of interest in conjunction with spacecraft orbit ephemerides.
All higher-level (and some lower-level) element types are derived from a base type that contains the attributes for the referencing mechanism and for UCD labeling.

#### 6.2.1 Referencing Mechanism

Xlink has been adopted as a general-purpose IVOA referencing mechanism. The standard is defined at http://www.w3.org/1999/xlink and the schema may be found at http://www.ivoa.net/xml/Xlink/xlink.xsd. Xlink allows an element to refer to any object anywhere. In the context of STC we restrict this to actual XML elements available on the web, using XPath syntax. We have therefore defined an STC basetype from which many types are derived that contains the following attributes:

| | |
|---|---|
| xlink:type | We only use the value “simple” |
| xlink:href | The URI where an element of the correct type is to be found |
| idref | A reference to an existing element in the document |
| id | An ID allowing the element to be referenced |
| ucd | An optional UCD string |
| ID_type | Optional attribute that may contain the names (comma-separated) of attributes that are of type ID |
| IDREF_type | Optional attribute that may contain the names (comma-separated) of attributes that are of type IDREF |

Consequently, the contents of an element may be specified in three ways:

1. Through the actual contents in its body
2. Through a reference to another element in the document (idref)
3. Through a reference to an element defined elsewhere (Xlink)

This referencing model has some consequences that one should be aware of. First, there may be multiple (and possibly contradictory) specifications of the contents of such an element; if this is the case, the order of precedence is the one provided in the above list. Second, since the attributes are necessarily optional and the element, by necessity nillable, there may not be any contents at all; if this is the case, the contents should be considered UNKNOWN. Finally, there is no obligation for the client to go and substitute an element that is referenced through Xlink (though it should always be permissible). We will provide some standard libraries (e.g., of common coordinate systems and observatory locations; see Appendix C) with a standard naming convention; a client that is familiar with such libraries may use its inside knowledge instead.

*Note: The use of ID and IDREF pairs has the advantage that the association is unambiguous, since IDs are required to be unique within each document. However, that also means that it may not be possible to concatenate elements from different documents into a single new document, since there is no guarantee that IDs will remain unique; this is particularly an issue for the coordinate system IDs, when the same coordinate system may be part of several included elements. One could turn to another association mechanism (e.g.,* unique *and* keyref*), but scope and uniqueness remain a problem. This is really an issue that requires a general IVOA-wide convention that ensures associations to be unambiguous within document scope, while still addressing the possibility of the same element occurring more than once in a document.*
*At this time STC provides support for allowing applications to identify ID and IDREF attributes through the optional attributes ID_type and IDREF_type.*

### 6.2.2 Versioning

The STC schema will be versioned using three numbers: *i.jk*. Here *i* is the major version number and *j* the minor version number. *k* is the patch number. All versions with identical major and minor version numbers SHALL be downward compatible. Users are encouraged to reference the STC version only by these numbers which will result in retrieving the highest patch number for that combination. For version 1.3, the link to 1.30 shall be considered the equivalent of 1.3.

### 6.3 STC-S

STC-S is a string, or command line, implementation, based on the XML schema. It allows one to specify an STC object in a simple phrase with sensible defaults and without redundancy. There are certain restrictions, though. The design of this specification is such that a one-to-one conversion to and from XML documents is possible. It is especially intended for applications like ADQL/S and Dublin Core descriptions. For details, see the document referenced in Section 1: http://www.ivoa.net/Documents/latest/STC-S.html.

## 7 Conclusion and Usage Notes

VO needs a stringent specification in order to not get into trouble later on when we want to, for instance:

- use far field data for near field objects
- mix data in equatorial coordinates with data in Jovian coordinates
- combine pulsar observations from ground-based, spacecraft, and lunar observatories

### 7.1 Methods

The metadata objects described in Section 4 provide sufficient and necessary information, and ensure self-consistency. As commented before, this document focuses on the structure of the design, not on the methods. However, we can make some general observations. First, most of the client-level methods will probably appear as methods on the coordinate transformation class. Second, robust design and implementation favors access to the various elements through the top-level objects. In other words, public methods should appear on the AstroCoordSystem, AstroCoords, and AstroCoordArea classes, while most methods on the component classes ought to be restricted to the friend classes in the whole of the STC design structure.
Examples of methods that are obviously required:

- AstroCoords: get {spatial, temporal, spectral, redshift} coordinates of object
- AstroCoords: return new AstroCoords object which references a specified (presumably different) AstroCoordSystem object (i.e., coordinate transformation)
- AstroCoordArea: same methods as for AstroCoords, above
- AstroCoordArea: return (measured) area and fill_factor of object
- AstroCoordArea: is a given AstroCoords point inside or outside the region?
- AstroCoordArea: constructor methods that implement the four operations on Region (negation, union, intersection, difference)

- AstroCoordArea: constructor methods that create the largest inscribing and smallest circumscribing circles and rectangles, with correct fill_factor

## 7.2 Usage

Here are some notes on how the various components are intended to be used in the different use contexts. The reader is reminded that the numeric components of the Coordinate elements may indicate a definite value or a range of values, depending on whether there is a single instance or a pair of instances of these components.

### 7.2.1 Resource Profile

The STC Resource Profile, or Coverage, describes what part of STC space is potentially covered by the datasets available from the resource. This is a potential coverage since there is no guarantee that the entire volume is covered (in particular, if the resource contains a collection of pointed observations); note that this is where the Region fill-factor is a useful attribute. Examples of resources are depositories of pointed observations, survey atlases, and catalogs. The profile consists of an AstroCoordSystem object that represents the resource's native coordinate system; an AstroCoordArea object that represents the actual coverage in the four standard STC frames; and an AstroCoords object that gives some specifics on the products in the resource. The purpose of the first two will be obvious, but the third may need some explanation. The AstroCoords object does not intend to convey a particular position, but the typical properties of datasets in the resource. The Coordinate objects will not contain a CoordValue, but CoordError, CoordResolution, CoordSize, and CoordPixSize represent typical values for the datasets, where CoordSize refers to the field-of-view. Additional AstroCoordSystem objects may be required if Custom Reference Frames or Reference Positions are used.

### 7.2.2 Query Constraint

A query constraint is, in a way, a mirror image of a resource profile, and its structure and the meaning of its components is very much the same: the AstroCoordArea defines the coordinate volume that the query requests to be filled, while the AstroCoords components (again, with missing CoordValue), as far as present, represent desirable values for errors, resolution, pixelsize, and field-of-view (if the whole area cannot be filled at once). Additional AstroCoordSystem objects may be required if Custom Reference Frames or Reference Positions are used.

### 7.2.3 Catalog Entry

The STC object that accompanies the return of catalog information (typically a table of selected catalog entries) is set up slightly differently. The AstroCoordSystem provides the usual coordinate system information, of course. The AstroCoordArea describes the coverage of the returned information; that will usually be the intersection of the query coverage with the resource (catalog) coverage. The AstroCoords object will generally refer to the catalog data that are being returned; i.e., its components are likely to refer to columns in the table that the STC object is associated with (such as galaxy positions, sizes, and redshifts) or have specific values if they happen to be constant for all entries (for instance, errors or resolution). Since Catalog objects may contain columns for multiple coordinate systems, or entries using different systems, we need to allow multiples of all three STC objects.

#### 7.2.4 Observational Data

In this context two STC objects are required. One describes the location of the observatory (AstroCoordSystem and AstroCoords; note that this may refer to an orbit ephemeris file for spacecraft). The other STC object describes the volume in coordinate space that is occupied by the observational data: AstroCoordSystem, AstroCoordArea (the field-of-view), and AstroCoords (to convey errors, resolution, and pixel size). The information about the observatory's location is essential, in particular for near-field/far-field spatial transformations and time transformations. Note that its velocity is also needed, either explicitly or implicitly. This is a typical case where metadata that are "obvious" (and often implied) in the context of a particular observatory's archive needs to be made explicit. Additional AstroCoordSystem objects may be required for either STC object if Custom Reference Frames or Reference Positions are used.
As noted in Section 5, there is a third Coordinate metadata element in the case of pixelated data. The PixelCoordFrame will define the transformation between the Observation Location and the Pixel Space.
One should be aware that in those cases the AstroCoordArea object associated with the Observation Locations defines the volume **occupied by the data**, and that this will be contained in, but may be smaller than, the CoordArea object associated with the Pixel Space.

# Appendix A: UML Representation

The following pages contain the STC design in UML diagrams.
The first four pages illustrate the design of the Coordinate System classes: an overview page and three pages emphasizing each of the three subclasses. The subsequent pages show the design of the Coordinates and Coordinate Area classes, though these represent the design of versions 1.2 and 1.3, and are therefore slightly outdated at this time; in particular, the Pixel Space is missing.

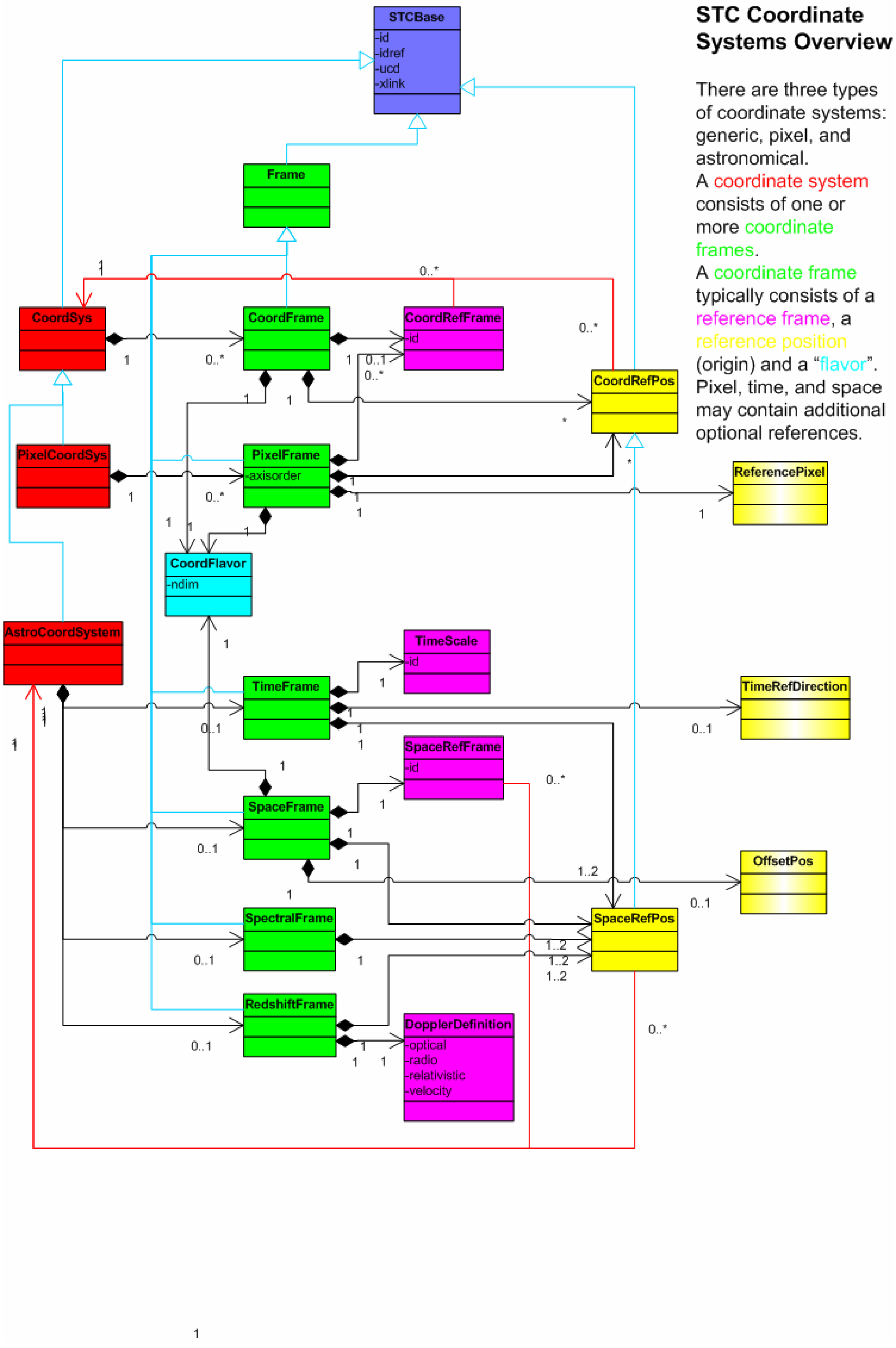
STC Coordinate Systems Overview
There are three types of coordinate systems: generic, pixel, and astronomical.
A coordinate system consists of one or more coordinate frames.
A coordinate frame typically consists of a reference frame, a reference position (origin) and a "flavor".
Pixel, time, and space may contain additional optional references.
STCBase
-id
-idref
-ucd
-xlink
Frame
CoordSys
CoordFrame
CoordRefFrame
-id
CoordRefPos
PixelCoordSys
PixelFrame
-axisorder
ReferencePixel
CoordFlavor
-ndim
AstroCoordSystem
TimeScale
-id
TimeFrame
TimeRefDirection
SpaceRefFrame
-id
SpaceFrame
OffsetPos
SpectralFrame
SpaceRefPos
RedshiftFrame
DopplerDefinition
-optical
-radio
-relativistic
-velocity

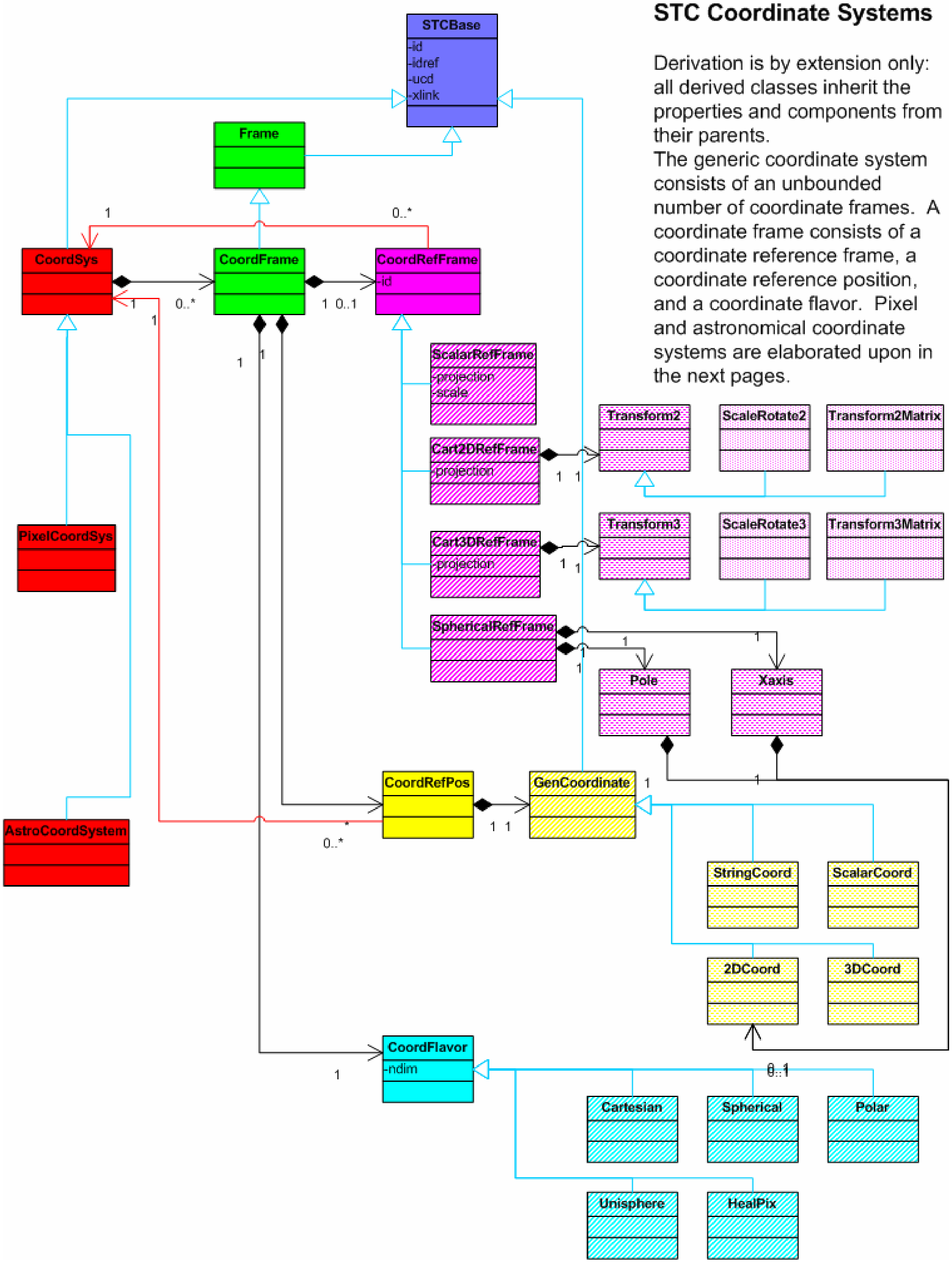

## STC Coordinate Systems

Derivation is by extension only: all derived classes inherit the properties and components from their parents.
The generic coordinate system consists of an unbounded number of coordinate frames. A coordinate frame consists of a coordinate reference frame, a coordinate reference position, and a coordinate flavor. Pixel and astronomical coordinate systems are elaborated upon in the next pages.

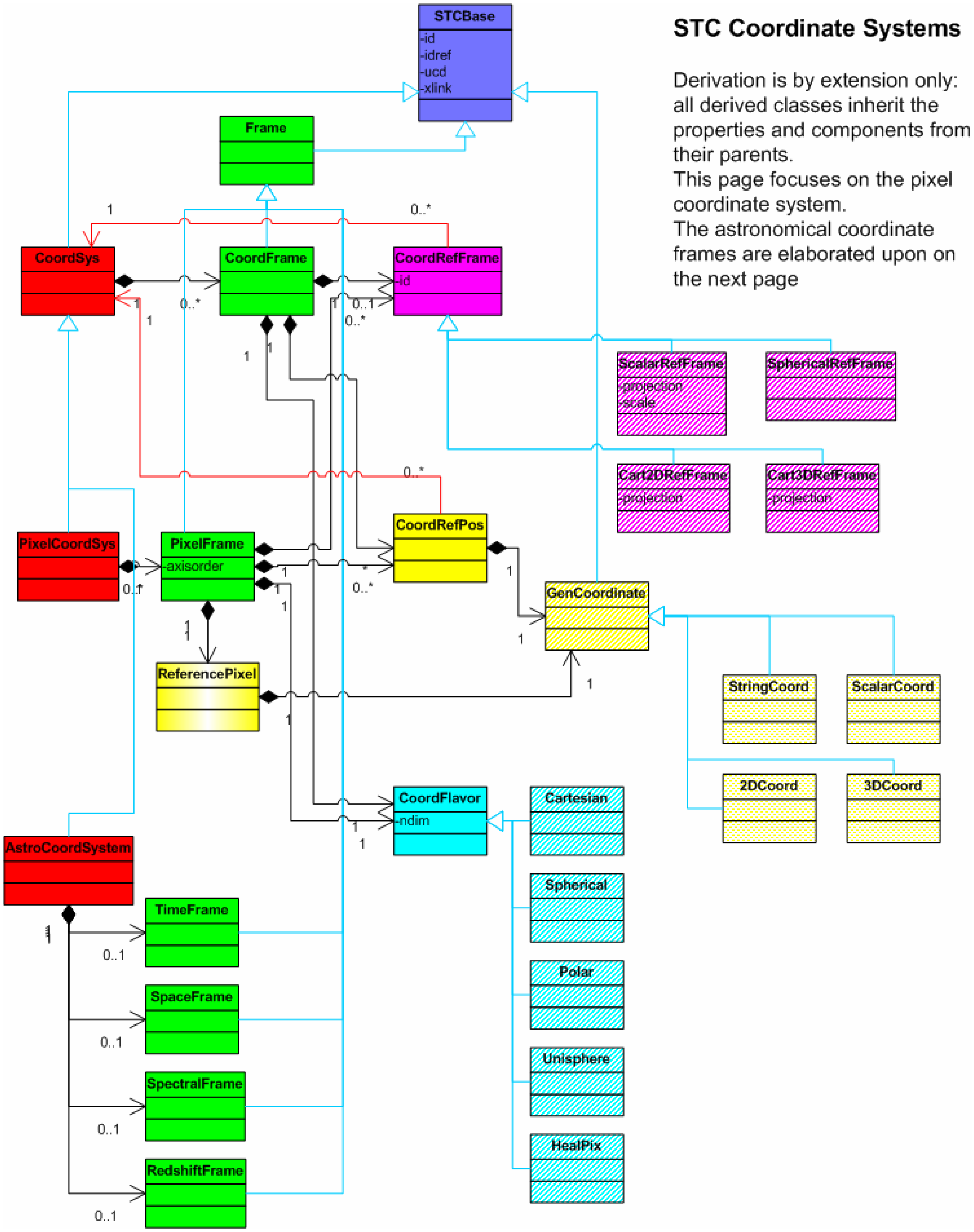
STC Coordinate Systems
Derivation is by extension only: all derived classes inherit the properties and components from their parents.
This page focuses on the pixel coordinate system.
The astronomical coordinate frames are elaborated upon on the next page
STCBase
-id
-idref
-ucd
-xlink
Frame
CoordSys
CoordFrame
CoordRefFrame
-id
ScalarRefFrame
-projection
-scale
SphericalRefFrame
Cart2DRefFrame
-projection
Cart3DRefFrame
-projection
PixelCoordSys
PixelFrame
-axisorder
CoordRefPos
GenCoordinate
ReferencePixel
StringCoord
ScalarCoord
2DCoord
3DCoord
CoordFlavor
-ndim
Cartesian
Spherical
Polar
Unisphere
HealPix
AstroCoordSystem
TimeFrame
SpaceFrame
SpectralFrame
RedshiftFrame

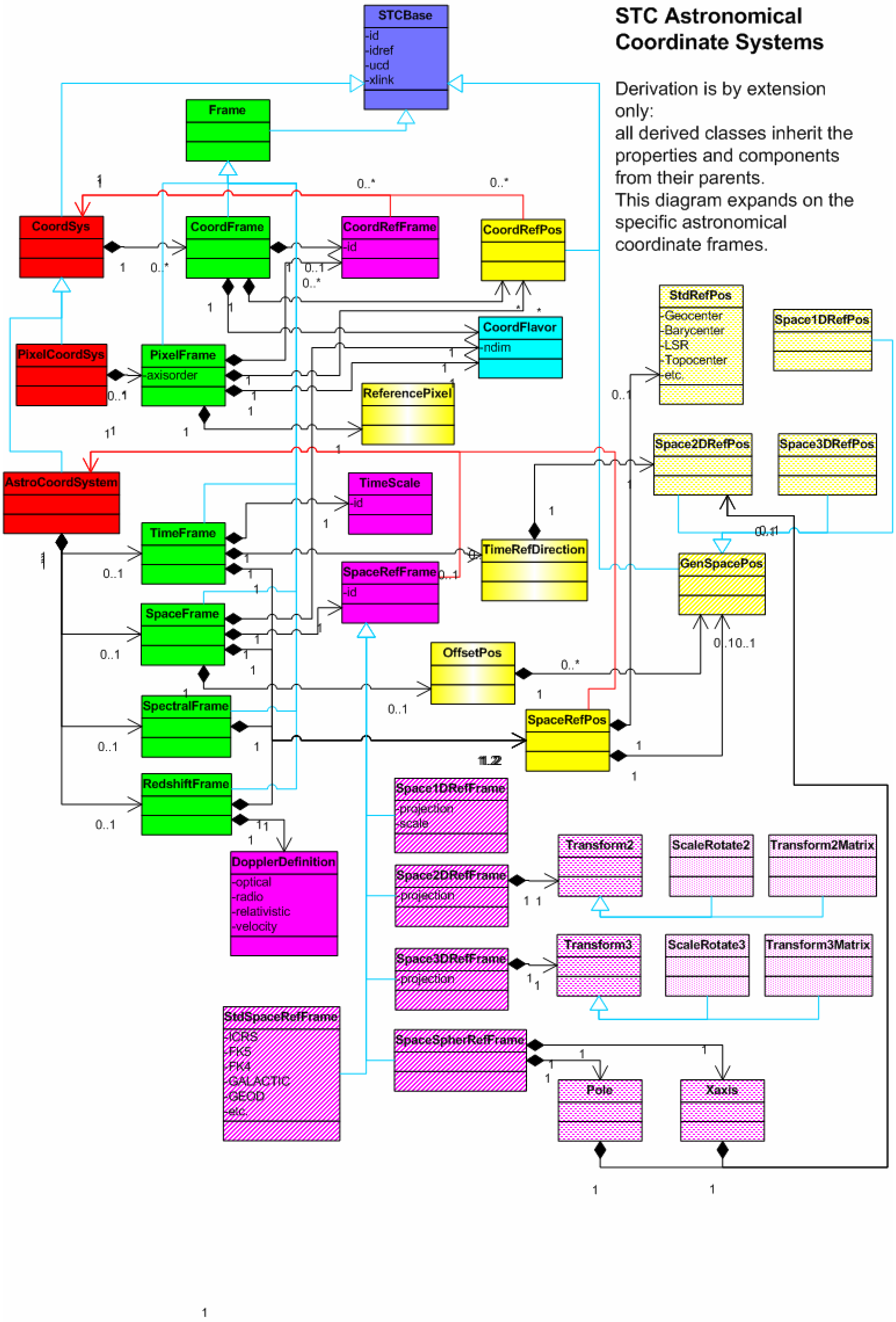
STC Astronomical Coordinate Systems
Derivation is by extension only:
all derived classes inherit the properties and components from their parents.
This diagram expands on the specific astronomical coordinate frames.
STCBase
-id
-idref
-ucd
-xlink
Frame
CoordSys
CoordFrame
CoordRefFrame
-id
CoordRefPos
CoordFlavor
-ndim
PixelCoordSys
PixelFrame
-axisorder
ReferencePixel
StdRefPos
-Geocenter
-Barycenter
-LSR
-Topocenter
-etc.
Space1DRefPos
Space2DRefPos
Space3DRefPos
AstroCoordSystem
TimeScale
-id
TimeFrame
TimeRefDirection
GenSpacePos
SpaceFrame
SpaceRefFrame
-id
OffsetPos
SpectralFrame
SpaceRefPos
RedshiftFrame
DopplerDefinition
-optical
-radio
-relativistic
-velocity
Space1DRefFrame
-projection
-scale
Space2DRefFrame
-projection
Transform2
ScaleRotate2
Transform2Matrix
Space3DRefFrame
-projection
Transform3
ScaleRotate3
Transform3Matrix
StdSpaceRefFrame
-ICRS
-FK5
-FK4
-GALACTIC
-GEOD
-etc.
SpaceSpherRefFrame
Pole
Xaxis

6 July, 2004

# Space-Time Coordinate Metadata for the Virtual Observatory

Arnold Rots, CfA/SAO

The UML diagrams on the following pages specify the design of the VO Space-Time Coordinate metadata. Specifically, these metadata describe the volume in coordinate space that is occupied by the data object that they are attached to (or the query, or the resource), with additional information on actual or desired properties of the data along those coordinate axes.
The coordinates that are covered by these metadata are:

- Time
- Spatial position (1-D, 2-D, 3-D; spherical, Cartesian, or unit vector coordinates)
- Spatial velocity (1-D, 2-D, 3-D)
- Redshift (or Doppler velocity)
- Spectrum

The reason for lumping these 5 coordinates together is that their reference frames are intertwined.
The coordinate axis properties included are:

- Name
- Value
- Error
- Resolution
- Size
- Pixel size

Note that mapping metadata objects are still required to tie pixels to these axes (and vice versa).
The basic building block classes (aggregated into all-encompassing STC class) are:

- ***AstroCoordSys***: defines the coordinate systems and reference frames for all coordinates in a fully general way, allowing inclusion of, e.g., terrestrial, solar system, and planetary frames. Note that AstroCoordSys is inherited from a generalized CoordSys which is a collection of one or more CoordFrames. The coordinate frames required in AstroCoordSys are all derived from CoordFrame.
- ***AstroCoords***: defines a position in coordinate space, referenced to an AstroCoordSys, with error, resolution, pixel, etc., information. Inherited from a generalized Coords.
- ***AstroCoordArea***: (derived from CoordArea) defines a volume in coordinate space, again tied to an AstroCoordSys; this includes the more sophisticated ***Region*** specification for 2-D spatial areas.

The following shapes are being used in these UML diagrams:

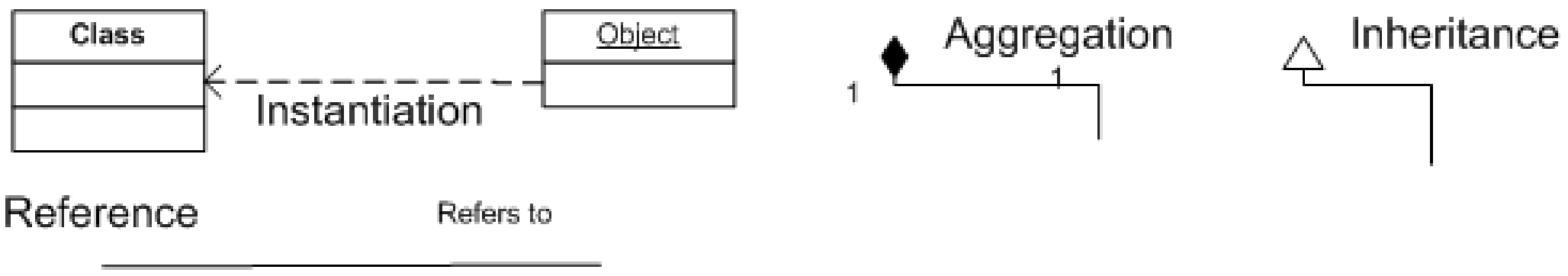

We grant that one might argue that some of the classes should be aggregates of objects, rather than aggregates of classes, but feel that this does not really affect the essence of this design which is, admittedly, somewhat fuzzy in this respect.
Note that for each of the Coordinate classes derived from the Generic Coordinate base class, the classes that are aggregated into the coordinate class (with the exception of CoordName) are constrained by the derived class. I.e., the CoordValue, CoordError, etc., for TimeCoord, SpatialCoord, RedshiftCoord should probably be classes that are multiply inherited from their generic classes as well as from the types that are shown in their UML diagrams. Similarly, one might wonder about the implied multiple inheritance of CoordInterval in the Coordinate Area diagram.

Space-Time Metadata Top-level Classes

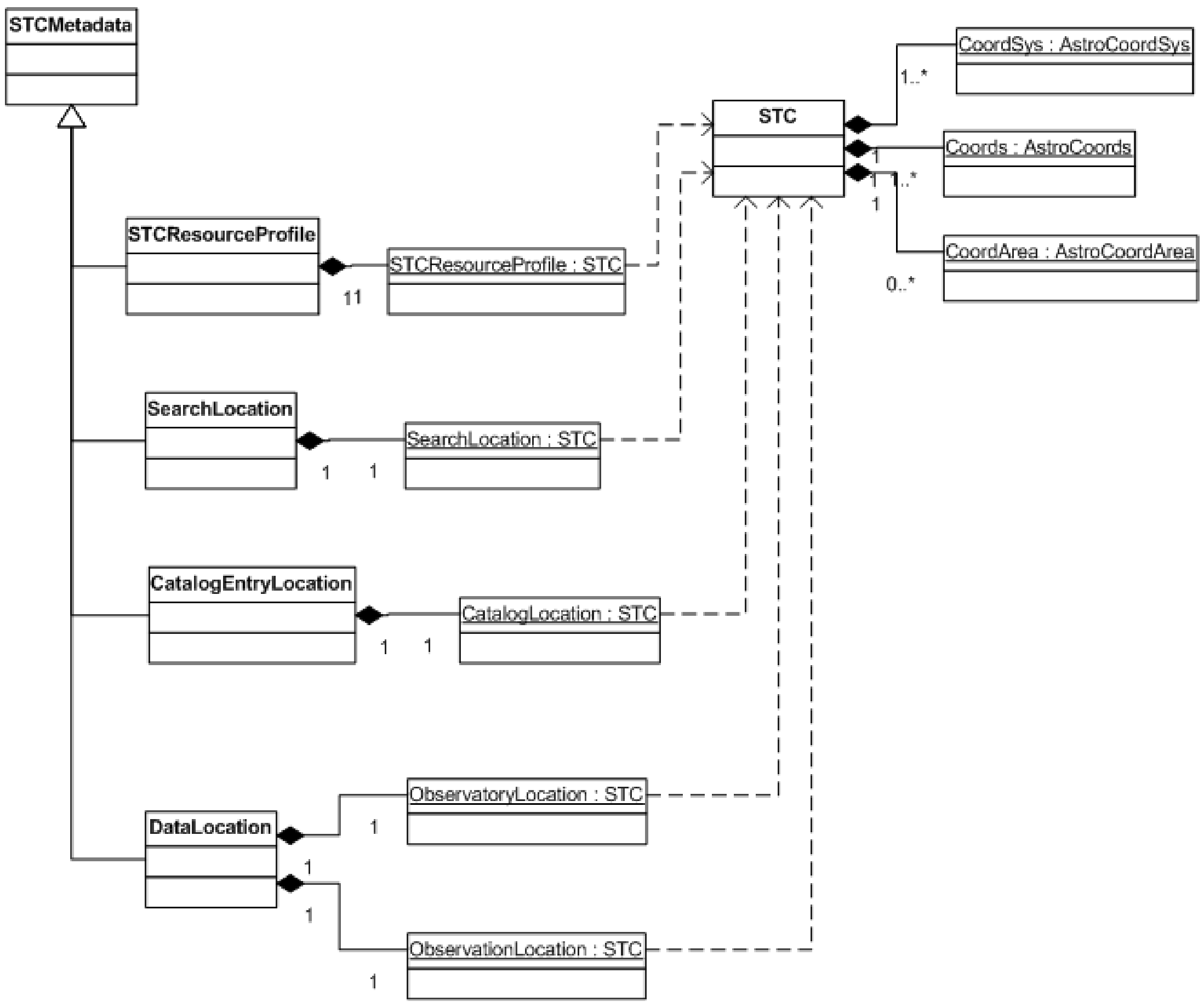

The ***STCMetadata*** class has four derived forms, each of which consists of one or two instantiations of ***STC***:

- ***STCResourceprofile***: describes the coverage of a resource.
- ***SearchLocation***: describes the coordinate volume requested in a query, as well as desired resolutions, errors, etc.
- ***CatalogEntryLocation***: describes the coordinate coverage of a set of catalog entries, as well as the coordinates, errors, etc., when these are part of the entries.
- ***DataLocation***: describes the coordinate volume occupied by a particular dataset (***ObservationLocation***), as well as the position of the observatory that produced the dataset (***ObservatoryLocation***); both are needed, in all coordinates, to facilitate transformations.

***STC*** contains one or more instantiations of ***CoordSys***, ***Coords***, and (optionally) ***CoordArea***.

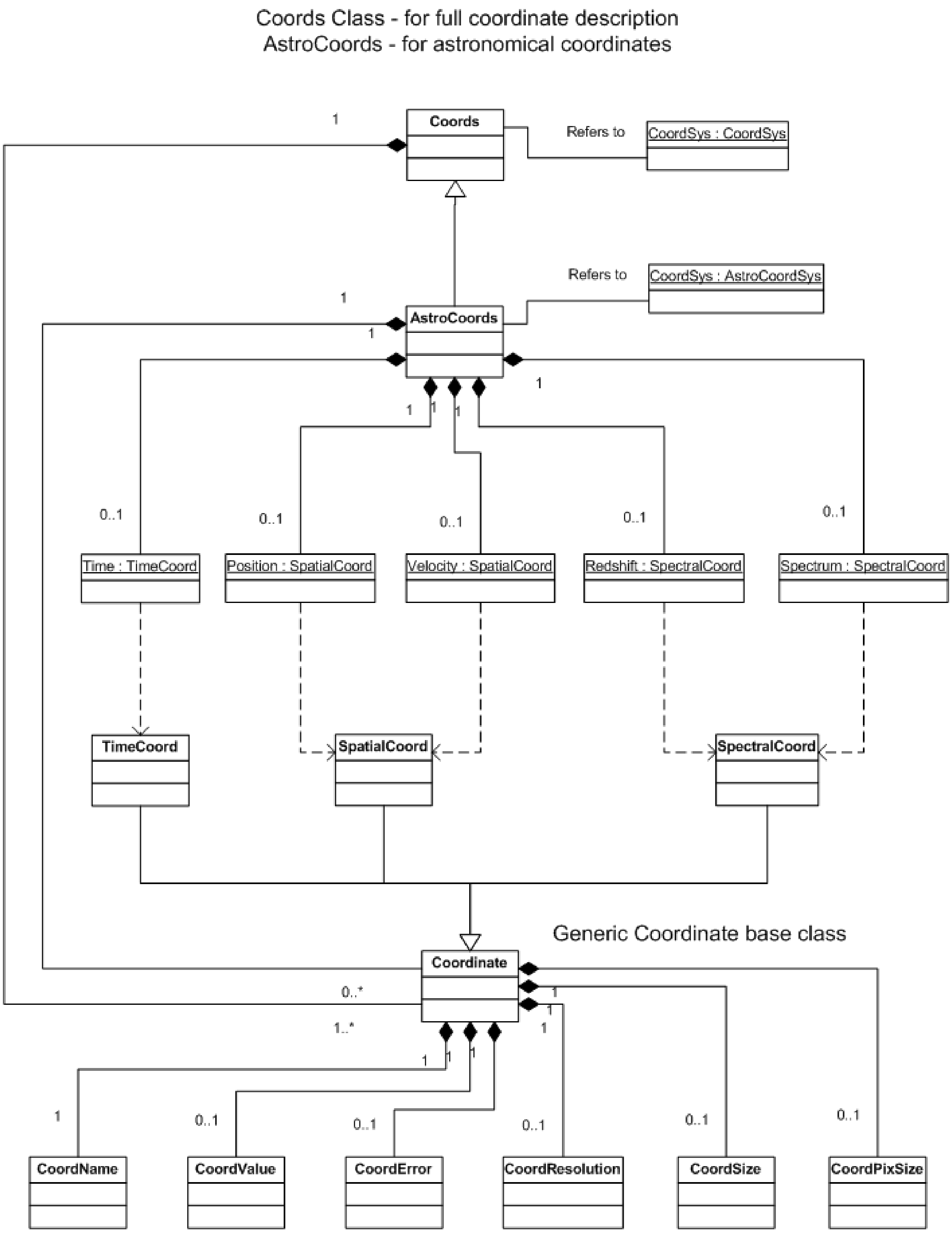

TimeCoord Class - for Time Coordinate

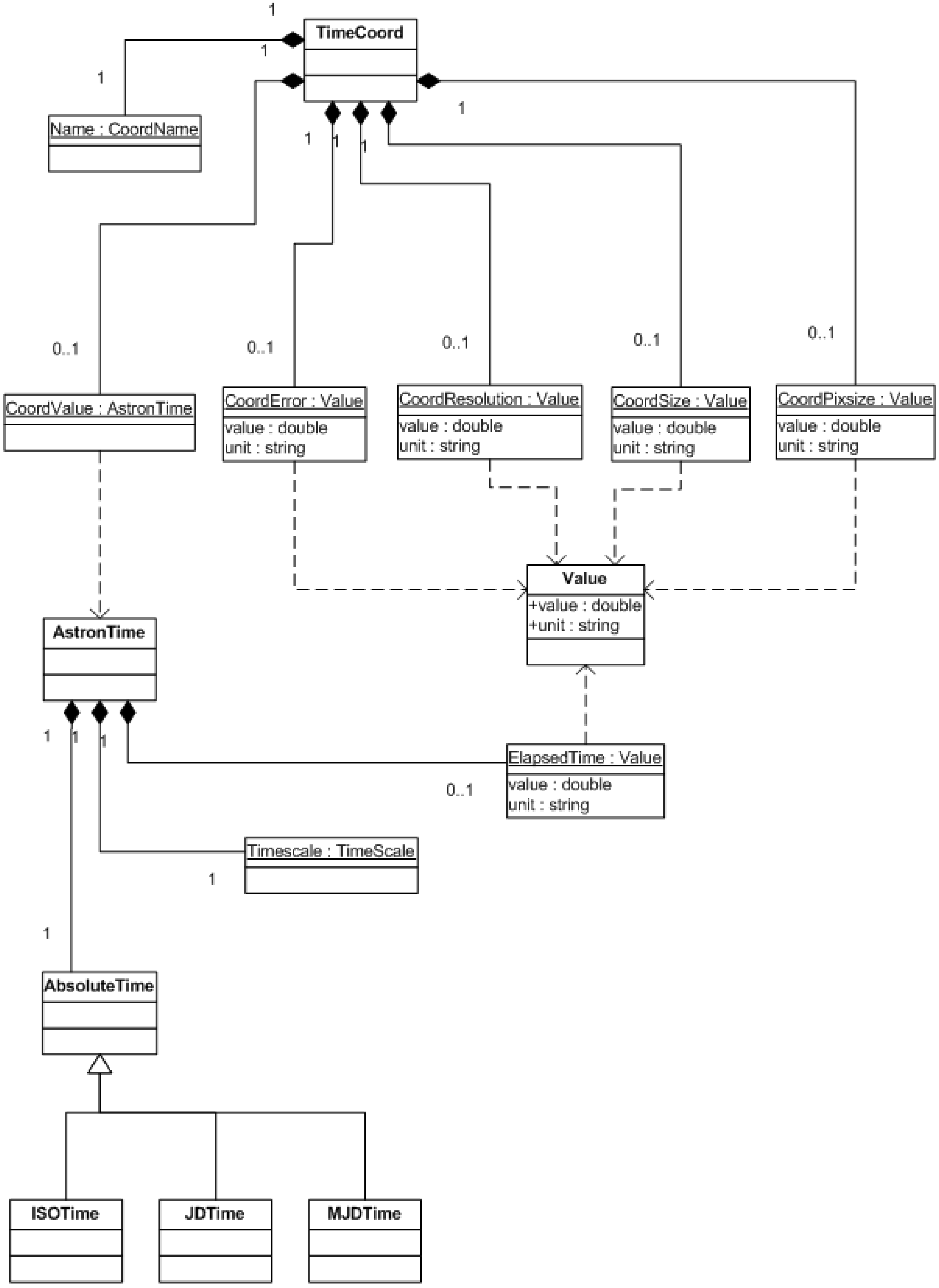

SpatialCoord Class - for Spatial Position and Velocity Coordinates

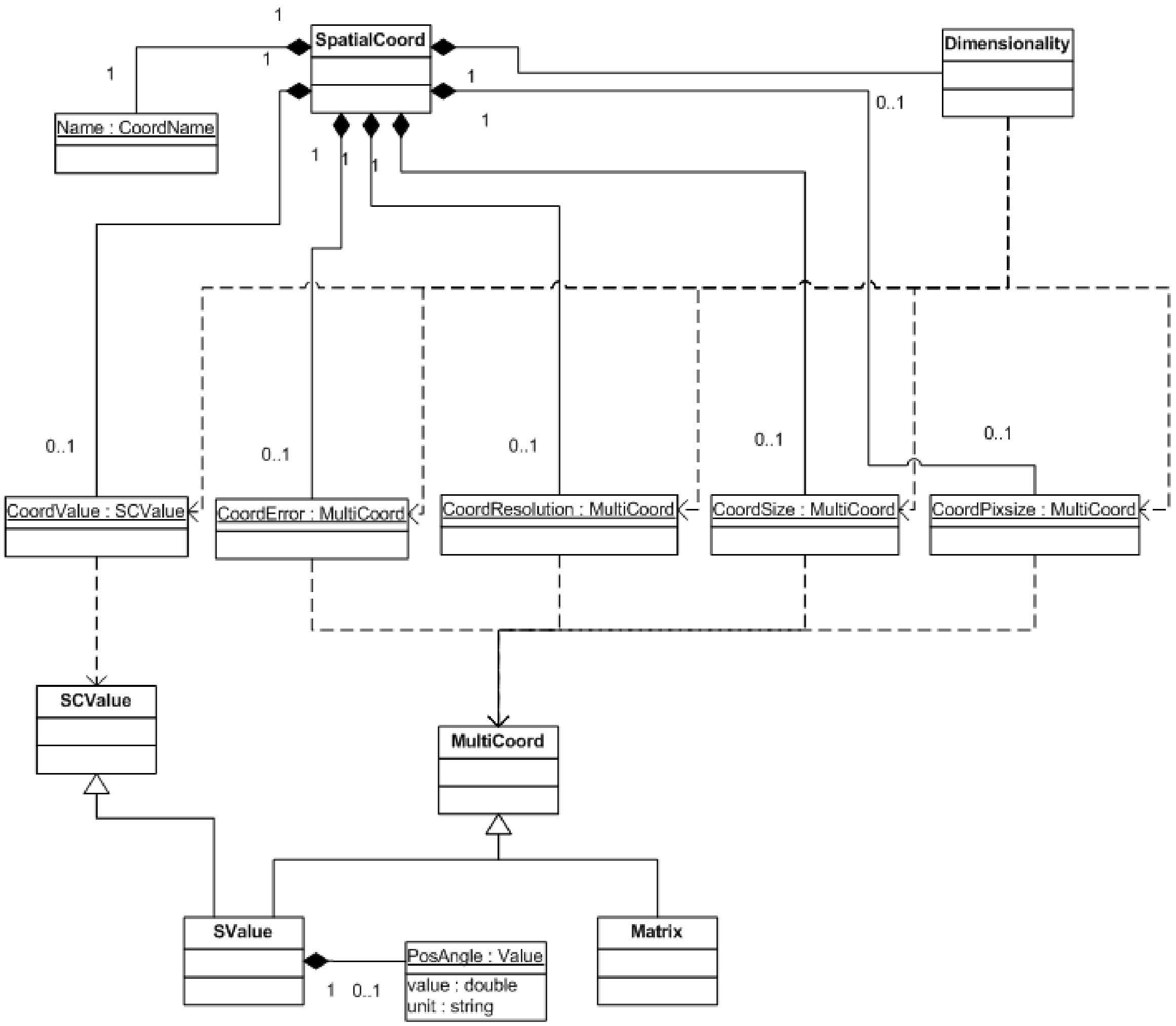

SpectralCoord Class - for Redshift and Spectral Coordinates

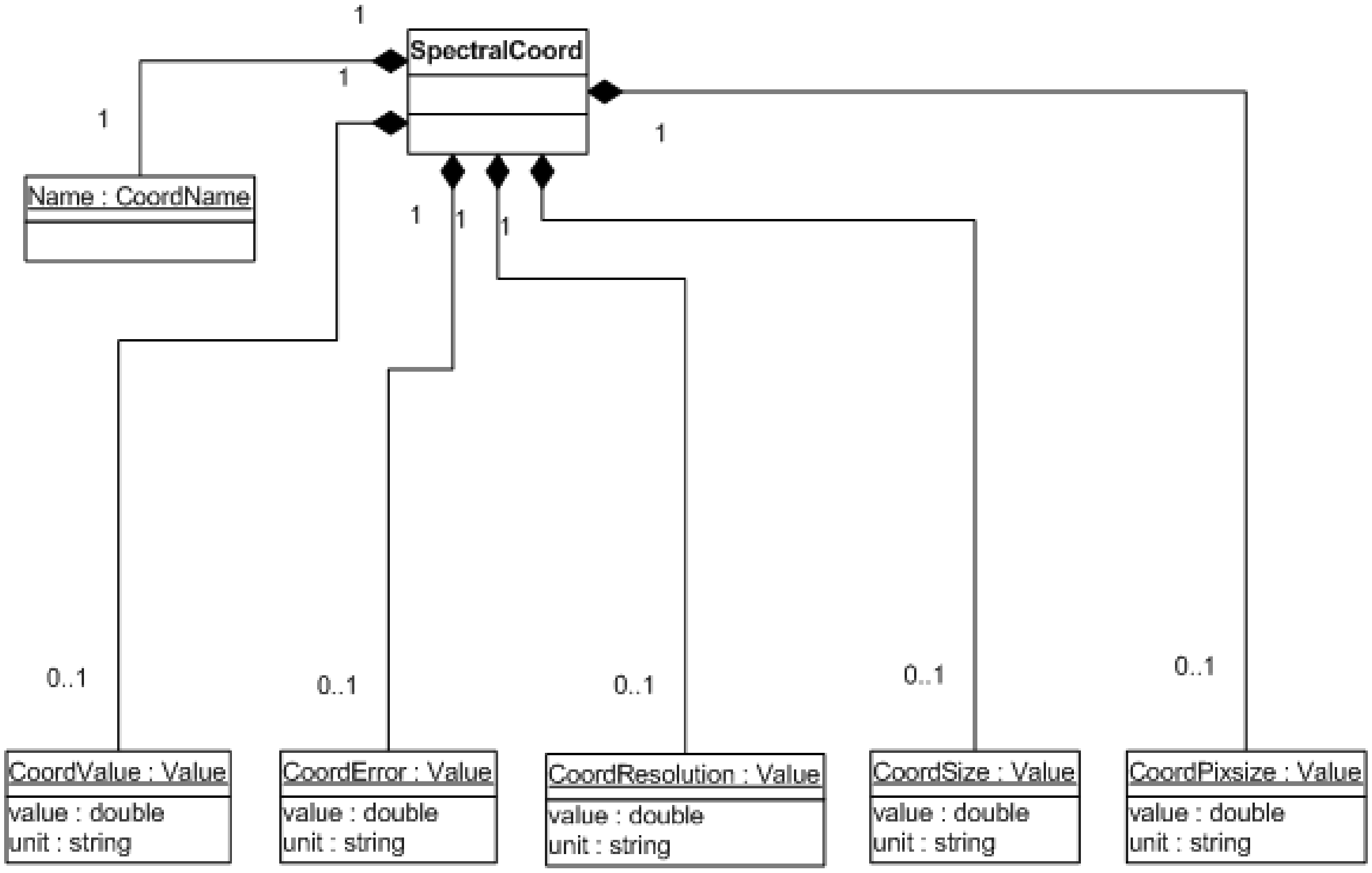

Coordinate Area - the coordinate volume that is taken up

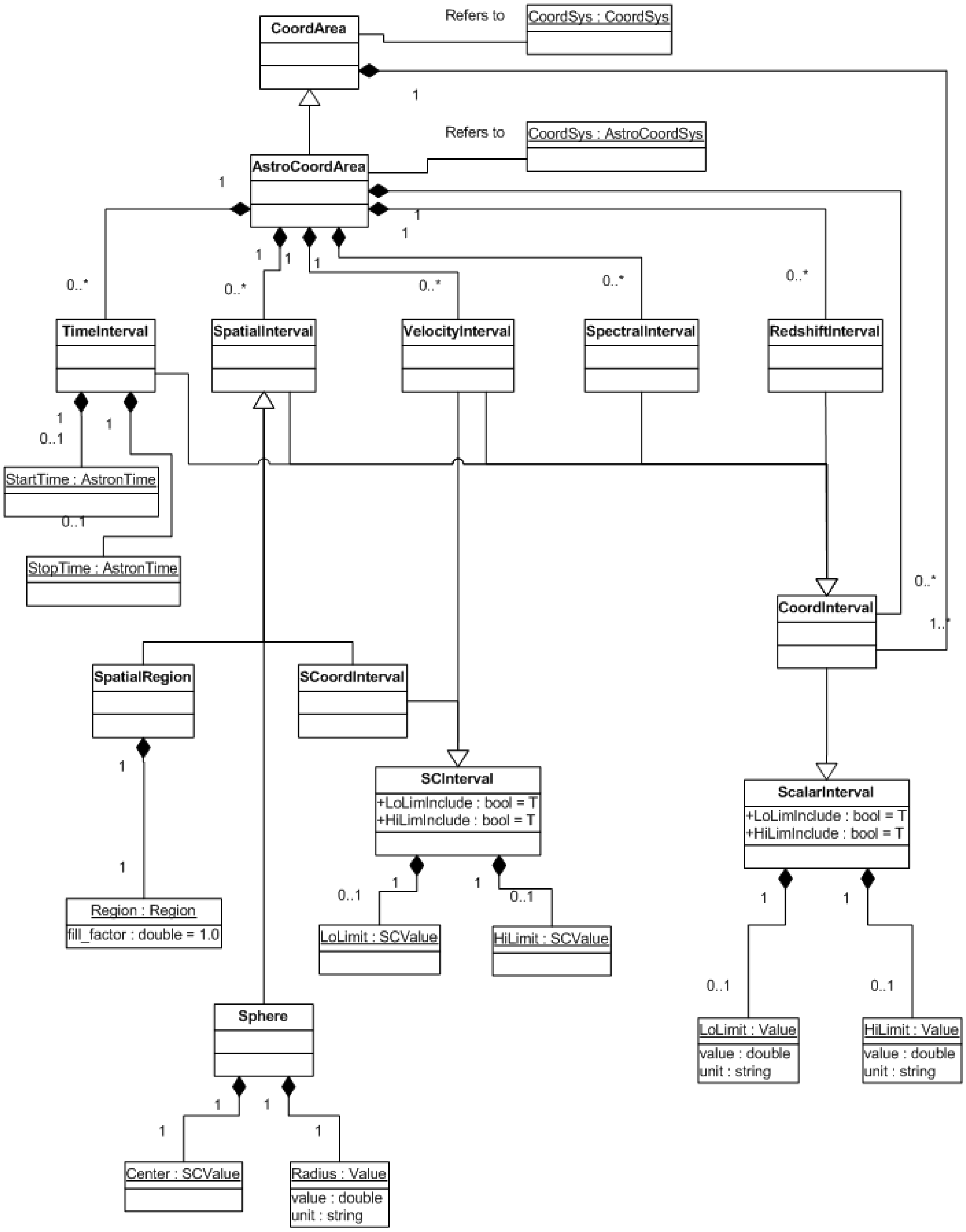

Region (spatial)

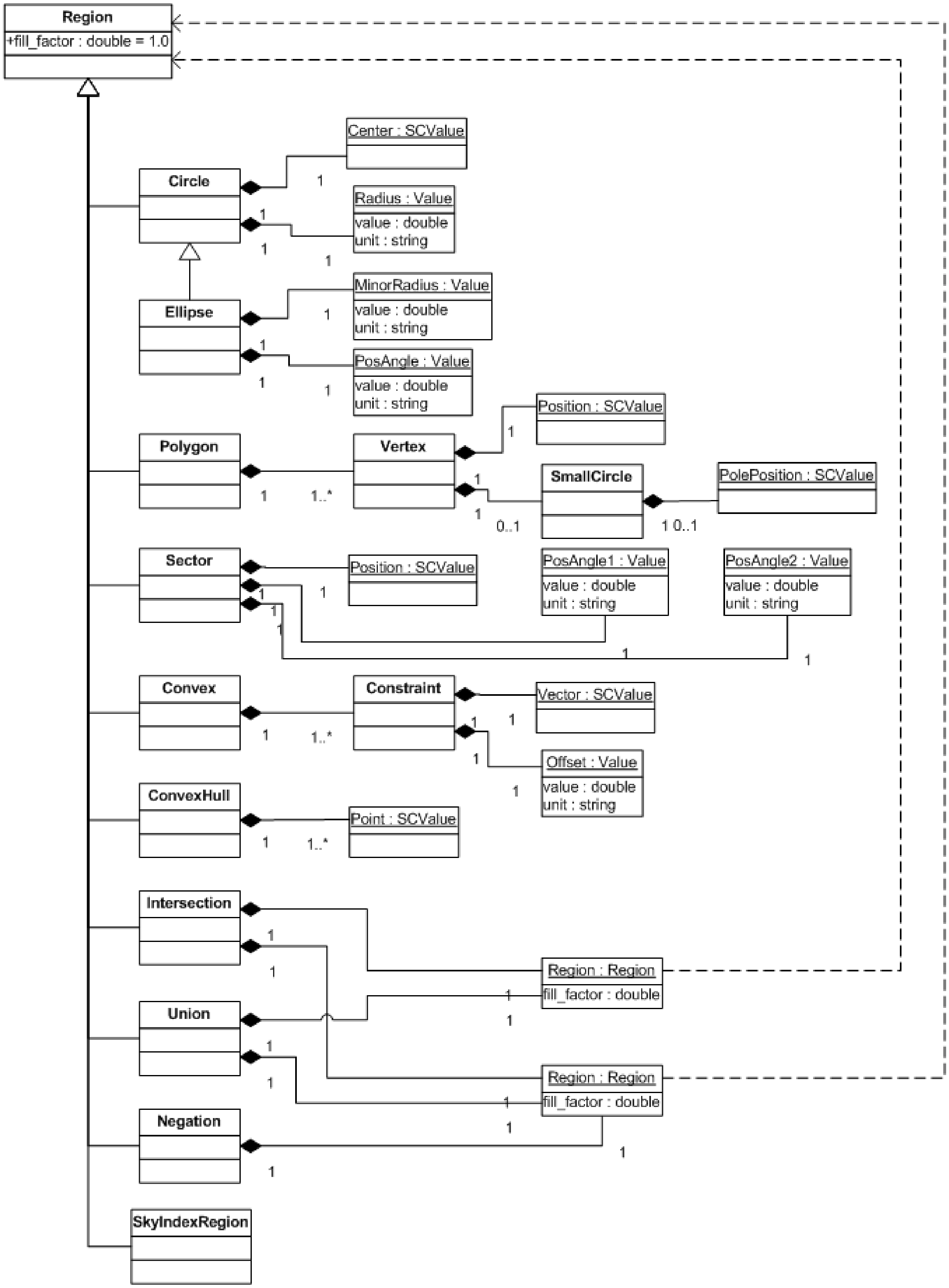

# Appendix B: Examples

## B.1 STCResourceProfile for Chandra X-ray Observatory Data Archive

```
<?xml version="1.0" encoding="UTF-8"?>
  <!--
    The Resource Profile for the Chandra archive
  -->
<STCResourceProfile xmlns="http://www.ivoa.net/xml/STC/stc-v1.30.xsd"
xmlns:xsi="http://www.w3.org/2001/XMLSchema-instance"
xsi:schemaLocation="http://www.ivoa.net/xml/STC/stc-v1.30.xsd
http://www.ivoa.net/xml/STC/stc-v1.30.xsd">
    <!--
      The native coordinate systes: TT, ICRS, and energy.
    -->
  <AstroCoordSystem id="TT-ICRS-CXO">
    <TimeFrame>
      <Name>Time</Name>
      <TimeScale>TT</TimeScale>
      <TOPOCENTER/>
    </TimeFrame>
    <SpaceFrame>
      <Name>Space</Name>
      <ICRS/>
      <TOPOCENTER/>
      <SPHERICAL coord_naxes="2"/>
    </SpaceFrame>
    <SpectralFrame>
      <Name>Energy</Name>
      <TOPOCENTER/>
    </SpectralFrame>
  </AstroCoordSystem>
    <!--
      The characteristics of the data in the archive along the
      coordinate axes.
    -->
  <AstroCoords coord_system_id="TT-ICRS-CXO">
      <!--
        Absolute timing error ranges between 5 and 100 microsec, time
        resolution between 16 microsec and 3 s, length of observation
        between 1 and 170 ks.
      -->
    <Time unit="s">
      <Name>Time</Name>
      <Error>0.000005</Error>
      <Error>0.0001</Error>
      <Resolution>0.000016</Resolution>
      <Resolution>3.0</Resolution>
      <Size>1000</Size>
      <Size>170000</Size>
    </Time>
      <!--
```

```
        Absolute spatial accuracy is 1 arcsec, resolution 0.5 arcsec;
        FOV ranges between 1000 and 4000 arcsec.
      -->
    <Position2D unit="arcsec">
      <Name>Position</Name>
      <Error2Radius>1.0</Error2Radius>
      <Resolution2Radius>0.5</Resolution2Radius>
      <Size2>
        <C1>1000</C1>
        <C2>1000</C2>
      </Size2>
      <Size2>
        <C1>4000</C1>
        <C2>4000</C2>
      </Size2>
    </Position2D>
      <!--
        Spectral information is expressed in keV, uncertainty of 0.1
        keV, resolution ranging from 0.02 to 2 keV, total bandwidth
        from 2 to 10 keV
      -->
    <Spectral unit="keV">
      <Name>Energy</Name>
      <Error>0.1</Error>
      <Resolution>0.02</Resolution>
      <Resolution>2.0</Resolution>
      <Size>2</Size>
      <Size>10</Size>
    </Spectral>
  </AstroCoords>
    <!--
      The volume in coordinate space that is occupied by the Chandra
      archive's data.
    -->
  <AstroCoordArea id="AllSky-CXO" coord_system_id="TT-ICRS-CXO">
      <!–
        The mission started in July 1999 and is still ongoing
      -->
    <TimeInterval>
      <StartTime>
        <Timescale>TT</Timescale>
        <ISOTime>1999-07-23T16:00:00</ISOTime>
      </StartTime>
    </TimeInterval>
      <!–
        Pointings are all over the sky, but the fill factor is only 2%.
      -->
    <AllSky fill_factor="0.02"/>
      <!–
        The spectral band covered ranges from 0.12 to 10.0 keV.
      -->
    <SpectralInterval unit="keV">
      <LoLimit>0.12</LoLimit>
      <HiLimit>10.0</HiLimit>
    </SpectralInterval>
  </AstroCoordArea>
</STCResourceProfile>
```

### B.2 Blue M81 Image from KPNO

```
<?xml version="1.0" encoding="UTF-8"?>
  <!--
    The ObsDataLocation element is the container for the complete
    metadata description of an observational dataset, in this case
    an image.
    It specifies the various required namespaces (for STC specifically:
    STC and Xlink, and the location of the STC schema).
    ObsDataLocation contains two main components: ObservatoryLocation
    and ObservationLocation.
  -->
<ObsDataLocation xmlns="http://www.ivoa.net/xml/STC/stc-v1.30.xsd"
xmlns:xlink="http://www.w3.org/1999/xlink"
xmlns:xsi="http://www.w3.org/2001/XMLSchema-instance"
xsi:schemaLocation="http://www.ivoa.net/xml/STC/stc-v1.30.xsd
http://www.ivoa.net/xml/STC/stc-v1.30.xsd">
    <!--
      The location of the observatory is specified by a set of
      Spatial coordinates and an associated coordinate system
      - geodetic, in this case.
    -->
  <ObservatoryLocation id="KPNO">
    <AstroCoordSystem id="TT-GEOD-TOPO">
      <TimeFrame>
        <TimeScale>TT</TimeScale>
        <TOPOCENTER/>
      </TimeFrame>
      <SpaceFrame>
        <GEO_D/>
        <TOPOCENTER/>
        <SPHERICAL coord_naxes="3"/>
      </SpaceFrame>
    </AstroCoordSystem>
    <AstroCoords coord_system_id="TT-GEOD-TOPO">
      <Position3D>
        <Value3>
          <C1 pos_unit="deg">248.4056</C1>
          <C2 pos_unit="deg">31.9586</C2>
          <C3 pos_unit="m">2158</C3>
        </Value3>
      </Position3D>
    </AstroCoords>
  </ObservatoryLocation>
    <!--
      The observation's location and the volume it occupies in
      coordinate space is specified by a set of coordinates and their
      properties (errors, resolutions, etc.) and the area or interval
      covered in each coordinate
      - with the associated coordinate system.
      Note that a generic coordinate frame has been defined to handle
      the brightness axis.
    -->
  <ObservationLocation id="M81">
```

```
    <AstroCoordSystem id="TT-ICRS-WAVELENGTH-TOPO">
      <CoordFrame ucd="em..." id="brightness">
        <Name>Brightness</Name>
        <ScalarRefFrame>
          <Scale>1.0</Scale>
        </ScalarRefFrame>
        <CARTESIAN coord_naxes="1"/>
      </CoordFrame>
      <TimeFrame>
        <TimeScale>TT</TimeScale>
        <TOPOCENTER/>
      </TimeFrame>
      <SpaceFrame id="spaceFrame">
        <ICRS/>
        <TOPOCENTER/>
        <SPHERICAL coord_naxes="2"/>
      </SpaceFrame>
      <SpectralFrame>
        <TOPOCENTER/>
      </SpectralFrame>
    </AstroCoordSystem>
    <AstroCoords coord_system_id="TT-ICRS-WAVELENGTH-TOPO">
      <ScalarCoordinate unit="mJy/arcsec**2" frame_id="brightness">
        <Error>0.001</Error>
      </ScalarCoordinate>
      <Time unit="s">
        <TimeInstant>
          <ISOTime>2004-07-15T08:23:56</ISOTime>
        </TimeInstant>
        <Resolution>1000</Resolution>
        <PixSize>1000</PixSize>
      </Time>
      <Position2D unit="deg">
        <Value2 id="Center">
          <C1>148.88821</C1>
          <C2>69.06529</C2>
        </Value2>
        <Error2Radius>0.0003 </Error2Radius>
        <Resolution2>
          <C1>0.00025</C1>
          <C2>0.00025</C2>
        </Resolution2>
        <PixSize2>
          <C1>0.0001</C1>
          <C2>0.0001</C2>
        </PixSize2>
      </Position2D>
      <Spectral unit="Angstrom">
        <Value>4600</Value>
        <Resolution>400</Resolution>
        <PixSize>400</PixSize>
      </Spectral>
    </AstroCoords>
    <AstroCoordArea id="M81Image"
    coord_system_id="TT-ICRS-WAVELENGTH-TOPO">
      <CoordScalarInterval frame_id="brightness">
        <LoLimit>0.0</LoLimit>
```

```
        <HiLimit>10.0</HiLimit>
      </CoordScalarInterval>
      <TimeInterval>
        <StartTime>
          <ISOTime>2004-07-15T08:17:36</ISOTime>
        </StartTime>
        <StopTime>
          <ISOTime>2004-07-15T08:30:16</ISOTime>
        </StopTime>
      </TimeInterval>
      <Position2VecInterval unit="deg">
        <LoLimit2Vec>
          <C1>148.18821</C1>
          <C2>68.81529</C2>
        </LoLimit2Vec>
        <HiLimit2Vec>
          <C1>149.58821</C1>
          <C2>69.31529</C2>
        </HiLimit2Vec>
      </Position2VecInterval>
      <SpectralInterval unit="Angstrom">
        <LoLimit>4400</LoLimit>
        <HiLimit>4800</HiLimit>
      </SpectralInterval>
    </AstroCoordArea>
  </ObservationLocation>
    <!--
      The image is represented as a pixel array.
      The Pixel Space specifies the properties of this pixel array
      and the way it is related to the world coordinates specified
      above.
      It consists of a pixel coordinate system that defines the pixel
      Axes and their transformation on the sky (WCS) and a pixel
      Coordinate area that defines the array's dimensions.
      Note the use of frame_id to tie pixel interval to the correct
      frame; in this case it is not strictly required since there is
      only one frame, but in the general case this is the mechanism
      used to avoid ambiguity.
    -->
  <PixelSpace>
    <PixelCoordSystem id="M81Pix">
      <PixelCoordFrame id="spacepix" axis1_order="1" axis2_order="2"
      ref_frame_id="spaceFrame">
        <Cart2DRefFrame projection="TAN">
          <Transform2 unit="deg">
            <C1>0.0001</C1>
            <C2>0.0001</C2>
          </Transform2>
        </Cart2DRefFrame>
        <CoordRefPos>
          <Vector2DCoordinate unit="deg">
            <Value2 idref="Center" xsi:nil="true"/>
          </Vector2DCoordinate>
        </CoordRefPos>
        <CARTESIAN coord_naxes="2"/>
        <ReferencePixel>
          <Pixel2D>
```

```
            <Name1>RA</Name1>
            <Name2>Dec</Name2>
            <Value2>
              <C1>512</C1>
              <C2>512</C2>
            </Value2>
          </Pixel2D>
        </ReferencePixel>
      </PixelCoordFrame>
    </PixelCoordSystem>
    <PixelCoordArea coord_system_id="M81Pix" id="M81PixImage">
      <PixelCoord2VecInterval frame_id="spacepix">
        <LoLimit2Vec>
          <C1>1</C1>
          <C2>1</C2>
        </LoLimit2Vec>
        <HiLimit2Vec>
          <C1>1024</C1>
          <C2>1024</C2>
        </HiLimit2Vec>
      </PixelCoord2VecInterval>
    </PixelCoordArea>
  </PixelSpace>
</ObsDataLocation>
```

## B.3 M81 Image Using Xlink References

```
<?xml version="1.0" encoding="UTF-8"?>
  <!--
    This example is similar to M81image.xml, but illustrates how the
    observatory location and the coordinate system definition may be
    taken from a standard library.
    For simplicity, the pixel space has been omitted.
  -->
<ObsDataLocation xmlns="http://www.ivoa.net/xml/STC/stc-v1.30.xsd"
xmlns:xlink="http://www.w3.org/1999/xlink"
xmlns:xsi="http://www.w3.org/2001/XMLSchema-instance"
xsi:schemaLocation="http://www.ivoa.net/xml/STC/stc-v1.30.xsd
http://www.ivoa.net/xml/STC/stc-v1.30.xsd">
  <ObservatoryLocation id="KPNO" xlink:type="simple"
xlink:href="ivo://STClib/Observatories#KPNO"/>
  <ObservationLocation>
    <AstroCoordSystem id="TT-ICRS-TOPO" xlink:type="simple"
xlink:href="ivo://STClib/CoordSys#TT-ICRS-TOPO"/>
    <AstroCoords coord_system_id="TT-ICRS-TOPO">
      <Time unit="s">
        <TimeInstant>
          <ISOTime>2004-07-15T08:23:56</ISOTime>
        </TimeInstant>
        <Error>2</Error>
      </Time>
      <Position2D unit="deg">
        <Value2>
          <C1>148.88821</C1>
          <C2>69.06529</C2>
        </Value2>
        <Error2Radius>0.03
        </Error2Radius>
      </Position2D>
    </AstroCoords>
    <AstroCoordArea coord_system_id="TT-ICRS-TOPO">
      <Polygon unit="deg">
        <Vertex>
          <Position><C1>148.88821</C1> <C2>68.81529</C2></Position>
        </Vertex>
        <Vertex>
          <Position><C1>148.18821</C1> <C2>69.01529</C2></Position>
        </Vertex>
        <Vertex>
          <Position><C1>148.88821</C1> <C2>69.31529</C2></Position>
        </Vertex>
        <Vertex>
          <Position><C1>149.58821</C1> <C2>69.01529</C2></Position>
        </Vertex>
      </Polygon>
    </AstroCoordArea>
  </ObservationLocation>
</ObsDataLocation>
```

## B.4 A Coordinate System with an Unusual Reference Position

```xml
<?xml version="1.0" encoding="UTF-8"?>
  <!--
    This example was proposed by Jonathan McDowell.
    It defines a coordinate system that is standard enough in time (TT)
    and space (ICRS), but puts the reference position of the spectral
    frame at redshift 4.5; note the use of frame_id to tie the
    reference position to the correct coordinate.
  -->
<STCResourceProfile id="JCMSystem" ucd="TBD"
xsi:schemaLocation="http://www.ivoa.net/xml/STC/stc-v1.30.xsd
http://www.ivoa.net/xml/STC/stc-v1.30.xsd"
xmlns="http://www.ivoa.net/xml/STC/stc-v1.30.xsd"
xmlns:xlink="http://www.w3.org/1999/xlink"
xmlns:xsi="http://www.w3.org/2001/XMLSchema-instance">
  <AstroCoordSystem id="JCMsystem">
    <TimeFrame>
      <TimeScale>TT</TimeScale>
      <TOPOCENTER/>
    </TimeFrame>
    <SpaceFrame>
      <ICRS/>
      <TOPOCENTER/>
      <SPHERICAL coord_naxes="2"/>
    </SpaceFrame>
    <SpectralFrame>
      <CoordRefPos>
        <Redshift frame_id="ZFrame">
          <Name>RestZ</Name>
          <Value>4.5</Value>
        </Redshift>
      </CoordRefPos>
    </SpectralFrame>
    <RedshiftFrame id="ZFrame" value_type="REDSHIFT">
      <DopplerDefinition>OPTICAL</DopplerDefinition>
      <BARYCENTER/>
    </RedshiftFrame>
  </AstroCoordSystem>
</STCResourceProfile>
```

## B.5 Query Specification

```
<?xml version="1.0" encoding="UTF-8"?>
  <!--
    This STC metadata description specifies a query:
      Find all images, within 2 degrees from a certain position, that
      have these properties:
        spatial resolution: between 0.0001 and 0.003 degrees
        size: between 30 and 40 arcmins on the side
        pixelsize: between 0.00005 and 0.00015 degrees
        covering a band of 300 to 600 Angstrom in the spectral range
          4000 to 7000 A
        taken since 1900.0
    AstroCoordSystem specifies the coordinate system used in this query
    AstroCoordArea specifies where to look in coordinate space
    AstroCoords specifies the (ranges of) properties along the
      coordinates
  -->
<SearchLocation xmlns="http://www.ivoa.net/xml/STC/stc-v1.30.xsd"
xmlns:xsi="http://www.w3.org/2001/XMLSchema-instance"
xsi:schemaLocation="http://www.ivoa.net/xml/STC/stc-v1.30.xsd
http://www.ivoa.net/xml/STC/stc-v1.30.xsd">
  <AstroCoordSystem id="TT-ICRS-BARY">
    <TimeFrame>
      <Name>Time</Name>
      <TimeScale>TT</TimeScale>
      <BARYCENTER/>
    </TimeFrame>
    <SpaceFrame>
      <Name>Equatorial</Name>
      <ICRS/>
      <BARYCENTER/>
      <SPHERICAL coord_naxes="2"/>
    </SpaceFrame>
    <SpectralFrame>
      <Name>Wavelength</Name>
      <BARYCENTER/>
    </SpectralFrame>
  </AstroCoordSystem>
  <AstroCoords coord_system_id="TT-ICRS-BARY">
    <Position2D unit="deg">
      <Name>RA,Dec</Name>
      <Resolution2>
        <C1>0.0001</C1>
        <C2>0.0001</C2>
      </Resolution2>
      <Resolution2>
        <C1>0.0003</C1>
        <C2>0.0003</C2>
      </Resolution2>
      <Size2>
        <C1>0.5</C1>
        <C2>0.5</C2>
      </Size2>
      <Size2>
```

```
        <C1>0.67</C1>
        <C2>0.67</C2>
      </Size2>
      <PixSize2>
        <C1>0.00005</C1>
        <C2>0.00005</C2>
      </PixSize2>
      <PixSize2>
        <C1>0.00015</C1>
        <C2>0.00015</C2>
      </PixSize2>
    </Position2D>
    <Spectral unit="Angstrom">
      <Name>Lambda</Name>
      <Resolution>300</Resolution>
      <Resolution>600</Resolution>
    </Spectral>
  </AstroCoords>
  <AstroCoordArea id="M81" coord_system_id="TT-ICRS-BARY">
    <TimeInterval>
      <StartTime>
        <ISOTime>1900-01-01T00:00:00</ISOTime>
      </StartTime>
    </TimeInterval>
    <Circle unit="deg">
      <Center>
        <C1>148.9</C1>
        <C2>69.1</C2>
      </Center>
      <Radius>2</Radius>
    </Circle>
    <SpectralInterval unit="Angstrom">
      <LoLimit>4000</LoLimit>
      <HiLimit>7000</HiLimit>
    </SpectralInterval>
  </AstroCoordArea>
</SearchLocation>
```

## B.6 Arecibo Survey Profile

```
<?xml version="1.0" encoding="UTF-8"?>
  <!--
    This document is the resource description of an Arecibo HI survey.
    It provides the volume in coordinate space that the survey
    occupies, as well as its properties along the coordinate axes.
    The survey's native spatial coordinates are Galactic, but the
    coordinate space volume is also provided in ICRS, as a service
    to ICRS queries.
  -->
<ObsDataLocation xmlns="http://www.ivoa.net/xml/STC/stc-v1.30.xsd"
xmlns:xlink="http://www.w3.org/1999/xlink"
xmlns:xsi="http://www.w3.org/2001/XMLSchema-instance"
xsi:schemaLocation="http://www.ivoa.net/xml/STC/stc-v1.30.xsd
http://www.ivoa.net/xml/STC/stc-v1.30.xsd">
    <!--
      The observatory location is specified through a reference to the
      library of observatory locations.
    -->
  <ObservatoryLocation id="Arecibo" xlink:type="simple"
  xlink:href="ivo://STClib/Observatories#Arecibo"/>
    <!--
      Here come the metadata describing the survey as observed from
      Arecibo.
    -->
  <ObservationLocation>
      <!--
        The native coordinate system of the survey: Galactic
        Coordinates and TT times as measured at the telescope, and
        Doppler velocities wrt LSR, calculated using the radio
        definition.
      -->
    <AstroCoordSystem id="TT-GAL-RADIO-LSR-TOPO">
      <TimeFrame>
        <TimeScale>TT</TimeScale>
        <TOPOCENTER/>
      </TimeFrame>
      <SpaceFrame id="GALFrame">
        <GALACTIC_II/>
        <TOPOCENTER/>
        <SPHERICAL coord_naxes="2"/>
      </SpaceFrame>
      <SpectralFrame>
        <TOPOCENTER/>
      </SpectralFrame>
      <RedshiftFrame id="VELFrame" value_type="VELOCITY">
        <DopplerDefinition>RADIO</DopplerDefinition>
        <LSR/>
      </RedshiftFrame>
    </AstroCoordSystem>
      <!--
        Alternate coordinates: ICRS reference frame for spatial
        coordinates.
      -->
```

```
    <AstroCoordSystem id="TT-ICRS-RADIO-LSR-TOPO">
      <TimeFrame>
        <TimeScale>TT</TimeScale>
        <TOPOCENTER/>
      </TimeFrame>
      <SpaceFrame>
        <ICRS/>
        <TOPOCENTER/>
        <SPHERICAL coord_naxes="2"/>
      </SpaceFrame>
      <SpectralFrame>
        <TOPOCENTER/>
      </SpectralFrame>
      <RedshiftFrame>
        <DopplerDefinition>RADIO</DopplerDefinition>
        <LSR/>
      </RedshiftFrame>
    </AstroCoordSystem>
      <!--
        The properties of the data along the coordinate axes, using
        Galactic
      -->
    <AstroCoords coord_system_id="TT-GAL-RADIO-LSR-TOPO">
        <!--
          The observations, at 100 s per pointing, were made during a
          period of one year, rendering the uncertainty in the
          observing date of individual pointings half a year.
        -->
      <Time unit="s">
        <TimeInstant>
          <ISOTime>2005-12-15T08:23:56</ISOTime>
        </TimeInstant>
        <Error>15000000</Error>
        <Resolution>100</Resolution>
        <PixSize>100</PixSize>
      </Time>
        <!--
          The uncertainty, resolution, and pixel size in spatial
          coordinates
        -->
      <Position2D unit="deg">
        <Error2Radius>0.002</Error2Radius>
        <Resolution2>
          <C1>0.05</C1>
          <C2>0.05</C2>
        </Resolution2>
        <PixSize2>
          <C1>0.01</C1>
          <C2>0.01</C2>
        </PixSize2>
      </Position2D>
        <!--
          The spectral coordinate, with size and pixel size zero, is
          the rest frequency.
        -->
      <Spectral unit="MHz">
        <Value>1420.405752</Value>
```

```
      <Size>0</Size>
      <PixSize>0</PixSize>
    </Spectral>
      <!--
        Doppler velocity characteristics: resolution and sampling
      -->
    <Redshift  unit="km" vel_time_unit="s">
      <Resolution>0.5</Resolution>
      <PixSize>0.25</PixSize>
    </Redshift>
  </AstroCoords>
    <!--
      The volume in coordinate space taken up by this dataset.
    -->
  <AstroCoordArea coord_system_id="TT-GAL-RADIO-LSR-TOPO">
      <!--
        The time interval during which the observations were made
      -->
    <TimeInterval>
      <StartTime>
        <ISOTime>2005-06-15T06:57:36</ISOTime>
      </StartTime>
      <StopTime>
        <ISOTime>2006-06-15T10:07:16</ISOTime>
      </StopTime>
    </TimeInterval>
    <!--
      The spatial bounds of the survey: 10x6 degrees
    -->
    <Position2VecInterval unit="deg">
      <LoLimit2Vec>
        <C1>10</C1>
        <C2>-3</C2>
      </LoLimit2Vec>
      <HiLimit2Vec>
        <C1>20</C1>
        <C2>3</C2>
      </HiLimit2Vec>
    </Position2VecInterval>
      <!--
        The Doppler velocity range covered: -200 to 200 km/s
      -->
    <RedshiftInterval unit="km" vel_time_unit="s">
      <LoLimit>-200</LoLimit>
      <HiLimit>200</HiLimit>
    </RedshiftInterval>
  </AstroCoordArea>
    <!--
      For the benefit of queries in ICRS/FK5, the spatial box's
      bounds are also provided in ICRS coordinates; note that now it
      is expressed as a polygon.
      Strictly speaking, two of the sides should be specified as
      small circles with the Galactic pole as pole, but we'll let
      that go, for now: it's close.
    -->
  <AstroCoordArea coord_system_id="TT-ICRS-RADIO-LSR-TOPO">
    <TimeInterval>
```

```
      <StartTime>
        <ISOTime>2005-06-15T06:57:36</ISOTime>
      </StartTime>
      <StopTime>
        <ISOTime>2006-06-15T10:07:16</ISOTime>
      </StopTime>
    </TimeInterval>
    <Polygon unit="deg">
      <Vertex>
        <Position>
          <C1>274.762</C1>
          <C2>-21.726</C2>
        </Position>
      </Vertex>
      <Vertex>
        <Position>
          <C1>269.171</C1>
          <C2>-18.810</C2>
        </Position>
      </Vertex>
      <Vertex>
        <Position>
          <C1>274.185</C1>
          <C2>-10.081</C2>
        </Position>
      </Vertex>
      <Vertex>
        <Position>
          <C1>279.606</C1>
          <C2>-12.871</C2>
        </Position>
      </Vertex>
    </Polygon>
    <RedshiftInterval unit="km" vel_time_unit="s">
      <LoLimit>-200</LoLimit>
      <HiLimit>200</HiLimit>
    </RedshiftInterval>
  </AstroCoordArea>
</ObservationLocation>
  <!--
    Here is the specification as to how the data is put into pixel
    space: a pixel coordinate system consisting of two frames and the
    pixel volume.
    We are dealing here with a 3-D cube.
  -->
<PixelSpace>
  <PixelCoordSystem id="G1Pix">
      <!--
        The first pixel frame is spatial; it is 2-D and maps to pixel
        axes 1 and 2.
        It is a CAR projection and contains a transformation, a
        reference position, and a reference pixel.
      -->
    <PixelCoordFrame id="PixGALFrame" ref_frame_id="GALFrame"
    axis1_order="1" axis2_order="2">
      <Cart2DRefFrame projection="CAR">
        <Transform2 unit="deg">
```

```
          <C1>-0.01</C1>
          <C2>0.01</C2>
        </Transform2>
      </Cart2DRefFrame>
      <CoordRefPos>
        <Vector2DCoordinate unit="deg" frame_id="GALFrame">
          <Value2>
            <C1>20</C1>
            <C2>-3</C2>
          </Value2>
        </Vector2DCoordinate>
      </CoordRefPos>
      <CARTESIAN coord_naxes="2"/>
      <ReferencePixel>
        <Pixel2D>
          <Name1>l</Name1>
          <Name2>b</Name2>
          <Value2>
            <C1>1</C1>
            <C2>1</C2>
          </Value2>
        </Pixel2D>
      </ReferencePixel>
    </PixelCoordFrame>
      <!--
        The second pixel frame is Doppler velocity and maps to the
        third pixel axis.
      -->
    <PixelCoordFrame id="PixVELFrame" ref_frame_id="VELFrame"
    axis1_order="3">
      <ScalarRefFrame>
        <Scale>0.25</Scale>
      </ScalarRefFrame>
      <CoordRefPos>
        <Redshift frame_id="VELFrame">
          <Value>-200</Value>
        </Redshift>
      </CoordRefPos>
      <CARTESIAN coord_naxes="1"/>
      <ReferencePixel>
        <Pixel1D>
          <Name>Velocity</Name>
          <Value>-200</Value>
        </Pixel1D>
      </ReferencePixel>
    </PixelCoordFrame>
  </PixelCoordSystem>
    <!--
      The volume in pixel space: 1001x601x1001 pixels.
      Note the use of frame_ids to tie things together.
    -->
  <PixelCoordArea coord_system_id="G1Pix">
    <Coord2VecInterval frame_id="PixGALFrame">
      <LoLimit2Vec>
        <C1>1</C1>
        <C2>1</C2>
      </LoLimit2Vec>
```

```
        <HiLimit2Vec>
          <C1>1001</C1>
          <C2>601</C2>
        </HiLimit2Vec>
      </Coord2VecInterval>
      <CoordScalarInterval frame_id="PixVELFrame">
        <LoLimit>1</LoLimit>
        <HiLimit>1001</HiLimit>
      </CoordScalarInterval>
    </PixelCoordArea>
  </PixelSpace>
</ObsDataLocation>
```

## B.7 Arecibo Survey Query

```
<?xml version="1.0" encoding="UTF-8"?>
  <!--
    This metadata document specifies a query to the Arecibo survey,
    requesting a subset of the survey.
  -->
<SearchLocation xmlns="http://www.ivoa.net/xml/STC/stc-v1.30.xsd"
xmlns:xlink="http://www.w3.org/1999/xlink"
xmlns:xsi="http://www.w3.org/2001/XMLSchema-instance"
xsi:schemaLocation="http://www.ivoa.net/xml/STC/stc-v1.30.xsd
http://www.ivoa.net/xml/STC/stc-v1.30.xsd">
    <!--
      The query comes in Galactic coordinates, radio-defined Doppler
      velocities wrt LSR.
    -->
  <AstroCoordSystem id="TT-GAL-RADIO-LSR-TOPO">
    <TimeFrame>
      <TimeScale>TT</TimeScale>
      <TOPOCENTER/>
    </TimeFrame>
    <SpaceFrame>
      <GALACTIC_II/>
      <TOPOCENTER/>
      <SPHERICAL coord_naxes="2"/>
    </SpaceFrame>
    <SpectralFrame>
      <TOPOCENTER/>
    </SpectralFrame>
    <RedshiftFrame>
      <DopplerDefinition>RADIO</DopplerDefinition>
      <LSR/>
    </RedshiftFrame>
  </AstroCoordSystem>
    <!--
      The desired characteristics along the coordinate axes
      (fortunately, identical to those of the survey)
    -->
  <AstroCoords coord_system_id="TT-GAL-RADIO-LSR-TOPO">
    <Position2D unit="deg">
      <Resolution2><C1>0.05</C1> <C2>0.05</C2></Resolution2>
      <PixSize2><C1>0.01</C1> <C2>0.01</C2></PixSize2>
    </Position2D>
    <Spectral unit="MHz">
      <Value>1420.405752</Value>
      <Size>0</Size>
      <PixSize>0</PixSize>
    </Spectral>
    <Redshift unit="km" vel_time_unit="s">
      <Resolution>0.5</Resolution>
      <PixSize>0.25</PixSize>
    </Redshift>
  </AstroCoords>
    <!--
      What is requested here is only a narrow strip (+/- 1 deg) along
```

```
      the Galactic plane.
    -->
  <AstroCoordArea coord_system_id="TT-GAL-RADIO-LSR-TOPO">
    <Position2VecInterval unit="deg">
      <LoLimit2Vec><C1>10</C1> <C2>-1</C2></LoLimit2Vec>
      <HiLimit2Vec><C1>20</C1> <C2>1</C2></HiLimit2Vec>
    </Position2VecInterval>
    <RedshiftInterval unit="km" vel_time_unit="s">
      <LoLimit>-200</LoLimit>
      <HiLimit>200</HiLimit>
    </RedshiftInterval>
  </AstroCoordArea>
</SearchLocation>
```

### B.8 A Galaxy Catalog

```
<?xml version="1.0" encoding="UTF-8"?>
  <!--
    The STC metadata to describe a galaxy catalog containing positions,
    errors, sizes, proper motions (I know this is silly), errors, and
    Doppler velocities in FK4-B1950 and in Super-Galactic coordinates.
    All based on optical data.
  -->
<CatalogEntryLocation xmlns="http://www.ivoa.net/xml/STC/stc-v1.30.xsd"
xmlns:xsi="http://www.w3.org/2001/XMLSchema-instance"
xmlns:xlink="http://www.w3.org/1999/xlink"
xsi:schemaLocation="http://www.ivoa.net/xml/STC/stc-v1.30.xsd
http://www.ivoa.net/xml/STC/stc-v1.30.xsd">
    <!--
      Specify the FK4 frame, with ET and optical barycentric Doppler
      velocities.
    -->
  <AstroCoordSystem id="B1950-OPTICAL-ET">
    <TimeFrame>
      <Name>Time</Name>
      <TimeScale>ET</TimeScale>
      <TOPOCENTER/>
    </TimeFrame>
    <SpaceFrame>
      <Name>PosEq</Name>
      <FK4>
        <Equinox>B1950.0</Equinox>
      </FK4>
      <BARYCENTER/>
      <SPHERICAL coord_naxes="2"/>
    </SpaceFrame>
    <SpectralFrame>
      <Name>Optical</Name>
      <TOPOCENTER/>
    </SpectralFrame>
    <RedshiftFrame value_type="VELOCITY">
      <Name>DopplerVelocity</Name>
      <DopplerDefinition>OPTICAL</DopplerDefinition>
      <BARYCENTER/>
    </RedshiftFrame>
  </AstroCoordSystem>
    <!--
      Specify the SGC frame
    -->
  <AstroCoordSystem id="SGC-OPTICAL-ET">
    <TimeFrame>
      <Name>Time</Name>
      <TimeScale>ET</TimeScale>
      <TOPOCENTER/>
    </TimeFrame>
    <SpaceFrame>
      <Name>SGC</Name>
      <SUPER_GALACTIC/>
      <BARYCENTER/>
```

```
      <SPHERICAL coord_naxes="2"/>
    </SpaceFrame>
    <SpectralFrame>
      <Name>Optical</Name>
      <TOPOCENTER/>
    </SpectralFrame>
    <RedshiftFrame value_type="VELOCITY">
      <Name>DopplerVelocity</Name>
      <DopplerDefinition>OPTICAL</DopplerDefinition>
      <GALACTIC_CENTER/>
    </RedshiftFrame>
  </AstroCoordSystem>
    <!--
      The data in FK5.
      The IDREFs say that the values are not contained in the actual
      elements but refer to columns in a table that should follow in
      this document.
    -->
  <AstroCoords coord_system_id="B1950-OPTICAL-ET">
    <Position2D unit="deg">
      <Name>RA,Dec</Name>
      <Value2>
        <C1 xsi:nil="true" idref="  Column02"/>
        <C2 xsi:nil="true" idref="  Column03"/>
      </Value2>
      <Error2>
        <C1 xsi:nil="true" idref="  Column04"/>
        <C2 xsi:nil="true" idref="  Column05"/>
      </Error2>
      <Size2>
        <C1 xsi:nil="true" idref="  Column06"/>
        <C2 xsi:nil="true" idref="  Column07"/>
      </Size2>
    </Position2D>
    <Velocity2D unit="arcsec" vel_time_unit="a">
      <Name>ProperMotion</Name>
      <Value2>
        <C1 xsi:nil="true" idref="  Column08"/>
        <C2 xsi:nil="true" idref="  Column09"/>
      </Value2>
      <Error2>
        <C1 xsi:nil="true" idref="  Column10"/>
        <C2 xsi:nil="true" idref="  Column11"/>
      </Error2>
    </Velocity2D>
    <Redshift unit="km" vel_time_unit="s">
      <Name>Vrad(barycenter)</Name>
      <Value xsi:nil="true" idref="  Column12"/>
      <Error xsi:nil="true" idref="  Column13"/>
    </Redshift>
  </AstroCoords>
    <!--
      The data columns with the data in SGC.
    -->
  <AstroCoords coord_system_id="SGC-OPTICAL-ET">
    <Position2D unit="deg">
      <Name>SGLong,SGLat</Name>
```

```
      <Value2>
        <C1 xsi:nil="true" idref="  Column14"/>
        <C2 xsi:nil="true" idref="  Column15"/>
      </Value2>
      <Error2>
        <C1 xsi:nil="true" idref="  Column16"/>
        <C2 xsi:nil="true" idref="  Column17"/>
      </Error2>
      <Size2>
        <C1 xsi:nil="true" idref="  Column18"/>
        <C2 xsi:nil="true" idref="  Column19"/>
      </Size2>
    </Position2D>
    <Redshift unit="km" vel_time_unit="s">
      <Name>Vrad(Galcenter)</Name>
      <Value xsi:nil="true" idref="  Column20"/>
      <Error xsi:nil="true" idref="  Column21"/>
    </Redshift>
  </AstroCoords>
    <!--
      This Coordinate Area element provides the boundaries of this
      catalog:
      The period during which the observations were made, the area on
      the sky that was covered, the spectral band used, and the Doppler
      velocity range.
    -->
  <AstroCoordArea coord_system_id="B1950-OPTICAL-ET"
  id="RA6-18hDec20-70deg">
    <TimeInterval>
      <StartTime>
        <Timescale>ET</Timescale>
        <JDTime>2440000</JDTime>
      </StartTime>
      <StopTime>
        <Timescale>ET</Timescale>
        <JDTime>2441000</JDTime>
      </StopTime>
    </TimeInterval>
    <Polygon unit="deg">
      <Vertex>
        <Position>
          <C1>270</C1>
          <C2>20</C2>
        </Position>
      </Vertex>
      <Vertex>
        <Position>
          <C1>90</C1>
          <C2>20</C2>
        </Position>
        <SmallCircle/>
      </Vertex>
      <Vertex>
        <Position>
          <C1>90</C1>
          <C2>70</C2>
        </Position>
```

```
      </Vertex>
      <Vertex>
        <Position>
          <C1>270</C1>
          <C2>70</C2>
        </Position>
        <SmallCircle/>
      </Vertex>
    </Polygon>
    <SpectralInterval unit="Angstrom">
      <LoLimit>5000</LoLimit>
      <HiLimit>6500</HiLimit>
    </SpectralInterval>
    <RedshiftInterval unit="km" vel_time_unit="s">
      <HiLimit>10000</HiLimit>
    </RedshiftInterval>
  </AstroCoordArea>
    <!–
      At this point the table columns 1-21 should be specified with
      ID attributes Column01 through Column21 and filled with data.
    -->
  </CatalogEntryLocation>
```

## B.9 Using a FITS File

```
<?xml version="1.0" encoding="UTF-8"?>
<!--
    The ObsDataLocation element is the container for the
    complete metadata description of an observational dataset,
    in this case an image.
    It specifies the various required namespaces (STC and Xlink),
    and the location of the STC schema.
    ObsDataLocation contains two main components: ObservatoryLocation
    and ObservationLocation.
    This example shows how to do this for a spacecraft observatory,
    using an orbit ephemeris file.
    The data file may be found at:
ftp://cdaftp.cfa.harvard.edu/pub/science/ao02/cat5/1952/primary/acisf01952N002_cntr_img2.fits.gz
  -->
<ObsDataLocation xmlns="http://www.ivoa.net/xml/STC/stc-v1.30.xsd"
xmlns:xlink="http://www.w3.org/1999/xlink"
xmlns:xsi="http://www.w3.org/2001/XMLSchema-instance"
xsi:schemaLocation="http://www.ivoa.net/xml/STC/stc-v1.30.xsd
http://www.ivoa.net/xml/STC/stc-v1.30.xsd">
  <!--
      The location of the observatory is specified by a Chandra orbit
      ephemeris file and an associated coordinate system – geocentric
      3-D Cartesian aligned with FK5
    -->
  <ObservatoryLocation id="CXO">
    <AstroCoordSystem id="TT-TOPO-FK5-GEO">
      <TimeFrame>
        <TimeScale>TT</TimeScale>
        <TOPOCENTER/>
      </TimeFrame>
      <SpaceFrame>
        <FK5>
          <Equinox>J2000.0</Equinox>
        </FK5>
        <GEOCENTER/>
        <CARTESIAN coord_naxes="3"/>
      </SpaceFrame>
    </AstroCoordSystem>
    <AstroCoords coord_system_id="TT-TOPO-FK5-GEO">
      <CoordFile>
        <FITSFile hdu_name="ORBITEPHEM" hdu_num="1">
         ftp://cda.cfa.harvard.edu/pub/bary/orbitf129297900N001_eph0.fits
        </FITSFile>
        <FITSTime>
          <Value>Time</Value>
        </FITSTime>
        <FITSPosition>
          <Value>X,Y,Z</Value>
        </FITSPosition>
        <FITSVelocity>
          <Value>Vx,Vy,Vz</Value>
        </FITSVelocity>
      </CoordFile>
```

```
    </AstroCoords>
  </ObservatoryLocation>
  <!--
      The observation's location and the volume it occupies in
      coordinate space is specified by a set of coordinates and their
      properties (errors, resolutions, etc.) and the area or interval
      covered in each coordinate - with the associated
      coordinate system.
      Note that this coordinate system includes a generic coordinate
      for brightness.
      This part of the document is similar to Appendix B.2.
    -->
  <ObservationLocation id="M81">
    <AstroCoordSystem id="TT-ICRS-ENERGY-TOPO">
      <CoordFrame id="counts">
        <Name>PhotonCounts</Name>
        <ScalarRefFrame>
          <Scale>1.0</Scale>
        </ScalarRefFrame>
        <CARTESIAN coord_naxes="1"/>
      </CoordFrame>
      <TimeFrame>
        <TimeScale>TT</TimeScale>
        <TOPOCENTER/>
      </TimeFrame>
      <SpaceFrame id="spaceFrame">
        <ICRS/>
        <TOPOCENTER/>
        <SPHERICAL coord_naxes="2"/>
      </SpaceFrame>
      <SpectralFrame>
        <TOPOCENTER/>
      </SpectralFrame>
    </AstroCoordSystem>
    <AstroCoords coord_system_id="TT-ICRS-ENERGY-TOPO">
      <ScalarCoordinate unit="count/s" frame_id="counts">
        <Resolution>1</Resolution>
      </ScalarCoordinate>
      <Time unit="s">
        <TimeInstant>
          <MJDTime>52311.266</MJDTime>
        </TimeInstant>
        <Resolution>49664</Resolution>
        <PixSize>49664</PixSize>
      </Time>
      <Position2D unit="deg">
        <Value2 id="Center">
          <C1>350.9171576</C1>
          <C2>58.79377795</C2>
        </Value2>
        <Error2Radius>0.0003 </Error2Radius>
        <Resolution2>
          <C1>0.00015</C1>
          <C2>0.00015</C2>
        </Resolution2>
        <PixSize2>
          <C1>0.0001367</C1>
```

```
        <C2>0.0001367</C2>
      </PixSize2>
    </Position2D>
    <Spectral unit="keV">
      <Value>2.5</Value>
      <Resolution>7</Resolution>
      <PixSize>7</PixSize>
    </Spectral>
  </AstroCoords>
  <AstroCoordArea id="Cas-A" coord_system_id="TT-ICRS-ENERGY-TOPO">
    <CoordScalarInterval frame_id="counts">
      <LoLimit>0.0</LoLimit>
    </CoordScalarInterval>
    <TimeInterval>
      <StartTime>
        <ISOTime>2002-02-06T06:23:17</ISOTime>
      </StartTime>
      <StopTime>
        <ISOTime>2002-02-06T20:39:48</ISOTime>
      </StopTime>
    </TimeInterval>
    <Polygon unit="deg">
      <Vertex>
        <Position>
          <C1>350.99584</C1>
          <C2>58.74772</C2>
        </Position>
      </Vertex>
      <Vertex>
        <Position>
          <C1>350.72609</C1>
          <C2>58.74760</C2>
        </Position>
      </Vertex>
      <Vertex>
        <Position>
          <C1>350.72532</C1>
          <C2>58.88741</C2>
        </Position>
      </Vertex>
      <Vertex>
        <Position>
          <C1>350.99616</C1>
          <C2>58.88753</C2>
        </Position>
      </Vertex>
    </Polygon>
    <SpectralInterval unit="keV">
      <LoLimit>0.5</LoLimit>
      <HiLimit>7.0</HiLimit>
    </SpectralInterval>
  </AstroCoordArea>
</ObservationLocation>
```

```
  <!--
      The image is represented as a pixel array.
      The Pixel Space specifies the properties of this pixel array
      and the way it is related to the world coordinates specified
      above.
      It consists of a pixel coordinate system that defines the pixel
      axes and their transformation on the sky (WCS) and a pixel
      coordinate area that defines the array's dimensions.
      Note the use of frame_id to tie pixel interval to the correct
      frame; in this case it is not strictly required since there is
      only one frame, but in the general case this is the mechanism
      used to avoid ambiguity.
    -->
  <PixelSpace>
    <PixelCoordSystem id="Cas-APix">
      <PixelCoordFrame id="spacepix" axis1_order="1" axis2_order="2"
        ref_frame_id="spaceFrame">
        <Cart2DRefFrame projection="TAN">
          <Transform2 unit="deg">
            <C1>0.0001367</C1>
            <C2>0.0001367</C2>
          </Transform2>
        </Cart2DRefFrame>
        <CoordRefPos>
          <Vector2DCoordinate unit="deg">
            <Value2 idref="Center" xsi:nil="true"/>
          </Vector2DCoordinate>
        </CoordRefPos>
        <CARTESIAN coord_naxes="2"/>
        <ReferencePixel>
          <Pixel2D>
            <Name1>RA</Name1>
            <Name2>Dec</Name2>
            <Value2>
              <C1>299.69</C1>
              <C2>337.84</C2>
            </Value2>
          </Pixel2D>
        </ReferencePixel>
      </PixelCoordFrame>
    </PixelCoordSystem>
    <PixelCoordArea coord_system_id="Cas-APix" id="Cas-APixImage">
      <PixelCoord2VecInterval frame_id="spacepix">
        <LoLimit2Vec>
          <C1>1</C1>
          <C2>1</C2>
        </LoLimit2Vec>
        <HiLimit2Vec>
          <C1>1025</C1>
          <C2>1024</C2>
        </HiLimit2Vec>
      </PixelCoord2VecInterval>
    </PixelCoordArea>
  </PixelSpace>
</ObsDataLocation>
```

## B.10 Orbit Ephemeris

The following example shows an elliptical orbit in three representations and a parabolic orbit.

```
<?xml version="1.0" encoding="UTF-8"?>

  <!—-
    Sample XML file containing the orbital information from
    MPEC 1998-F10: 1998 EP4 in three representations
  -->
<CatalogEntryLocation xmlns="http://www.ivoa.net/xml/STC/stc-v1.30.xsd"
xmlns:xlink="http://www.w3.org/1999/xlink"
xmlns:xsi="http://www.w3.org/2001/XMLSchema-instance"
xsi:schemaLocation="http://www.ivoa.net/xml/STC/stc-v1.30.xsd
http://www.ivoa.net/xml/STC/stc-v1.30.xsd">

    <!—-
      First define the three coordinate systems we are going to use
    -->
  <AstroCoordSystem xsi:nil="true" xlink:type="simple"
    xlink:href="ivo://STClib/CoordSys#TDB-ECLIPTIC-BARY"
    id="TDB-ECLIPTIC-BARY"/>
  <AstroCoordSystem xsi:nil="true" xlink:type="simple"
    xlink:href="ivo://STClib/CoordSys#TDB-UNIT-ECLIPTIC-BARY"
    id="TDB-UNIT-ECLIPTIC-BARY"/>
  <AstroCoordSystem xsi:nil="true" xlink:type="simple"
    xlink:href="ivo://STClib/CoordSys#TDB-ICRS-BARY"
    id="TDB-ICRS-BARY"/>

    <!—-
      Next the actual coordinate values
      First, the orbit defined by orbital elements
    -->
  <AstroCoords coord_system_id="TDB-ECLIPTIC-BARY">
    <Time>
      <TimeInstant><ISOTime>1998-03-08T00:00:00</ISOTime></TimeInstant>
    </Time>
    <Orbit>
      <a unit="AU">1.5610990</a>
      <e>0.4412673</e>
      <i unit="deg">7.21282</i>
      <Node unit="deg">353.14214</Node>
      <Aop unit="deg">265.00121</Aop>
      <M unit="deg">319.73232</M>
      <T><ISOTime>1998-03-08T00:00:00</ISOTime></T>
    </Orbit>
  </AstroCoords>

    <!—-
      Then the pole and pericenter vectors of the orbit
    -->
  <AstroCoords coord_system_id="TDB-UNIT-ECLIPTIC-BARY">
    <Time>
      <TimeInstant><ISOTime>1998-03-08T00:00:00</ISOTime></TimeInstant>
```

```
    </Time>
    <Position3D>
      <Name>P</Name>
      <Value3>
        <C1>-0.18756117</C1><C2>-0.84401876</C2><C3>-0.50243968</C3>
      </Value3>
    </Position3D>
  </AstroCoords>
  <AstroCoords coord_system_id="TDB-UNIT-ECLIPTIC-BARY" >
    <Time>
      <TimeInstant><ISOTime>1998-03-08T00:00:00</ISOTime></TimeInstant>
    </Time>
    <Position3D>
      <Name>Q</Name>
      <Value3>
        <C1>+0.98213469</C1><C2>-0.16897175</C2><C3>-0.08282516</C3>
      </Value3>
    </Position3D>
  </AstroCoords>

    <!--
      Third, a list of RAs and Decs at 5-day intervals
     -->
  <AstroCoords coord_system_id="TDB-ICRS-BARY" >
    <Time>
      <TimeInstant><ISOTime>1998-03-18T00:00:00</ISOTime></TimeInstant>
    </Time>
    <Position2D unit="deg">
      <Value2><C1>179.52</C1><C2>-3.933</C2></Value2>
    </Position2D>
  </AstroCoords>
  <AstroCoords coord_system_id="TDB-ICRS-BARY" >
    <Time>
      <TimeInstant><ISOTime>1998-03-23T00:00:00</ISOTime></TimeInstant>
    </Time>
    <Position2D unit="deg">
      <Value2><C1>176.53</C1><C2>-7.713</C2></Value2>
    </Position2D>
  </AstroCoords>
  <AstroCoords coord_system_id="TDB-ICRS-BARY" >
    <Time>
      <TimeInstant><ISOTime>1998-03-28T00:00:00</ISOTime></TimeInstant>
    </Time>
    <Position2D unit="deg">
      <Value2><C1>171.91</C1><C2>-13.343</C2></Value2>
    </Position2D>
  </AstroCoords>
  <AstroCoords coord_system_id="TDB-ICRS-BARY" >
    <Time>
      <TimeInstant><ISOTime>1998-04-02T00:00:00</ISOTime></TimeInstant>
    </Time>
    <Position2D unit="deg">
      <Value2><C1>164.30</C1><C2>-22.002</C2></Value2>
    </Position2D>
  </AstroCoords>
  <AstroCoords coord_system_id="TDB-ICRS-BARY" >
    <Time>
```

```
      <TimeInstant><ISOTime>1998-04-07T00:00:00</ISOTime></TimeInstant>
    </Time>
    <Position2D unit="deg">
      <Value2><C1>150.28</C1><C2>-34.952</C2></Value2>
    </Position2D>
  </AstroCoords>
  <AstroCoords coord_system_id="TDB-ICRS-BARY" >
    <Time>
      <TimeInstant><ISOTime>1998-04-12T00:00:00</ISOTime></TimeInstant>
    </Time>
    <Position2D unit="deg">
      <Value2><C1>121.76</C1><C2>-49.342</C2></Value2>
    </Position2D>
  </AstroCoords>
  <AstroCoords coord_system_id="TDB-ICRS-BARY" >
    <Time>
      <TimeInstant><ISOTime>1998-04-17T00:00:00</ISOTime></TimeInstant>
    </Time>
    <Position2D unit="deg">
      <Value2><C1>79.64</C1><C2>-52.920</C2></Value2>
    </Position2D>
  </AstroCoords>
  <AstroCoords coord_system_id="TDB-ICRS-BARY" >
    <Time>
      <TimeInstant><ISOTime>1998-04-22T00:00:00</ISOTime></TimeInstant>
    </Time>
    <Position2D unit="deg">
      <Value2><C1>51.43</C1><C2>-45.930</C2></Value2>
    </Position2D>
  </AstroCoords>
  <AstroCoords coord_system_id="TDB-ICRS-BARY" >
    <Time>
      <TimeInstant><ISOTime>1998-04-27T00:00:00</ISOTime></TimeInstant>
    </Time>
    <Position2D unit="deg">
      <Value2><C1>37.82</C1><C2>-38.272</C2></Value2>
    </Position2D>
  </AstroCoords>
  <AstroCoords coord_system_id="TDB-ICRS-BARY" >
    <Time>
      <TimeInstant><ISOTime>1998-05-02T00:00:00</ISOTime></TimeInstant>
    </Time>
    <Position2D unit="deg">
      <Value2><C1>30.69</C1><C2>-32.217</C2></Value2>
    </Position2D>
  </AstroCoords>
  <AstroCoords coord_system_id="TDB-ICRS-BARY" >
    <Time>
      <TimeInstant><ISOTime>1998-05-07T00:00:00</ISOTime></TimeInstant>
    </Time>
    <Position2D unit="deg">
      <Value2><C1>26.62</C1><C2>-27.443</C2></Value2>
    </Position2D>
  </AstroCoords>
```

```
    <!—-
      Finally, the parabolic orbit from IAUC 8653: C/2006 A1
    -->
  <AstroCoords coord_system_id="TDB-ECLIPTIC-BARY" >
    <Time>
      <TimeInstant><ISOTime>2006-01-05T00:00:00</ISOTime></TimeInstant>
    </Time>
    <Orbit>
      <a unit="AU">0.56748</a>
      <e>1.0</e>
      <i unit="deg">93.230</i>
      <Node unit="deg">212.275</Node>
      <Aop unit="deg">350.499</Aop>
      <T><ISOTime>2006-02-22T16:06:14</ISOTime></T>
    </Orbit>
  </AstroCoords>

</CatalogEntryLocation>
```

## B.11 Coordinate Table

The following example shows how one can easily write a custom schema – in this case for creating lists of coordinates – that automatically pulls in complete coordinate metadata from STC.
First the example schema:

```
<?xml version="1.0" encoding="UTF-8"?>
<xs:schema xmlns="stcTab.xsd"
    xmlns:xs="http://www.w3.org/2001/XMLSchema"
    xmlns:stc="http://www.ivoa.net/xml/STC/stc-v1.30.xsd"
    xmlns:xlink="http://www.w3.org/1999/xlink"
    targetNamespace="stcTab.xsd" elementFormDefault="qualified">
  <xs:import namespace="http://www.ivoa.net/xml/STC/stc-v1.30.xsd"
    schemaLocation="http://www.ivoa.net/xml/STC/stc-v1.30.xsd"/>
  <!--
      The root element of a stcTab document is an STCTable which
      May contain tabulated coordinate data, including generic
      ones or relative times.
      The STCTable takes an STCmetadata element from STC for the
      Coordinate metadata (this may be a CatalogEntryLocation,
      ResourceProfile, SearchLocation, or ObsDataLocation element),
      followed by a Table consisting of a header row and at least
      one row of data; the data are doubles, integers, strings, or
      ISO-8601 datatime strings.
      The table is easily convertible to HTML.
    -->
  <xs:element name="STCTable" type="stcTableType"/>
  <xs:complexType name="stcTableType">
    <xs:sequence>
      <xs:element ref="stc:STCmetadata"/>
      <xs:element name="Table" type="tableType"/>
    </xs:sequence>
  </xs:complexType>
  <xs:complexType name="tableType">
    <xs:sequence>
      <xs:element name="Header" type="headerType"/>
      <xs:element name="TR" type="rowType" maxOccurs="unbounded"/>
    </xs:sequence>
  </xs:complexType>
  <xs:complexType name="headerType">
    <xs:sequence>
      <xs:element name="TH" type="columnType" maxOccurs="unbounded"/>
    </xs:sequence>
  </xs:complexType>
  <xs:complexType name="columnType">
    <xs:simpleContent>
      <xs:extension base="xs:string">
        <xs:attribute name="id" type="xs:ID"/>
      </xs:extension>
    </xs:simpleContent>
  </xs:complexType>
  <xs:complexType name="rowType">
    <xs:choice maxOccurs="unbounded">
      <xs:element name="TD" type="xs:double"/>
```

```
      <xs:element name="TDI" type="xs:int"/>
      <xs:element name="TDT" type="xs:dateTime"/>
      <xs:element name="TDS" type="xs:string"/>
    </xs:choice>
  </xs:complexType>
</xs:schema>
```

And here is a list of positions, based on stcTab.xsd:

```
<?xml version="1.0" encoding="UTF-8"?>
  <!--
    The STC metadata to describe a list of ICRS positions
    -->
<STCTable xmlns="stcTab.xsd"
    xmlns:xsi="http://www.w3.org/2001/XMLSchema-instance"
    xsi:schemaLocation="stcTab.xsd
    http://hea-www.harvard.edu/~arots/nvometa/v1.30/stcTab.xsd">
  <CatalogEntryLocation
    xmlns="http://www.ivoa.net/xml/STC/stc-v1.30.xsd"
    xmlns:xlink="http://www.w3.org/1999/xlink"
    xsi:schemaLocation="http://www.ivoa.net/xml/STC/stc-v1.30.xsd
      http://www.ivoa.net/xml/STC/stc-v1.30.xsd">
      <!--
          Specify the ICRS frame, referenced to the barycenter
        -->
    <AstroCoordSystem id="ICRS">
      <SpaceFrame>
        <ICRS/>
        <BARYCENTER/>
        <SPHERICAL coord_naxes="2"/>
      </SpaceFrame>
    </AstroCoordSystem>
    <AstroCoords coord_system_id="ICRS">
      <Position2D unit="deg">
        <Name>RA,Dec</Name>
        <Value2>
          <C1 xsi:nil="true" idref="RA"/>
          <C2 xsi:nil="true" idref="Dec"/>
        </Value2>
      </Position2D>
    </AstroCoords>
  </CatalogEntryLocation>
  <!--
      And here is the table
    -->
  <Table>
    <Header><TH id="RA">RA</TH><TH id="Dec">Dec</TH></Header>
    <TR><TD>123.4</TD><TD>56.7</TD></TR>
    <TR><TD>133.4</TD><TD>46.7</TD></TR>
    <TR><TD>143.4</TD><TD>36.7</TD></TR>
    <TR><TD>153.4</TD><TD>26.7</TD></TR>
    <TR><TD>163.4</TD><TD>16.7</TD></TR>
  </Table>
</STCTable>
```

## B.12 Time Series

The stcTab.xsd schema from the previous example turns out to be very useful and versatile. Here is a time series based on that schema:

```
<?xml version="1.0" encoding="UTF-8"?>
<!--  The STC metadata to describe a list of ICRS positions  -->
  <STCTable xmlns="stcTab.xsd"
    xmlns:xsi="http://www.w3.org/2001/XMLSchema-instance"
    xsi:schemaLocation="stcTab.xsd
    http://hea-www.harvard.edu/~arots/nvometa/v1.30/stcTab.xsd">
  <ObsDataLocation xmlns="http://www.ivoa.net/xml/STC/stc-v1.30.xsd"
    xmlns:xlink="http://www.w3.org/1999/xlink"
    xsi:schemaLocation="http://www.ivoa.net/xml/STC/stc-v1.30.xsd
      http://www.ivoa.net/xml/STC/stc-v1.30.xsd">
   <!--
       Specify the ICRS frame, referenced to the barycenter
       and define a flux density scale
     -->
    <ObservatoryLocation xsi:nil="true" xlink:type="simple"
      xlink:href="ivo://STClib/Observatories#GEO"/>
    <ObservationLocation>
      <AstroCoordSystem id="TT-ICRS-FluxDen">
        <CoordFrame>
          <ScalarRefFrame id="FD">
            <Name>FluxDensity</Name>
            <Scale>1.0</Scale>
          </ScalarRefFrame>
          <CARTESIAN coord_naxes="1"/>
        </CoordFrame>
        <TimeFrame>
          <TimeScale>TT</TimeScale>
          <GEOCENTER/>
        </TimeFrame>
        <SpaceFrame>
          <ICRS/>
          <GEOCENTER/>
          <SPHERICAL coord_naxes="2"/>
        </SpaceFrame>
        <SpectralFrame>
          <GEOCENTER/>
        </SpectralFrame>
      </AstroCoordSystem>
        <!--
            Refer flux density and time to table columns
          -->
      <AstroCoords coord_system_id="TT-ICRS-FluxDen">
        <ScalarCoordinate frame_id="FD" unit="mJy">
          <Name>Flux Density</Name>
          <Value xsi:nil="true" idref="FDVal"/>
          <Error xsi:nil="true" idref="FDErr"/>
        </ScalarCoordinate>
        <Time unit="s">
          <Name>Time</Name>
```

```
          <TimeInstant>
            <TimeOffset xsi:nil="true" idref="Time"/>
            <ISOTime>2007-09-27T12:34:56</ISOTime>
          </TimeInstant>
        </Time>
        <Position2D unit="deg">
          <Name>RA,Dec</Name>
          <Value2>
            <C1>123.45</C1>
            <C2>67.89</C2>
          </Value2>
        </Position2D>
        <Spectral unit="GHz">
          <Value>5.0</Value>
          <Size>0.07</Size>
        </Spectral>
      </AstroCoords>
    </ObservationLocation>
  </ObsDataLocation>
  <Table>
      <!--
          Table header
        -->
    <Header>
      <TH id="Time">Time</TH>
      <TH id="FDVal">FluxDensity</TH>
      <TH id="FDErr">FD Error</TH>
    </Header>
      <!--
          Table data
        -->
     <TR>
      <TD>123.4</TD>
      <TD>56.7</TD>
      <TD>.98</TD>
    </TR>
    <TR>
      <TD>133.4</TD>
      <TD>46.7</TD>
      <TD>.98</TD>
    </TR>
    <TR>
      <TD>143.4</TD>
      <TD>36.7</TD>
      <TD>.98</TD>
    </TR>
    <TR>
      <TD>153.4</TD>
      <TD>26.7</TD>
    </TR>
    <TR>
      <TD>163.4</TD>
      <TD>16.7</TD>
    </TR>
  </Table>
</STCTable>
```

## *B.13 Event List*

A photon event list extracted from Chandra data can also be expressed on the basis of the stcTab.xsd schema from example B.11:

```
<?xml version="1.0" encoding="UTF-8"?>
<!--
    Chandra event list using the STC-based table format
  -->
  <STCTable xmlns="stcTab.xsd"
    xmlns:xsi="http://www.w3.org/2001/XMLSchema-instance"
    xsi:schemaLocation="stcTab.xsd
    http://hea-www.harvard.edu/~arots/nvometa/v1.30/stcTab.xsd">
  <ObsDataLocation xmlns="http://www.ivoa.net/xml/STC/stc-v1.30.xsd"
    xmlns:xlink="http://www.w3.org/1999/xlink"
    xsi:schemaLocation="http://www.ivoa.net/xml/STC/stc-v1.30.xsd
      http://www.ivoa.net/xml/STC/stc-v1.30.xsd">
    <!--
        Observatory location: Chandra orbit ephemeris file;
        this could be turned into a standard library item
      -->
    <ObservatoryLocation id="CXO">
      <AstroCoordSystem id="TT-TOPO-FK5-GEO">
        <TimeFrame>
          <TimeScale>TT</TimeScale>
          <TOPOCENTER/>
        </TimeFrame>
        <SpaceFrame>
          <FK5>
            <Equinox>J2000.0</Equinox>
          </FK5>
          <GEOCENTER/>
          <CARTESIAN coord_naxes="3"/>
        </SpaceFrame>
      </AstroCoordSystem>
      <!--
          The CoordFile option in AstrCoords allows one to specify
          coordinate values through pointers to individual columns
          in a FITS file
        -->
      <AstroCoords coord_system_id="TT-TOPO-FK5-GEO">
        <CoordFile>
          <FITSFile hdu_name="ORBITEPHEM" hdu_num="1">
        ftp://cda.cfa.harvard.edu/pub/arcftp/bary/ephem/orbitf233323500N001_eph1.fits
        </FITSFile>
          <FITSTime>
            <Value>Time</Value>
          </FITSTime>
          <FITSPosition>
            <Value>X,Y,Z</Value>
          </FITSPosition>
          <FITSVelocity>
            <Value>Vx,Vy,Vz</Value>
          </FITSVelocity>
        </CoordFile>
```

```
      </AstroCoords>
    </ObservatoryLocation>
    <!--
        The observation location becomes rather involved as it is
        expressed in three different coordinate frames
      -->
    <ObservationLocation>
      <!--
          TT-ICRS-ENERGY-TOPO is the World Coordinate System
          CHIP is the coordinate frame that describes
          chip-related items, chip and node id, but also PHA and
          grades; the origins are not particularly well described,
          but doing it rigorously gets too laborious without any
          real gain
        -->
      <!--
          DETECTOR defines the detector coordinate system
          in the focal plane
        -->
      <AstroCoordSystem id="TT-ICRS-ENERGY-TOPO">
        <TimeFrame>
          <TimeScale>TT</TimeScale>
          <TOPOCENTER/>
        </TimeFrame>
        <SpaceFrame id="RADecICRS">
          <ICRS/>
          <TOPOCENTER/>
          <SPHERICAL coord_naxes="2"/>
        </SpaceFrame>
        <SpectralFrame>
          <TOPOCENTER/>
        </SpectralFrame>
      </AstroCoordSystem>
      <AstroCoordSystem id="CHIP">
        <CoordFrame id="CCD_ID">
          <Name>ccd_id</Name>
          <ScalarRefFrame>
            <Scale>1.0</Scale>
          </ScalarRefFrame>
          <CARTESIAN coord_naxes="1"/>
        </CoordFrame>
        <CoordFrame id="NODE_ID">
          <Name>node_id</Name>
          <ScalarRefFrame>
            <Scale>1.0</Scale>
          </ScalarRefFrame>
          <CARTESIAN coord_naxes="1"/>
        </CoordFrame>
        <CoordFrame id="PHA">
          <Name>pha</Name>
          <ScalarRefFrame>
            <Scale>1.0</Scale>
          </ScalarRefFrame>
          <CARTESIAN coord_naxes="1"/>
        </CoordFrame>
        <CoordFrame id="PHA_RO">
          <Name>pha_ro</Name>
```

```
          <ScalarRefFrame>
            <Scale>1.0</Scale>
          </ScalarRefFrame>
          <CARTESIAN coord_naxes="1"/>
        </CoordFrame>
        <CoordFrame id="PI">
          <Name>pi</Name>
          <ScalarRefFrame>
            <Scale>1.0</Scale>
          </ScalarRefFrame>
          <CARTESIAN coord_naxes="1"/>
        </CoordFrame>
        <CoordFrame id="FLTGRADE">
          <Name>fltgrade</Name>
          <ScalarRefFrame>
            <Scale>1.0</Scale>
          </ScalarRefFrame>
          <CARTESIAN coord_naxes="1"/>
        </CoordFrame>
        <CoordFrame id="GRADE">
          <Name>grade</Name>
          <ScalarRefFrame>
            <Scale>1.0</Scale>
          </ScalarRefFrame>
          <CARTESIAN coord_naxes="1"/>
        </CoordFrame>
        <SpaceFrame id="CHIPFRAME">
          <Name>Chipframe</Name>
          <Cart2DRefFrame ref_frame_id="CXO">
            <Transform2>
              <C1>0.0</C1>
              <C2>0.0</C2>
            </Transform2>
          </Cart2DRefFrame>
          <TOPOCENTER/>
          <CARTESIAN coord_naxes="2" handedness="left"/>
        </SpaceFrame>
      </AstroCoordSystem>
      <AstroCoordSystem id="DETECTOR">
        <SpaceFrame id="DETECTORFRAME">
          <Name>Detectorframe</Name>
          <Cart2DRefFrame ref_frame_id="CXO">
            <Transform2>
              <C1>0.0</C1>
              <C2>0.0</C2>
            </Transform2>
          </Cart2DRefFrame>
          <TOPOCENTER/>
          <POLAR coord_naxes="2" handedness="left"/>
        </SpaceFrame>
      </AstroCoordSystem>
      <!--
          For some coordinates we provide resolution information,
          for others pointers to the table columns where the values
          may be found (through IDREF)
        -->
      <AstroCoords coord_system_id="TT-ICRS-ENERGY-TOPO">
```

```
        <Time unit="s">
          <Name>Time</Name>
          <TimeInstant>
            <TimeOffset xsi:nil="true" idref="Time"/>
            <MJDTime>50814.0</MJDTime>
          </TimeInstant>
          <Error>0.0001</Error>
          <Resolution>3.241</Resolution>
          <PixSize>3.241</PixSize>
        </Time>
        <Position2D unit="deg">
          <Name>RA,Dec</Name>
          <Resolution2Radius pos_angle_unit="arcsec">
            0.5
          </Resolution2Radius>
        </Position2D>
        <Spectral unit="keV">
          <Value xsi:nil="true" idref="energy"/>
          <Resolution>0.07</Resolution>
        </Spectral>
      </AstroCoords>
      <AstroCoords coord_system_id="CHIP">
        <ScalarCoordinate frame_id="CCD_ID" idref="ccd_id"
          xsi:nil="true"/>
      </AstroCoords>
      <AstroCoords coord_system_id="CHIP">
        <ScalarCoordinate frame_id="CCD_ID" idref="ccd_id"
          xsi:nil="true"/>
      </AstroCoords>
      <AstroCoords coord_system_id="CHIP">
        <ScalarCoordinate frame_id="NODE_ID" idref="node_id"
          xsi:nil="true"/>
      </AstroCoords>
      <AstroCoords coord_system_id="CHIP">
        <ScalarCoordinate frame_id="PHA" idref="pha" xsi:nil="true"/>
      </AstroCoords>
      <AstroCoords coord_system_id="CHIP">
        <ScalarCoordinate frame_id="PI" idref="pi" xsi:nil="true"/>
      </AstroCoords>
      <AstroCoords coord_system_id="CHIP">
        <ScalarCoordinate frame_id="FLTGRADE" idref="fltgrade"
          xsi:nil="true"/>
      </AstroCoords>
      <AstroCoords coord_system_id="CHIP">
        <ScalarCoordinate frame_id="GRADE" idref="grade"
          xsi:nil="true"/>
      </AstroCoords>
      <!--
          The volume in World Coordinate space represented by
          this event list
        -->
      <AstroCoordArea coord_system_id="TT-ICRS-ENERGY-TOPO">
        <!--
            This is the full time interval of the observation,
            although this sample table only contains the first
            10 events
          -->
```

```
      <TimeInterval>
        <StartTime>
          <ISOTime>2005-05-26T03:47:25</ISOTime>
        </StartTime>
        <StopTime>
          <ISOTime>2005-05-26T07:29:06</ISOTime>
        </StopTime>
      </TimeInterval>
      <!--
          This polygon is a placeholder for the true FOV
        -->
      <Polygon>
        <Vertex>
          <Position>
            <C1>148.69</C1>
            <C2>68.965</C2>
          </Position>
        </Vertex>
        <Vertex>
          <Position>
            <C1>148.69</C1>
            <C2>69.165</C2>
          </Position>
        </Vertex>
        <Vertex>
          <Position>
            <C1>149.09</C1>
            <C2>69.165</C2>
          </Position>
        </Vertex>
        <Vertex>
          <Position>
            <C1>149.09</C1>
            <C2>68.965</C2>
          </Position>
        </Vertex>
      </Polygon>
      <SpectralInterval unit="keV">
        <LoLimit>0.2</LoLimit>
        <HiLimit>10.0</HiLimit>
      </SpectralInterval>
    </AstroCoordArea>
  </ObservationLocation>
  <!--
      In addition to the coordinate systems above, there are
      4 pixel coordinate frames
    -->
  <PixelSpace>
    <!--
        Chip pixel coordinates, tiled detector pixel coordinates,
        detector (focal plane) pixel coordinates, and pixels on
        the sky; except for the tiled coordinates, they are all
        tied to one of the coordinate frames defined in the
        AstroCoordSystem
      -->
    <PixelCoordSystem id="EventPix">
      <PixelCoordFrame id="ChipPix" axis1_order="1" axis2_order="2"
```

```
          ref_frame_id="CHIPFRAME">
          <Cart2DRefFrame>
            <Transform2 unit="mm">
              <C1>0.023987</C1>
              <C2>0.023987</C2>
            </Transform2>
          </Cart2DRefFrame>
          <CoordRefPos>
            <Vector2DCoordinate unit="mm">
              <Value2>
                <C1>0.0</C1>
                <C2>0.0</C2>
              </Value2>
            </Vector2DCoordinate>
          </CoordRefPos>
          <CARTESIAN coord_naxes="2"/>
          <ReferencePixel>
            <Pixel2D>
              <Name1>ChipX</Name1>
              <Name2>ChipY</Name2>
              <Value2>
                <C1>0.5</C1>
                <C2>0.5</C2>
              </Value2>
            </Pixel2D>
          </ReferencePixel>
        </PixelCoordFrame>
        <PixelCoordFrame id="TdetPix" axis1_order="1" axis2_order="2"
          ref_frame_id="CHIPFRAME">
          <Cart2DRefFrame>
            <Transform2 unit="mm">
              <C1>0.023987</C1>
              <C2>0.023987</C2>
            </Transform2>
          </Cart2DRefFrame>
          <CoordRefPos>
            <Vector2DCoordinate unit="deg">
              <Value2>
                <C1>0.0</C1>
                <C2>0.0</C2>
              </Value2>
            </Vector2DCoordinate>
          </CoordRefPos>
          <CARTESIAN coord_naxes="2"/>
          <ReferencePixel>
            <Pixel2D>
              <Name1>DetX</Name1>
              <Name2>DetY</Name2>
              <Value2>
                <C1>4096.5</C1>
                <C2>4096.5</C2>
              </Value2>
            </Pixel2D>
          </ReferencePixel>
        </PixelCoordFrame>
        <PixelCoordFrame id="DetPix" axis1_order="1" axis2_order="2"
          ref_frame_id="DETECTORFRAME">
```

```
        <Cart2DRefFrame projection="TAN">
          <Transform2 unit="deg">
            <C1>0.000136667</C1>
            <C2>0.000136667</C2>
          </Transform2>
        </Cart2DRefFrame>
        <CoordRefPos>
          <Vector2DCoordinate unit="deg">
            <Value2>
              <C1>0.0</C1>
              <C2>0.0</C2>
            </Value2>
          </Vector2DCoordinate>
        </CoordRefPos>
        <CARTESIAN coord_naxes="2"/>
        <ReferencePixel>
          <Pixel2D>
            <Name1>DetX</Name1>
            <Name2>DetY</Name2>
            <Value2>
              <C1>4096.5</C1>
              <C2>4096.5</C2>
            </Value2>
          </Pixel2D>
        </ReferencePixel>
      </PixelCoordFrame>
      <PixelCoordFrame id="SkyPix" axis1_order="1" axis2_order="2"
        ref_frame_id="RADecICRS">
        <Cart2DRefFrame projection="TAN">
          <Transform2 unit="deg">
            <C1>-0.000136667</C1>
            <C2>0.000136667</C2>
          </Transform2>
        </Cart2DRefFrame>
        <CoordRefPos>
          <Vector2DCoordinate unit="deg">
            <Value2>
              <C1>148.88172623167</C1>
              <C2>69.055951213634</C2>
            </Value2>
          </Vector2DCoordinate>
        </CoordRefPos>
        <CARTESIAN coord_naxes="2"/>
        <ReferencePixel>
          <Pixel2D>
            <Name1>X</Name1>
            <Name2>Y</Name2>
            <Value2>
              <C1>4096.5</C1>
              <C2>4096.5</C2>
            </Value2>
          </Pixel2D>
        </ReferencePixel>
      </PixelCoordFrame>
    </PixelCoordSystem>
    <!--
        The pixel coordinates are referenced to specific table
```

```
        Columns
      -->
    <PixelCoords coord_system_id="EventPix">
      <Pixel2D frame_id="ChipPix">
        <Name1>ChipX</Name1>
        <Name2>ChipY</Name2>
        <Value2>
          <C1 xsi:nil="true" idref="chipx"/>
          <C2 xsi:nil="true" idref="chipy"/>
        </Value2>
      </Pixel2D>
    </PixelCoords>
    <PixelCoords coord_system_id="EventPix">
      <Pixel2D frame_id="TdetPix">
        <Name1>TdetX</Name1>
        <Name2>TdetY</Name2>
        <Value2>
          <C1 xsi:nil="true" idref="tdetx"/>
          <C2 xsi:nil="true" idref="tdety"/>
        </Value2>
      </Pixel2D>
    </PixelCoords>
    <PixelCoords coord_system_id="EventPix">
      <Pixel2D frame_id="DetPix">
        <Name1>DetX</Name1>
        <Name2>DetY</Name2>
        <Value2>
          <C1 xsi:nil="true" idref="detx"/>
          <C2 xsi:nil="true" idref="dety"/>
        </Value2>
      </Pixel2D>
    </PixelCoords>
    <PixelCoords coord_system_id="EventPix">
      <Pixel2D frame_id="SkyPix">
        <Name1>X</Name1>
        <Name2>Y</Name2>
        <Value2>
          <C1 xsi:nil="true" idref="x"/>
          <C2 xsi:nil="true" idref="y"/>
        </Value2>
      </Pixel2D>
    </PixelCoords>
    <!--
        The PixelCoordAreas provide the ranges for all pixel
        Coordinates
      -->
    <PixelCoordArea coord_system_id="EventPix">
      <PixelCoord2VecInterval frame_id="ChipPix">
        <LoLimit2Vec>
          <C1>1</C1>
          <C2>1</C2>
        </LoLimit2Vec>
        <HiLimit2Vec>
          <C1>1024</C1>
          <C2>1024</C2>
        </HiLimit2Vec>
      </PixelCoord2VecInterval>
```

```
      <PixelCoord2VecInterval frame_id="DetPix">
        <LoLimit2Vec>
          <C1>1</C1>
          <C2>1</C2>
        </LoLimit2Vec>
        <HiLimit2Vec>
          <C1>8092</C1>
          <C2>8092</C2>
        </HiLimit2Vec>
      </PixelCoord2VecInterval>
      <PixelCoord2VecInterval frame_id="SkyPix">
        <LoLimit2Vec>
          <C1>1</C1>
          <C2>1</C2>
        </LoLimit2Vec>
        <HiLimit2Vec>
          <C1>8092</C1>
          <C2>8092</C2>
        </HiLimit2Vec>
      </PixelCoord2VecInterval>
    </PixelCoordArea>
  </PixelSpace>
</ObsDataLocation>
<!--
    The table with the photon events
  -->
<Table>
  <!--
      First, the table header
    -->
  <Header>
    <TH id="Time">Time</TH>
    <TH id="ccd_id">CCD</TH>
    <TH id="node_id">Node</TH>
    <TH id="chipx">Chip X Coord</TH>
    <TH id="chipy">Chip Y Coord</TH>
    <TH id="tdetx">Tiled det X Coord</TH>
    <TH id="tdety">Tiled det Y Coord</TH>
    <TH id="detx">Detector X Coord</TH>
    <TH id="dety">Detector Y Coord</TH>
    <TH id="x">Sky X Coord</TH>
    <TH id="y">Sky Y Coord</TH>
    <TH id="pha">Pulse height</TH>
    <TH id="pha_ro">Read-out pulse height</TH>
    <TH id="energy">Nominal energy</TH>
    <TH id="pi">Pulse-invariant energy</TH>
    <TH id="fltgrade">Flight grade</TH>
    <TH id="grade">Grade</TH>
  </Header>
  <!--
      Then, any number of table rows
    -->
  <TR>
    <TD>2.334684078389896E+08</TD>
    <TDI>2</TDI>
    <TDI>0</TDI>
    <TDI>173</TDI>
```

```
      <TDI> 99</TDI>
      <TDI>3160</TDI>
      <TDI>3913</TDI>
      <TD>3.1353450E+03</TD>
      <TD>2.3983059E+03</TD>
      <TD>5.8374014E+03</TD>
      <TD> 3.2439470E+03</TD>
      <TDI>2160</TDI>
      <TDI>2169</TDI>
      <TD> 8.2779902E+03</TD>
      <TDI> 567</TDI>
      <TDI>0</TDI>
      <TDI>0</TDI>
    </TR>
    <TR>
      <TD>2.334684078389896E+08</TD>
      <TDI>2</TDI>
      <TDI>3</TDI>
      <TDI>903</TDI>
      <TDI>254</TDI>
      <TDI>3315</TDI>
      <TDI>3183</TDI>
      <TD>3.2914187E+03</TD>
      <TD>3.1255881E+03</TD>
      <TD>5.1011631E+03</TD>
      <TD> 3.3500186E+03</TD>
      <TDI>3776</TDI>
      <TDI>3736</TDI>
      <TD> 1.4497490E+04</TD>
      <TDI> 993</TDI>
      <TDI>0</TDI>
      <TDI>0</TDI>
    </TR>
    <TR>
      <TD>2.334684078389896E+08</TD>
      <TDI>2</TDI>
      <TDI>2</TDI>
      <TDI>544</TDI>
      <TDI>304</TDI>
      <TDI>3365</TDI>
      <TDI>3542</TDI>
      <TD>3.3413210E+03</TD>
      <TD>2.7677412E+03</TD>
      <TD>5.4547695E+03</TD>
      <TD> 3.4242280E+03</TD>
      <TDI>2301</TDI>
      <TDI>2217</TDI>
      <TD> 8.6658369E+03</TD>
      <TDI> 594</TDI>
      <TDI>0</TDI>
      <TDI>0</TDI>
    </TR>
    <TR>
      <TD>2.334684078389896E+08</TD>
      <TDI>2</TDI>
      <TDI>1</TDI>
      <TDI>349</TDI>
```

```
      <TDI>575</TDI>
      <TDI>3636</TDI>
      <TDI>3737</TDI>
      <TD>3.6109248E+03</TD>
      <TD>2.5733503E+03</TD>
      <TD>5.6303062E+03</TD>
      <TD> 3.7064707E+03</TD>
      <TDI>3086</TDI>
      <TDI>2927</TDI>
      <TD> 1.1672590E+04</TD>
      <TDI> 800</TDI>
      <TDI>0</TDI>
      <TDI>0</TDI>
    </TR>
    <TR>
      <TD>2.334684078389896E+08</TD>
      <TDI>2</TDI>
      <TDI>3</TDI>
      <TDI>909</TDI>
      <TDI>856</TDI>
      <TDI>3917</TDI>
      <TDI>3177</TDI>
      <TD>3.8928091E+03</TD>
      <TD>3.1323667E+03</TD>
      <TD>5.0533545E+03</TD>
      <TD> 3.9495439E+03</TD>
      <TDI> 451</TDI>
      <TDI> 370</TDI>
      <TD> 1.7327786E+03</TD>
      <TDI> 119</TDI>
      <TDI>0</TDI>
      <TDI>0</TDI>
    </TR>
    <TR>
      <TD>2.334684078800296E+08</TD>
      <TDI>7</TDI>
      <TDI>1</TDI>
      <TDI>344</TDI>
      <TDI>117</TDI>
      <TDI>4261</TDI>
      <TDI>1819</TDI>
      <TD>4.2339629E+03</TD>
      <TD>4.4894370E+03</TD>
      <TD>3.6761646E+03</TD>
      <TD> 4.1972803E+03</TD>
      <TDI>2614</TDI>
      <TDI>2601</TDI>
      <TD> 1.2488189E+04</TD>
      <TDI> 856</TDI>
      <TDI>8</TDI>
      <TDI>3</TDI>
    </TR>
    <TR>
      <TD>2.334684078800296E+08</TD>
      <TDI>7</TDI>
      <TDI>1</TDI>
      <TDI>416</TDI>
```

```
    <TDI>156</TDI>
    <TDI>4333</TDI>
    <TDI>1858</TDI>
    <TD>4.3064810E+03</TD>
    <TD>4.4505332E+03</TD>
    <TD>3.7100283E+03</TD>
    <TD> 4.2722847E+03</TD>
    <TDI>2595</TDI>
    <TDI>2578</TDI>
    <TD> 1.2342618E+04</TD>
    <TDI> 846</TDI>
    <TDI> 11</TDI>
    <TDI>6</TDI>
  </TR>
  <TR>
    <TD>2.334684078800296E+08</TD>
    <TDI>7</TDI>
    <TDI>2</TDI>
    <TDI>707</TDI>
    <TDI>357</TDI>
    <TDI>4624</TDI>
    <TDI>2059</TDI>
    <TD>4.5971719E+03</TD>
    <TD>4.2501328E+03</TD>
    <TD>3.8901213E+03</TD>
    <TD> 4.5759751E+03</TD>
    <TDI>3391</TDI>
    <TDI>3381</TDI>
    <TD> 1.6004620E+04</TD>
    <TDI>1024</TDI>
    <TDI> 64</TDI>
    <TDI>2</TDI>
  </TR>
  <TR>
    <TD>2.334684078800296E+08</TD>
    <TDI>7</TDI>
    <TDI>1</TDI>
    <TDI>378</TDI>
    <TDI>373</TDI>
    <TDI>4295</TDI>
    <TDI>2075</TDI>
    <TD>4.2681963E+03</TD>
    <TD>4.2340381E+03</TD>
    <TD>3.9286318E+03</TD>
    <TD> 4.2488652E+03</TD>
    <TDI>1949</TDI>
    <TDI>1935</TDI>
    <TD> 9.3225713E+03</TD>
    <TDI> 639</TDI>
    <TDI>104</TDI>
    <TDI>6</TDI>
  </TR>
  <TR>
    <TD>2.334684078800296E+08</TD>
    <TDI>7</TDI>
    <TDI>1</TDI>
    <TDI>510</TDI>
```

```
      <TDI>394</TDI>
      <TDI>4427</TDI>
      <TDI>2096</TDI>
      <TD>4.4002852E+03</TD>
      <TD>4.2134824E+03</TD>
      <TD>3.9401243E+03</TD>
      <TD> 4.3820493E+03</TD>
      <TDI>2737</TDI>
      <TDI>2725</TDI>
      <TD> 1.2950969E+04</TD>
      <TDI> 888</TDI>
      <TDI>104</TDI>
      <TDI>6</TDI>
    </TR>
  </Table>
</STCTable>
```

# Appendix C: Standard Libraries

## *C.1 Standard Coordinate Systems*

We will maintain a library of common and standard coordinate system descriptions that can be referred to in STC-compliant documents through Xlink. The library has IVOA identifier `ivo://STClib/CoordSys` and individual systems may be referred to through a Xpath identifier to the particular element, such as: `ivo://STClib/CoordSys#TT-ICRS-TOPO`
The recommended referencing syntax is:

```
<AstroCoordSystem id="TT-ICRS-TOPO" xlink:type="simple"
     xlink:href="ivo://STClib/CoordSys#TT-ICRS-TOPO"/>
```

The current list of available coordinate systems in this library is provided in Table 6.

Table 6. Library of Standard Coordinate Systems

| Xlink Tag | Time | | Space | | | Spectral | | Redshift | |
|---|---|---|---|---|---|---|---|---|---|
| ivo://STClib/CoordSys# | Time Scale | Ref Pos | Ref Frame | Ref Pos | Flavor | Unit | Ref Pos | Redshift/ Doppler | Ref Pos |
| TT-ICRS-TOPO | TT | TOPO | ICRS | TOPO | Sphere | | | | |
| TT-FK5-TOPO | TT | TOPO | FK5 | TOPO | Sphere | | | | |
| UTC-ICRS-TOPO | UTC | TOPO | ICRS | TOPO | Sphere | | | | |
| UTC-FK5-TOPO | UTC | TOPO | FK5 | TOPO | Sphere | | | | |
| TT-ICRS-GEO | TT | GEO | ICRS | GEO | Sphere | | | | |
| TT-FK5-GEO | TT | GEO | FK5 | GEO | Sphere | | | | |
| UTC-ICRS-GEO | UTC | GEO | ICRS | GEO | Sphere | | | | |
| UTC-FK5-GEO | UTC | GEO | FK5 | GEO | Sphere | | | | |
| TDB-ICRS-BARY | TDB | BARY | ICRS | BARY | Sphere | | | | |
| TDB-FK5-BARY | TDB | BARY | FK5 | BARY | Sphere | | | | |
| TT-ICRS-BARY | TT | BARY | ICRS | BARY | Sphere | | | | |
| TDB-ECLIPTIC-BARY | TDB | BARY | ECLIPTIC | BARY | Sphere | | | | |
| TDB-UNIT-ECLIPTIC-BARY | TDB | BARY | ECLIPTIC | BARY | Unit Sphere | | | | |
| UTC-HPC-TOPO | UTC | TOPO | HPC | TOPO | Cart 2 | | | | |
| UTC-HPR-TOPO | UTC | TOPO | HPR | TOPO | Polar | | | | |
| UTC-HGS-TOPO | UTC | TOPO | HGS | TOPO | Sphere | | | | |
| UTC-HGC-TOPO | UTC | TOPO | HGC | TOPO | Sphere | | | | |
| TT-ICRS-HZ-TOPO | TT | TOPO | ICRS | TOPO | Sphere | Hz | TOPO | | |
| TT-ICRS-OPT-BARY-TOPO | TT | TOPO | ICRS | TOPO | Sphere | | | DoppOptic | BARY |
| TT-ICRS-RADIO-LSR-TOPO | TT | TOPO | ICRS | TOPO | Sphere | | | DoppRadio | LSR |

Anyone can define one's own set of pre-defined coordinate systems and Xlink to them, using a valid URL, though one should be careful not to create a chaotic situation where clients may be utterly confused by hundreds of "standard" coordinate systems that are subtly different in ways that may not be immediately apparent.
Sample file:

```
<?xml version="1.0" encoding="UTF-8"?>
  <!--
    This document contains the library of standard coordinate system
    descriptions.
    Usage:
      <AstroCoordSystem id="{your choice}" xlink:type="simple"
        xlink:href="ivo://STClib/CoordSys#{element name}"/>
      where {element name} is the name of the element in this file and
      {your choice} is, in principle, any valid ID, but we recommend
      using the element name
    Interpretation:
      One should interpret the xlink reference as if one inserted the
      body of the element that is pointed to in xlink:href in the
      referencing element, if its body is empty, and omitted the two
      xlink: attributes.
  -->
<AstroCoordSystemLib xsi:type="xsi:anyType"
xmlns="http://www.ivoa.net/xml/STC/stc-v1.30.xsd"
xmlns:xlink="http://www.w3.org/1999/xlink"
xmlns:xsi="http://www.w3.org/2001/XMLSchema-instance"
xsi:schemaLocation="http://www.ivoa.net/xml/STC/stc-v1.30.xsd
http://www.ivoa.net/xml/STC/stc-v1.30.xsd">
  <TT-GEOD-TOPO xsi:type="astroCoordSystemType" id="TT-GEOD-TOPO">
    <TimeFrame>
      <TimeScale>TT</TimeScale>
      <TOPOCENTER/>
    </TimeFrame>
    <SpaceFrame>
      <GEO_D/>
      <TOPOCENTER/>
      <SPHERICAL coord_naxes="3"/>
    </SpaceFrame>
  </TT-GEOD-TOPO>
</AstroCoordSystemLib>
```

## C.2 Observatory Locations

We also maintain a library of observatory locations which may be used in a similar fashion at `ivo://STClib/Observatories` with recommended syntax:

```
<ObservatoryLocation id="KPNO" xlink:type="simple"
      xlink:href="ivo://STClib/Observatories#KPNO"/>
```

The present list is provided in Table 7.

Table 7. Library of Observatory Locations

| Xlink Tag | Longitude (deg) | Latitude (deg) | Altitude (m) | Other |
|---|---|---|---|---|
| KPNO | 248.4056 | 31.9586 | 2158 | |
| Arecibo | 293.2469 | 18.3435 | 497 | |

The same comments that were made in the previous section apply here as well. Sample file:

```
<?xml version="1.0" encoding="UTF-8"?>
  <!--
    This document contains the library of standard coordinate system
    descriptions.
    Usage:
      <ObservatoryLocation id="{your choice}" xlink:type="simple"
        xlink:href="ivo://STClib/Observatories#{element name}"/>
      where {element name} is the name of the element in this file and
      {your choice} is, in principle, any valid ID, but we recommend
      using the element name
    Interpretation:
      One should interpret the xlink reference as if one inserted the
      body of the element that is pointed to in xlink:href in the
      referencing element, if its body is empty, and omitted the two
      xlink: attributes.
  -->
<ObservatoryLib xsi:type="xsi:anyType"
xmlns="http://www.ivoa.net/xml/STC/stc-v1.30.xsd"
xmlns:xlink="http://www.w3.org/1999/xlink"
xmlns:xsi="http://www.w3.org/2001/XMLSchema-instance"
xsi:schemaLocation="http://www.ivoa.net/xml/STC/stc-v1.30.xsd
http://www.ivoa.net/xml/STC/stc-v1.30.xsd">
  <Arecibo xsi:type="observatoryLocationType" id="Arecibo">
    <AstroCoordSystem id="TT-GEOD-TOPO">
      <TimeFrame>
        <TimeScale>TT</TimeScale>
        <TOPOCENTER/>
      </TimeFrame>
      <SpaceFrame>
        <GEO_D/>
        <TOPOCENTER/>
        <SPHERICAL coord_naxes="3"/>
```

```
      </SpaceFrame>
    </AstroCoordSystem>
    <AstroCoords coord_system_id="TT-GEOD-TOPO">
      <Position3D>
        <Value3>
          <C1 pos_unit="deg">293.2469</C1>
          <C2 pos_unit="deg">18.3435</C2>
          <C3 pos_unit="m">497</C3>
        </Value3>
      </Position3D>
    </AstroCoords>
  </Arecibo>
  <KPNO xsi:type="observatoryLocationType" id="KPNO">
    <AstroCoordSystem id="TT-GEOD-TOPO">
      <TimeFrame>
        <TimeScale>TT</TimeScale>
        <TOPOCENTER/>
      </TimeFrame>
      <SpaceFrame>
        <GEO_D/>
        <TOPOCENTER/>
        <SPHERICAL coord_naxes="3"/>
      </SpaceFrame>
    </AstroCoordSystem>
    <AstroCoords coord_system_id="TT-GEOD-TOPO">
      <Position3D>
        <Value3>
          <C1 pos_unit="deg">248.4056</C1>
          <C2 pos_unit="deg">31.9586</C2>
          <C3 pos_unit="m">2158</C3>
        </Value3>
      </Position3D>
    </AstroCoords>
  </KPNO>
</ObservatoryLib>
```

## Appendix D: Changes from Previous Versions

From V0.5 to V0.51:

- Added STCResourceProfile example in Appendix B.

From V0.51 to V0.6:

- Finished schemas resulting in some tweaks to the design.

From V0.6 to V1.0:

- Added solar and planetary reference frames.
- Enumerated spatial reference positions, reference frames, and time scales in tables.
- Added mechanism to reference coordinate values in FITS files.
- Added examples
- Release as Working Draft

From V1.0 to V1.10:

- Added Pixel Space
- Allow *nillable* (specifically unknown) elements; for some an `UNKNOWN` element has been added
- Allow definite values as well as ranges (note that this means there may be 0, 1, or 2 Error, Resolution, Size, PixSize elements: 0 means N/A, 1 is a definite value, 2 define a range)
- Specify rest frequencies (through the Spectral axis, in the presence of a Redshift axis)
- General clarifications
- AllSky was added to Region as a convenient mnemonic for "all"
- Added rest frequency specification

From V1.10 to V1.20:

- Distinguished LSRK and LSRD; replaced SUPER_GALACTIC_CENTER by LOCAL_GROUP_CENTER
- Added Box shape to Region
- Require polygon sides to be less than 180°; added explanation on defining inside and outside
- Added LinearSTC implementation
- Subtle changes to the schemata to force validation under Xerces-J2; this did not change anything but made the schemata a little uglier
- Changed units "micron" and "A" to "um" and "Angstrom", respectively, consistent with the FITS standard
- Added timescale TEB, with references
- Added warning on use of ISOTime and restricted its use explicitly
- Added an explicit definition of Position Angle; in the XML implementation it is now an element, rather than an attribute (see, e.g., Ellipse)

- Added a reference spheroid to GEO_D, changed GEO to GEO_C
- Added an explicit discussion of planetographic and planetocentric coordinate systems, with references, as well as a discussion of right-handed and left-handed systems
- Removed GEOGRAPHIC CoordFlavor, replacing it by 3-dimensional SPHERICAL, and added POLAR

From V1.20 to V1.21:

- Corrected the catalog entry example in B.2, adding proper motion columns
- General clarifications
- Changed name of LinearSTC to STC-S; XML implementation is STC-X
- Reformatted to new template
- Promoted to Proposed Recommendation

From V1.21 to V1.30 (including schema changes)

- *Name* element will be optional in *Coordinate* element
- The required *Frames* in *AstroCoordSystem* will only be Time and Space
- *TimeScale* will be optional in *astronTimeType*
- *AstroCoordArea* will be optional in *ObservationLocation*
- 2-D error and size circles will get their own definition, to simplify things
- 3-D spherical coordinates will allow units “deg deg Mpc”
- An optional attribute *epoch* has been added to *Position* and *PositionInterval* (including *Region*)
- An optional attribute *ID* has been added to the leaves of the coordinate elements, to allow reverse referencing
- *Ellipse* is derived directly from *Shape*, rather than from *Circle*, to maintain proper inheritance while retaining a simple *Circle* specification
- The Region Shape *Constraint* has been renamed *Halfspace*
- *ObservatoryLocation* is allowed to be an *IDREF*, to facilitate reuse
- The components of *coordsType* are allowed optional *unit*s
- A new general referencing mechanism is introduced, based on Xlink, involving a baseType and attribute group, and an xlink schema is added
- A number of descriptions were elucidated
- Examples in Appendix B were updated
- The UML diagrams in Appendix A were updated
- The baseType has been given a UCD attribute, inherited by many types
- The definitions of generic coordinates and coordinate frames have been improved and links are properly defined
- Most derivations in the schema are now by extension, rather than restriction; Pixel and Astro both are derived by extension from their Generic counterparts
- Coordinate transformation options are added to all types of coordinates, not just spherical; projections have been added; the mechanisms are consistent with FITS WCS

From V1.30 to V1.31 (including schema changes)
- Orbital parameter specification completed and documented
- More examples in Appendix B

From V1.31 to V1.33 (including schema changes)
- Added references, made editorial corrections and clarifications
- Added support for specifying handedness of coordinate systems
- Added support for identifying ID/IDREF pairs in documents
- Added GPS time scale
- Added one more example to Appendix B